\pdfoutput=1
\documentclass[12pt,a4paper]{article}
\usepackage{ifthen} %
\newboolean{pdflatex}
\setboolean{pdflatex}{true} %
\usepackage{multirow} %
\newboolean{articletitles}
\setboolean{articletitles}{true} %

\newboolean{uprightparticles}
\setboolean{uprightparticles}{false} %

\def\Vcb{\ensuremath{\lvert V_{\cquark\bquark}\rvert}\xspace}
\newcommand{\etaEW}{\ensuremath{\eta_{\rm EW}}\xspace}
\newcommand{\Fnorm}{\ensuremath{h_{A_1}(1)}\xspace}
\newcommand{\Rone}{\ensuremath{R_1(1)}\xspace}
\newcommand{\Rtwo}{\ensuremath{R_2(1)}\xspace}
\newcommand{\Gnorm}{\ensuremath{\mathcal{G}(0)}\xspace}
\newcommand{\diff}{\ensuremath{{\rm d}}\xspace}
\newcommand{\ext}{\aunit{(ext)}\xspace}

\newcommand{\BuorBd}{\ensuremath{\B^{+/0}}\xspace}
\newcommand{\BorBs}{\ensuremath{\B^{0}_{(\squark)}}\xspace}
\newcommand{\DorDs}{\ensuremath{\D_{(\squark)}}\xspace}
\newcommand{\DSorDsS}{\ensuremath{\D_{(\squark)}^{*}}\xspace}
\newcommand{\DorDsm}{\ensuremath{\D_{(\squark)}^-}\xspace}
\newcommand{\DSorDsSm}{\ensuremath{\D_{(\squark)}^{*-}}\xspace}
\newcommand{\DorDsp}{\ensuremath{\D_{(\squark)}^+}\xspace}

\newcommand{\DS}{\ensuremath{\D^{*}}\xspace}

\newcommand{\DSm}{\ensuremath{\D^{*-}}\xspace}
\newcommand{\DsSm}{\ensuremath{\D_{\squark}^{*-}}\xspace}
\newcommand{\DSp}{\ensuremath{\D^{*+}}\xspace}

\newcommand{\DsorDsS}{\ensuremath{\D^{\scalebox{0.4}{(}*\scalebox{0.4}{)}}_{\squark}}\xspace}
\newcommand{\DorDS}{\ensuremath{\D^{\scalebox{0.4}{(}*\scalebox{0.4}{)}}}\xspace}
\newcommand{\DsorDsSm}{\ensuremath{\D^{\scalebox{0.4}{(}*\scalebox{0.4}{)}-}_{\squark}}\xspace}
\newcommand{\DorDSm}{\ensuremath{\D^{\scalebox{0.4}{(}*\scalebox{0.4}{)}-}}\xspace}

\newcommand{\DorDSp}{\ensuremath{\D^{\scalebox{0.4}{(}*\scalebox{0.4}{)}+}}\xspace}

\newcommand{\DorDsorSm}{\ensuremath{\D^{\scalebox{0.4}{(}*\scalebox{0.4}{)}-}_{(\squark)}}\xspace}

\newcommand{\bsDs}{\mbox{\ensuremath{\Bs\to\Dsm\mup\neum}}\xspace}
\newcommand{\bsDsS}{\mbox{\ensuremath{\Bs\to\DsSm\mup\neum}}\xspace}
\newcommand{\bdD}{\mbox{\ensuremath{\Bd\to\Dm\mup\neum}}\xspace}
\newcommand{\bdDS}{\mbox{\ensuremath{\Bd\to\DSm\mup\neum}}\xspace}

\newcommand{\pperp}{\ensuremath{p_{\perp}}\xspace}
\newcommand{\pperpDs}{\ensuremath{p_{\perp}\!(\Dsm)}\xspace}

\newcommand{\pperpD}{\ensuremath{p_{\perp}\!(\Dm)}\xspace}
\newcommand{\pperpDorDs}{\ensuremath{p_{\perp}\!(\DorDsm)}\xspace}
\newcommand{\mcorr}{\ensuremath{m_{\text{corr}}}\xspace}

\newcommand{\RD}{\ensuremath{\mathcal{R}}\xspace}
\newcommand{\RDS}{\ensuremath{\mathcal{R}^*}\xspace}
\newcommand{\RDorDS}{\ensuremath{\mathcal{R}^{\scalebox{0.4}{(}*\scalebox{0.4}{)}}}\xspace}

\newcommand{\nDorDS}{\ensuremath{\mathcal{N}^{\scalebox{0.4}{(}*\scalebox{0.4}{)}}}\xspace}

\newcommand{\kDorDS}{\ensuremath{\mathcal{K}^{\scalebox{0.4}{(}*\scalebox{0.4}{)}}}\xspace}

\newcommand{\nsigDorDS}{\ensuremath{N_{\text{sig}}^{\scalebox{0.4}{(}*\scalebox{0.4}{)}}}\xspace}
\newcommand{\nrefD}{\ensuremath{N_{\text{ref}}}\xspace}
\newcommand{\nrefDS}{\ensuremath{N_{\text{ref}}^{*}}\xspace}
\newcommand{\nrefDorDS}{\ensuremath{N_{\text{ref}}^{\scalebox{0.4}{(}*\scalebox{0.4}{)}}}\xspace}

\newcommand{\effratioDorDS}{\ensuremath{\xi^{\scalebox{0.4}{(}*\scalebox{0.4}{)}}}\xspace}

\newcommand{\rsq}{\ensuremath{\rho^2}\xspace}
\newcommand{\rsqD}{\ensuremath{\rho^2(\Dm)}\xspace}
\newcommand{\rsqDS}{\ensuremath{\rho^2(\DSm)}\xspace}
\newcommand{\rsqDs}{\ensuremath{\rho^2(\Dsm)}\xspace}
\newcommand{\rsqDsS}{\ensuremath{\rho^2(\DsSm)}\xspace}

\newcommand{\rsqDsResult}{\ensuremath{1.27}\xspace}%
\newcommand{\rsqDsStatErr}{\ensuremath{0.05}\xspace}%

\newcommand{\rsqDsExtErr}{\ensuremath{0.00}\xspace}

\newcommand{\GnormCLNResult}{\ensuremath{1.102}\xspace}%
\newcommand{\GnormCLNStatErr}{\ensuremath{0.034}\xspace}%

\newcommand{\GnormCLNExtErr}{\ensuremath{0.004}\xspace}

\newcommand{\rsqDsSResult}{\ensuremath{1.23}\xspace}%
\newcommand{\rsqDsSStatErr}{\ensuremath{0.17}\xspace}%

\newcommand{\rsqDsSExtErr}{\ensuremath{0.01}\xspace}

\newcommand{\RoneResult}{\ensuremath{1.34}\xspace}%
\newcommand{\RoneStatErr}{\ensuremath{0.25}\xspace}%

\newcommand{\RoneExtErr}{\ensuremath{0.02}\xspace}

\newcommand{\RtwoResult}{\ensuremath{0.83}\xspace}%
\newcommand{\RtwoStatErr}{\ensuremath{0.16}\xspace}%

\newcommand{\RtwoExtErr}{\ensuremath{0.01}\xspace}

\newcommand{\doneResult}{\ensuremath{-0.017}\xspace}%
\newcommand{\doneStatErr}{\ensuremath{0.007}\xspace}%

\newcommand{\doneExtErr}{\ensuremath{0.001}\xspace}

\newcommand{\dtwoResult}{\ensuremath{-0.26}\xspace}%
\newcommand{\dtwoStatErr}{\ensuremath{0.05}\xspace}%

\newcommand{\dtwoExtErr}{\ensuremath{0.00}\xspace}

\newcommand{\GnormBGLResult}{\ensuremath{1.097}\xspace}%
\newcommand{\GnormBGLStatErr}{\ensuremath{0.034}\xspace}%

\newcommand{\GnormBGLExtErr}{\ensuremath{0.001}\xspace}

\newcommand{\boneResult}{\ensuremath{-0.06}\xspace}%
\newcommand{\boneStatErr}{\ensuremath{0.07}\xspace}%

\newcommand{\boneExtErr}{\ensuremath{0.01}\xspace}

\newcommand{\coneResult}{\ensuremath{0.0031}\xspace}%
\newcommand{\coneStatErr}{\ensuremath{0.0022}\xspace}%

\newcommand{\coneExtErr}{\ensuremath{0.0006}\xspace}

\newcommand{\azeroResult}{\ensuremath{0.037}\xspace}%
\newcommand{\azeroStatErr}{\ensuremath{0.009}\xspace}%

\newcommand{\azeroExtErr}{\ensuremath{0.001}\xspace}

\newcommand{\aoneResult}{\ensuremath{0.28}\xspace}%
\newcommand{\aoneStatErr}{\ensuremath{0.26}\xspace}%

\newcommand{\aoneExtErr}{\ensuremath{0.08}\xspace}

\newcommand{\RDsResult}{\ensuremath{1.09}\xspace}%
\newcommand{\RDsStatErr}{\ensuremath{0.05}\xspace}%
\newcommand{\RDsSystErr}{\ensuremath{0.06}\xspace}
\newcommand{\RDsExtErr}{\ensuremath{0.05}\xspace}

\newcommand{\RDsSResult}{\ensuremath{1.06}\xspace}%
\newcommand{\RDsSStatErr}{\ensuremath{0.05}\xspace}%
\newcommand{\RDsSSystErr}{\ensuremath{0.07}\xspace}
\newcommand{\RDsSExtErr}{\ensuremath{0.05}\xspace}

\newcommand{\VcbUnits}{\ensuremath{10^{-3}}\xspace}

\newcommand{\VcbResultCLN}{\ensuremath{41.4}\xspace}

\newcommand{\VcbStatErrCLN}{\ensuremath{0.6}\xspace}
\newcommand{\VcbSystErrCLN}{\ensuremath{0.9}\xspace}
\newcommand{\VcbExtErrCLN}{\ensuremath{1.2}\xspace}

\newcommand{\VcbResultBGL}{\ensuremath{42.3}\xspace}

\newcommand{\VcbStatErrBGL}{\ensuremath{0.8}\xspace}
\newcommand{\VcbSystErrBGL}{\ensuremath{0.9}\xspace}
\newcommand{\VcbExtErrBGL}{\ensuremath{1.2}\xspace}

\newcommand{\BFUnits}{\ensuremath{10^{-2}}\xspace}

\newcommand{\BFDsResult}{\ensuremath{2.49}\xspace}
\newcommand{\BFDsStatErr}{\ensuremath{0.12}\xspace}
\newcommand{\BFDsSystErr}{\ensuremath{0.14}\xspace}
\newcommand{\BFDsExtErr}{\ensuremath{0.16}\xspace}

\newcommand{\BFDsSResult}{\ensuremath{5.38}\xspace}
\newcommand{\BFDsSStatErr}{\ensuremath{0.25}\xspace}
\newcommand{\BFDsSSystErr}{\ensuremath{0.46}\xspace}
\newcommand{\BFDsSExtErr}{\ensuremath{0.30}\xspace}

\newcommand{\mypaperversion}{}

\newcommand{\mylhcbpapernumber}{LHCb-PAPER-2019-041}
\newcommand{\mycernpapernumber}{CERN-EP-2019-282}

\def\paperauthors{LHCb collaboration} %
\def\paperasciititle{Measurement of V_cb with Bs0 -> Ds(*)-mu+nu_mu decays} %
\def\papertitle{Measurement of \Vcb with $\Bs\to\DsorDsSm\mup\neum$ decays} %
\def\paperkeywords{{High Energy Physics}, {LHCb}} %
\def\papercopyright{\the\year\ CERN for the benefit of the LHCb collaboration} %
\def\paperlicence{CC-BY-4.0 licence}
\def\paperlicenceurl{https://creativecommons.org/licenses/by/4.0/}
\usepackage[top=1in, bottom=1.25in, left=1in, right=1in]{geometry}

\usepackage{microtype}
\usepackage{lineno}  %
\usepackage{xspace} %
\usepackage{caption} %

\usepackage{graphicx}  %
\usepackage{color}
\usepackage{colortbl}
\graphicspath{{./figs/}{./figs/paper/}{./figs/supplementary/}} %
\DeclareGraphicsExtensions{.pdf,.PDF,png,.PNG}

\usepackage{amsmath} %
\usepackage{amssymb}
\usepackage{amsfonts}
\usepackage{upgreek} %
\usepackage{overpic}
\newcommand*\patchAmsMathEnvironmentForLineno[1]{%
\expandafter\let\csname old#1\expandafter\endcsname\csname #1\endcsname
\expandafter\let\csname oldend#1\expandafter\endcsname\csname
end#1\endcsname
 \renewenvironment{#1}%
   {\linenomath\csname old#1\endcsname}%
   {\csname oldend#1\endcsname\endlinenomath}%
}
\newcommand*\patchBothAmsMathEnvironmentsForLineno[1]{%
  \patchAmsMathEnvironmentForLineno{#1}%
  \patchAmsMathEnvironmentForLineno{#1*}%
}
\AtBeginDocument{%
\patchBothAmsMathEnvironmentsForLineno{equation}%
\patchBothAmsMathEnvironmentsForLineno{align}%
\patchBothAmsMathEnvironmentsForLineno{flalign}%
\patchBothAmsMathEnvironmentsForLineno{alignat}%
\patchBothAmsMathEnvironmentsForLineno{gather}%
\patchBothAmsMathEnvironmentsForLineno{multline}%
\patchBothAmsMathEnvironmentsForLineno{eqnarray}%
}

\usepackage{hyperxmp}

\usepackage[pdftex,
            pdfauthor={\paperauthors},
            pdftitle={\paperasciititle},
            pdfkeywords={\paperkeywords},
            pdfcopyright={Copyright (C) \papercopyright},
            pdflicenseurl={\paperlicenceurl}]{hyperref}

\usepackage[all]{hypcap} %

\usepackage{xspace} 
\usepackage{upgreek}

\newcommand{\offsetoverline}[2][0.1em]{\kern #1\overline{\kern -#1 #2}}%

\def\lhcb   {\mbox{LHCb}\xspace}

\def\MagUp {\mbox{\em Mag\kern -0.05em Up}\xspace}

\ifthenelse{\boolean{uprightparticles}}%
{

 \def\Pmu         {\ensuremath{\upmu}\xspace}                 
 \def\Pnu         {\ensuremath{\upnu}\xspace}                 
                  
 \def\Ppi         {\ensuremath{\uppi}\xspace}

 \def\Ptau        {\ensuremath{\uptau}\xspace}

 \def\Ppsi        {\ensuremath{\uppsi}\xspace}

 \def\PDelta      {\ensuremath{\Delta}\xspace}                 
 \def\PXi         {\ensuremath{\Xi}\xspace}                 
 \def\PLambda     {\ensuremath{\Lambda}\xspace}                 
 \def\PSigma      {\ensuremath{\Sigma}\xspace}                 
 \def\POmega      {\ensuremath{\Omega}\xspace}                 
 \def\PUpsilon    {\ensuremath{\Upsilon}\xspace}

 \def\PB      {\ensuremath{\mathrm{B}}\xspace}                 
                  
 \def\PD      {\ensuremath{\mathrm{D}}\xspace}

 \def\PJ      {\ensuremath{\mathrm{J}}\xspace}                 
 \def\PK      {\ensuremath{\mathrm{K}}\xspace}

 \def\PW      {\ensuremath{\mathrm{W}}\xspace}

 \def\Pb      {\ensuremath{\mathrm{b}}\xspace}                 
 \def\Pc      {\ensuremath{\mathrm{c}}\xspace}                 
 \def\Pd      {\ensuremath{\mathrm{d}}\xspace}

 \def\Pi      {\ensuremath{\mathrm{i}}\xspace}

 \def\Pp      {\ensuremath{\mathrm{p}}\xspace}

 \def\Ps      {\ensuremath{\mathrm{s}}\xspace}                 
                  
 \def\Pu      {\ensuremath{\mathrm{u}}\xspace}

}
{

 \def\Pmu         {\ensuremath{\mu}\xspace}                 
 \def\Pnu         {\ensuremath{\nu}\xspace}                 
                  
 \def\Ppi         {\ensuremath{\pi}\xspace}

 \def\Ptau        {\ensuremath{\tau}\xspace}

 \def\Ppsi        {\ensuremath{\psi}\xspace}                 
                  
 \mathchardef\PDelta="7101
 \mathchardef\PXi="7104
 \mathchardef\PLambda="7103
 \mathchardef\PSigma="7106
 \mathchardef\POmega="710A
 \mathchardef\PUpsilon="7107
                  
 \def\PB      {\ensuremath{B}\xspace}                 
                  
 \def\PD      {\ensuremath{D}\xspace}

 \def\PJ      {\ensuremath{J}\xspace}                 
 \def\PK      {\ensuremath{K}\xspace}

 \def\PW      {\ensuremath{W}\xspace}

 \def\Pb      {\ensuremath{b}\xspace}                 
 \def\Pc      {\ensuremath{c}\xspace}                 
 \def\Pd      {\ensuremath{d}\xspace}

 \def\Pi      {\ensuremath{i}\xspace}

 \def\Pp      {\ensuremath{p}\xspace}

 \def\Ps      {\ensuremath{s}\xspace}                 
                  
 \def\Pu      {\ensuremath{u}\xspace}

}

\makeatletter
\ifcase \@ptsize \relax%
  \newcommand{\miniscule}{\@setfontsize\miniscule{4}{5}}%
\or%
  \newcommand{\miniscule}{\@setfontsize\miniscule{5}{6}}%
\or%
  \newcommand{\miniscule}{\@setfontsize\miniscule{5}{6}}%
\fi
\makeatother

\DeclareRobustCommand{\optbar}[1]{\shortstack{{\miniscule (\rule[.5ex]{1.25em}{.18mm})}
  \\ [-.7ex] $#1$}}

\def\mup        {{\ensuremath{\Pmu^+}}\xspace}
\def\mun        {{\ensuremath{\Pmu^-}}\xspace} %

\def\taup       {{\ensuremath{\Ptau^+}}\xspace}

\def\ellp       {{\ensuremath{\ell^+}}\xspace}

\def\neu        {{\ensuremath{\Pnu}}\xspace}
\def\neub       {{\ensuremath{\overline{\Pnu}}}\xspace}
\def\neum       {{\ensuremath{\neu_\mu}}\xspace}
\def\neumb      {{\ensuremath{\neub_\mu}}\xspace}

\def\neutb      {{\ensuremath{\neub_\tau}}\xspace}
\def\neul       {{\ensuremath{\neu_\ell}}\xspace}

\def\W      {{\ensuremath{\PW}}\xspace}
\def\Wp     {{\ensuremath{\PW^+}}\xspace}

\def\uquark    {{\ensuremath{\Pu}}\xspace}

\def\dquark    {{\ensuremath{\Pd}}\xspace}

\def\squark    {{\ensuremath{\Ps}}\xspace}

\def\cquark    {{\ensuremath{\Pc}}\xspace}
\def\cquarkbar {{\ensuremath{\overline \cquark}}\xspace}

\def\bquark    {{\ensuremath{\Pb}}\xspace}
\def\bquarkbar {{\ensuremath{\overline \bquark}}\xspace}

\def\pion   {{\ensuremath{\Ppi}}\xspace}

\def\pip    {{\ensuremath{\pion^+}}\xspace}
\def\pim    {{\ensuremath{\pion^-}}\xspace}

\def\kaon    {{\ensuremath{\PK}}\xspace}
  \def\Kbar    {{\kern 0.2em\overline{\kern -0.2em \PK}{}}\xspace}

\def\KorKbar {\kern 0.18em\optbar{\kern -0.18em K}{}\xspace}
\def\Kz      {{\ensuremath{\kaon^0}}\xspace}
\def\Kzb     {{\ensuremath{\Kbar{}^0}}\xspace}
\def\Kp      {{\ensuremath{\kaon^+}}\xspace}
\def\Km      {{\ensuremath{\kaon^-}}\xspace}

  \def\Dbar    {{\kern 0.2em\overline{\kern -0.2em \PD}{}}\xspace}
\def\D       {{\ensuremath{\PD}}\xspace}

\def\DorDbar {\kern 0.18em\optbar{\kern -0.18em D}{}\xspace}
\def\Dz      {{\ensuremath{\D^0}}\xspace}

\def\Dp      {{\ensuremath{\D^+}}\xspace}
\def\Dm      {{\ensuremath{\D^-}}\xspace}

\def\Dstarz  {{\ensuremath{\D^{*0}}}\xspace}

\def\Dstarm  {{\ensuremath{\D^{*-}}}\xspace}

\def\theDstarm{{\ensuremath{\D^{*}(2010)^{-}}}\xspace}

\def\Dsp     {{\ensuremath{\D^+_\squark}}\xspace}
\def\Dsm     {{\ensuremath{\D^-_\squark}}\xspace}

\def\B       {{\ensuremath{\PB}}\xspace}
\def\Bbar    {{\ensuremath{\kern 0.18em\overline{\kern -0.18em \PB}{}}}\xspace}

\def\BorBbar    {\kern 0.18em\optbar{\kern -0.18em B}{}\xspace}
\def\Bz      {{\ensuremath{\B^0}}\xspace}

\def\Bu      {{\ensuremath{\B^+}}\xspace}
\def\Bub     {{\ensuremath{\B^-}}\xspace}
\def\Bp      {{\ensuremath{\Bu}}\xspace}
\def\Bm      {{\ensuremath{\Bub}}\xspace}

\def\Bd      {{\ensuremath{\B^0}}\xspace}
\def\Bs      {{\ensuremath{\B^0_\squark}}\xspace}

\def\Bc      {{\ensuremath{\B_\cquark^+}}\xspace}

\def\jpsi     {{\ensuremath{{\PJ\mskip -3mu/\mskip -2mu\Ppsi\mskip 2mu}}}\xspace}

\def\Y#1S{\ensuremath{\PUpsilon{(#1S)}}\xspace}

\def\proton      {{\ensuremath{\Pp}}\xspace}

\def\Lz          {{\ensuremath{\PLambda}}\xspace}

\def\LorLbar     {\kern 0.18em\optbar{\kern -0.18em \PLambda}{}\xspace}

\def\Lc          {{\ensuremath{\Lz^+_\cquark}}\xspace}

\def\Lb           {{\ensuremath{\Lz^0_\bquark}}\xspace}

\def\BF         {{\ensuremath{\mathcal{B}}}\xspace}

\newcommand{\decay}[2]{\mbox{\ensuremath{#1\!\to #2}}\xspace}         %

\def\to                 {\ensuremath{\rightarrow}\xspace}

\def\qsq       {{\ensuremath{q^2}}\xspace}

\def\Vub  {{\ensuremath{V_{\uquark\bquark}}}\xspace}
\def\Vcb  {\ensuremath{|V_{\cquark\bquark}|}\xspace}

\def\AT#1     {\ensuremath{A_{\mathrm{T}}^{#1}}\xspace}           %

\def\C#1      {\ensuremath{\mathcal{C}_{#1}}\xspace}                       %
\def\Cp#1     {\ensuremath{\mathcal{C}_{#1}^{'}}\xspace}                    %
\def\Ceff#1   {\ensuremath{\mathcal{C}_{#1}^{\mathrm{(eff)}}}\xspace}        %
\def\Cpeff#1  {\ensuremath{\mathcal{C}_{#1}^{'\mathrm{(eff)}}}\xspace}       %
\def\Ope#1    {\ensuremath{\mathcal{O}_{#1}}\xspace}                       %
\def\Opep#1   {\ensuremath{\mathcal{O}_{#1}^{'}}\xspace}                    %

\newcommand{\nospaceunit}[1]{\ensuremath{\text{#1}}}       
\newcommand{\aunit}[1]{\ensuremath{\text{\,#1}}}       
\newcommand{\tev}{\aunit{Te\kern -0.1em V}\xspace}
\newcommand{\gev}{\aunit{Ge\kern -0.1em V}\xspace}
\newcommand{\mev}{\aunit{Me\kern -0.1em V}\xspace}
\newcommand{\kev}{\aunit{ke\kern -0.1em V}\xspace}
\newcommand{\ev}{\aunit{e\kern -0.1em V}\xspace}
\newcommand{\mevc}{\ensuremath{\aunit{Me\kern -0.1em V\!/}c}\xspace}
\newcommand{\gevc}{\ensuremath{\aunit{Ge\kern -0.1em V\!/}c}\xspace}
\newcommand{\mevcc}{\ensuremath{\aunit{Me\kern -0.1em V\!/}c^2}\xspace}
\newcommand{\gevcc}{\ensuremath{\aunit{Ge\kern -0.1em V\!/}c^2}\xspace}

\def\mum  {\ensuremath{\,\upmu\nospaceunit{m}}\xspace}

\def\fb   {\ensuremath{\aunit{fb}}\xspace}
\def\invfb   {\ensuremath{\fb^{-1}}\xspace}

\def\ps   {\ensuremath{\aunit{ps}}\xspace}

\newcommand{\stat}{\aunit{(stat)}\xspace}
\newcommand{\syst}{\aunit{(syst)}\xspace}

\newcommand{\chisq}{\ensuremath{\chi^2}\xspace}

\def\gsim{{~\raise.15em\hbox{$>$}\kern-.85em
          \lower.35em\hbox{$\sim$}~}\xspace}
\def\lsim{{~\raise.15em\hbox{$<$}\kern-.85em
          \lower.35em\hbox{$\sim$}~}\xspace}

\def\pt         {\ensuremath{p_{\mathrm{T}}}\xspace}

\def\ptot       {\ensuremath{p}\xspace}

\def\evtgen     {\mbox{\textsc{EvtGen}}\xspace}

\def\geant      {\mbox{\textsc{Geant4}}\xspace}

\def\photos     {\mbox{\textsc{Photos}}\xspace}

\def\pythia     {\mbox{\textsc{Pythia}}\xspace}

\xspace

\def\tell1  {TELL1\xspace}
\def\ukl1   {UKL1\xspace}

\newcommand{\eg}{\mbox{\itshape e.g.}\xspace}
\newcommand{\ie}{\mbox{\itshape i.e.}\xspace}

\newcommand{\vs}{\mbox{\itshape vs.}\xspace}
\usepackage{cite} %
\usepackage{mciteplus}

\usepackage{rotating}
\usepackage{booktabs} % end input ./preamble.tex
 
\begin{document}
\renewcommand{\thefootnote}{\fnsymbol{footnote}}
\setcounter{footnote}{1}
%
% start input ./title-LHCb-PAPER.tex
%
%
%
%
%
%

%
%
%
\begin{titlepage}
\pagenumbering{roman}

\vspace*{-1.5cm}
\centerline{\large EUROPEAN ORGANIZATION FOR NUCLEAR RESEARCH (CERN)}
\vspace*{1.5cm}
\noindent
\begin{tabular*}{\linewidth}{lc@{\extracolsep{\fill}}r@{\extracolsep{0pt}}}
\ifthenelse{\boolean{pdflatex}}%
% inside_import 
% before {\vspace*{-1.5cm}\mbox{\!\!\!
% ignored \!
% args width=.14\textwidth
% full_filename figs/lhcb-logo.pdf
% after } & &}%
{\vspace*{-1.5cm}\mbox{\!\!\!\includegraphics[width=.14\textwidth]{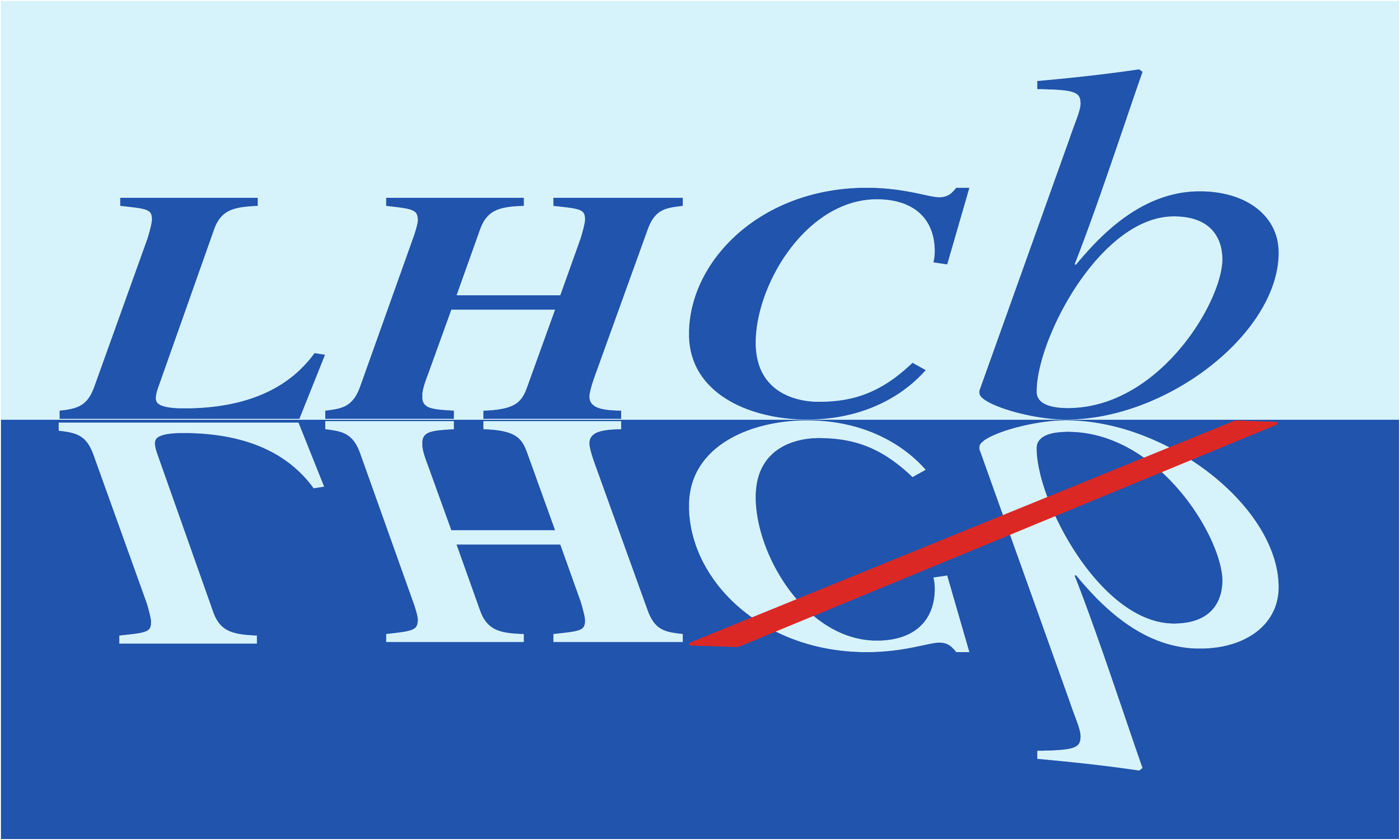}} & &}%
% inside_import 
% before {\vspace*{-1.2cm}\mbox{\!\!\!
% ignored \!
% args width=.12\textwidth
% full_filename figs/lhcb-logo.eps
% after } & &}%
{\vspace*{-1.2cm}\mbox{\!\!\!\includegraphics[width=.12\textwidth]{figs/lhcb-logo.eps}} & &}%
\\
 & & \mycernpapernumber \\  %
 & & \mylhcbpapernumber \\  %
 & & January 9, 2020 \\ %
 & & \mypaperversion \\
\end{tabular*}

\vspace*{4.0cm}

{\normalfont\bfseries\boldmath\huge
\begin{center}
  \papertitle 
\end{center}
}

\vspace*{2.0cm}

\begin{center}
\paperauthors\footnote{Authors are listed at the end of this paper.}
\end{center}

\vspace{\fill}

\begin{abstract}
%
% start input ./abstract.tex
\noindent The element \Vcb of the Cabibbo--Kobayashi--Maskawa matrix is measured using semileptonic \Bs decays produced in proton-proton collision data collected with the LHCb detector at center-of-mass energies of 7 and 8\tev, corresponding to an integrated luminosity of 3\invfb. Rates of  \bsDs and \bsDsS decays are analyzed using hadronic form-factor parametrizations derived either by Caprini, Lellouch and Neubert (CLN) or by Boyd, Grinstein and Lebed (BGL). The measured values of \Vcb are ${(\VcbResultCLN \pm \VcbStatErrCLN \pm \VcbSystErrCLN \pm \VcbExtErrCLN)\times \VcbUnits}$ and ${(\VcbResultBGL \pm \VcbStatErrBGL \pm \VcbSystErrBGL \pm \VcbExtErrBGL)\times \VcbUnits}$ in the CLN and BGL parametrization, respectively.  The first uncertainty is statistical, the second systematic, and the third is due to the external inputs used in the measurement.  These results are in agreement with those obtained from decays of \Bu and \Bd mesons. They are the first determinations of \Vcb at a hadron-collider experiment and the first using \Bs meson decays. % end input ./abstract.tex
 \end{abstract}

\vspace*{2.0cm}

\begin{center}
  Published in 
  Phys.~Rev.~D101 (2020) 072004

\end{center}

\vspace{\fill}

{\footnotesize 
\centerline{\copyright~\papercopyright. \href{\paperlicenceurl}{\paperlicence}.}}
\vspace*{2mm}

\end{titlepage}

\newpage
\setcounter{page}{2}
\mbox{~}

\cleardoublepage

 % end input ./title-LHCb-PAPER.tex
 
\renewcommand{\thefootnote}{\arabic{footnote}}
\setcounter{footnote}{0}

\pagestyle{plain} %
\setcounter{page}{1}
\pagenumbering{arabic}

%
%
%

%
% start input ./introduction.tex
\section{Introduction}\label{sec:introduction}

The semileptonic quark-level transition $\bquarkbar\to\cquarkbar\,\ellp\neul$, where $\ell$ is an electron or a muon, provides the cleanest way to access the strength of the coupling between the \bquark and \cquark quarks, expressed by the element \Vcb of the Cabibbo--Kobayashi--Maskawa (CKM) matrix.\footnote{The inclusion of charge-conjugate processes is implied throughout this paper.} Two complementary methods have been used to determine \Vcb. One measures the decay rate by looking at \emph{inclusive} \bquark-hadron decays to final states made of a \cquark-flavored hadron and a charged lepton; the other measures the rate of a specific (\emph{exclusive}) decay, such as $\Bd\to\theDstarm\mup\neum$ or $\Bd\to\Dm\mup\neum$. The average of the inclusive method yields \mbox{$\Vcb=(42.19\pm0.78)\times\VcbUnits$}, while the exclusive determinations give \mbox{$\Vcb=(39.25\pm0.56)\times\VcbUnits$}~\cite{HFLAV18}. The two values are approximately three standard deviations apart, and this represents a long-standing puzzle in flavor physics.

Exclusive determinations rely on a parametrization of strong-interaction effects in the hadronic current of the quarks bound in mesons, the so-called form factors. These are Lorentz-invariant functions of the squared mass \qsq of the virtual \Wp emitted in the $\bquarkbar \to \cquarkbar$ transition  and are calculated using nonperturbative quantum chromodynamics (QCD) techniques, such as lattice QCD (LQCD) or QCD sum rules. Several parametrizations have been proposed to model the form factors~\cite{Caprini:1997mu,Boyd:1994tt,Boyd:1995sq,Boyd:1997kz,Bourrely:2008za,Bernlochner:2017xyx}. The parametrization derived by Caprini, Lellouch and Neubert (CLN)~\cite{Caprini:1997mu} has been the reference model for the exclusive determinations of \Vcb. The approximations adopted in this parametrization have been advocated as a possible explanation for the discrepancy with the inclusive measurement~\cite{Bigi:2017njr,Grinstein:2017nlq,Jaiswal:2017rve,Colangelo:2018cnj}. A more general model by Boyd, Grinstein and Lebed (BGL)~\cite{Boyd:1994tt,Boyd:1995sq,Boyd:1997kz} has been used in recent high-precision measurements of \Vcb~\cite{Abdesselam:2018nnh,Dey:2019bgc} to overcome the CLN limitations. However, no significant difference in the \Vcb values measured with the two parametrizations has been found and the issue remains open~\cite{Gambino:2019sif,Bordone:2019vic,Bordone:2019guc,King:2019rvk}.  

All exclusive measurements of \Vcb performed so far make use of decays of \Bu and \Bd mesons. The study of other \bquark-hadron decays, which are potentially subject to different sources of uncertainties, can provide complementary information and may shed light on this puzzle. In particular, semileptonic \Bs decays, which are abundant at the LHC, have not yet been exploited to measure \Vcb. Exclusive semileptonic \Bs decays are more advantageous from a theoretical point of view. The larger mass of the valence \squark quark compared to \uquark or \dquark quarks makes LQCD calculations of the form factors for \Bs decays less computationally expensive than those for \Bu or \Bd decays, thus possibly allowing for a more precise determination of \Vcb~\cite{McLean:2019sds,Harrison:2017fmw,Atoui:2013zza,Flynn:2016vej}. Calculations of the form factor over the full \qsq spectrum are available for $\Bs \to \Dsm \ellp \neul$ decays~\cite{Monahan:2017uby,McLean:2019qcx} and can be used along with experimental data to measure \Vcb. Exclusive $\Bs\to\Dsm\ellp\neul$ and $\Bs\to\DsSm\ellp \neul$ decays are also experimentally appealing because background contamination from partially reconstructed decays is expected to be less severe than for their \BuorBd counterparts. Indeed, the majority of the excited states of the \Dsm meson (other than \DsSm) are expected to decay dominantly into $D^{\scalebox{0.4}{(}*\scalebox{0.4}{)}}K$ final states.

This paper presents the first determination of \Vcb from the exclusive decays \bsDs and \bsDsS. The analysis uses proton-proton collision data collected with the LHCb detector at center-of-mass energies of 7 and 8\tev, and corresponding to an integrated luminosity of 3\invfb. In both decays, only the $\Dsm\mup$ final state is reconstructed using the Cabibbo-favored mode $\Dsm\to[\Kp\Km]_\phi\pim$, where the kaon pair is required to have invariant mass in the vicinity of the $\phi(1020)$ resonance. The photon or the neutral pion emitted along with the \Dsm in the \DsSm decay is not reconstructed. The value of \Vcb is determined from the observed yields of \Bs decays normalized to those of \emph{reference} \Bd decays after correcting for the relative reconstruction and selection efficiencies. The reference decays are chosen to be \bdD and \bdDS, where the \Dm meson is reconstructed in the Cabibbo-suppressed mode $\Dm\to[\Kp\Km]_\phi\pim$. Hereafter the symbol \DSm refers to the \theDstarm meson. Signal and reference decays thus have identical final states and similar kinematic properties. This choice results in a reference sample of smaller size than that of the signal, but allows suppressing systematic uncertainties that affect the calculation of the efficiencies. Using the \Bd decays as a reference, the determination of \Vcb needs in input the measured branching fractions of these decays and the ratio of \Bs- to \Bd-meson production fractions. The latter is measured by LHCb using an independent sample of semileptonic decays with respect to that exploited in this analysis~\cite{LHCb-PAPER-2018-050}, and it assumes universality of the semileptonic decay width of \bquark hadrons~\cite{Bigi:2011gf}. %
The ratios of the branching fractions of signal and reference decays, 
\begin{align} 
\label{eq:RD}
\RD &\equiv \frac{\BF(\bsDs)}{\BF(\bdD)},\\
\label{eq:RDS}
\RDS &\equiv \frac{\BF(\bsDsS)}{\BF(\bdDS)}
\end{align}
are also determined from the same analysis. From the measured branching fractions of the reference decays, the branching fractions of \bsDs and \bsDsS decays are determined for the first time. 

This analysis uses either the CLN or the BGL parametrization to model the form factors, with parameters determined by analyzing the decay rates using a novel method: instead of approximating \qsq, which cannot be determined precisely because of the undetected neutrino, a variable that can be reconstructed fully from the final-state particles and that preserves information on the form factors is used. This variable is the component of the \Dsm momentum perpendicular to the \Bs flight direction, denoted as \pperpDs. The \pperpDs variable is highly correlated with the \qsq value of the \bsDs and \bsDsS decays, and, to a minor extent, with the helicity angles of the \bsDsS decay. When used together with the corrected mass, \mcorr, it also helps in determining the sample composition. The corrected mass is calculated from the mass of the reconstructed particles, $m(\Dsm\mup)$, and from the momentum of the $\Dsm\mup$ system transverse to the \Bs flight direction, $p_\perp\!(\Dsm\mup)$, as
\begin{equation}\label{eq:m-cor}
\mcorr \equiv \sqrt{m^2(\Dsm\mup)+p_\perp^2\!(\Dsm\mup)}+p_\perp\!(\Dsm\mup).
\end{equation}
Signal and background decays accumulate in well-separated regions of the two-dimensional space spanned by \mcorr and \pperpDs. A fit to the data distribution in the \mcorr \vs \pperpDs plane identifies the \bsDs and \bsDsS signal decays and simultaneously provides a measurement of \Vcb and of the form factors.

The paper is structured as follows. The formalism describing the semileptonic \BorBs decays and the parametrization of their form factors is outlined in Sec.~\ref{sec:formalism}. Section~\ref{sec:detector} gives a brief description of the LHCb detector and of the simulation software. The selection and the expected composition of the signal and reference samples are presented in Sec.~\ref{sec:selection}. Section~\ref{sec:fit} describes the method used to measure \Vcb and the other parameters of interest. The determination of the reference \Bd-decay yields is reported in Sec.~\ref{sec:reference}, and the analysis of the signal \Bs decays is discussed in Sec.~\ref{sec:signal}. Section~\ref{sec:systematics} describes the systematic uncertainties affecting the measurements and Sec.~\ref{sec:results} presents the final results, before concluding. % end input ./introduction.tex
 %
% start input ./formalism.tex
\section{Formalism} \label{sec:formalism}
The formalism used to describe the decay rate of a \B meson into a semileptonic final state with a pseudoscalar or a vector \D meson is outlined here. In this Section, the notation $\B\to\DorDS\mu\nu$ is used to identify both $\Bd\to\DorDSm\mup\neum$ and $\Bs\to\DsorDsSm\mup\neum$ decays, clarifying when the distinction is relevant.

\begin{figure}[b]
\centering
% inside_import 
% before 
% ignored 
% args width=0.6\textwidth
% full_filename figs/AnglesDefinition
% after \\
\includegraphics[width=0.6\textwidth]{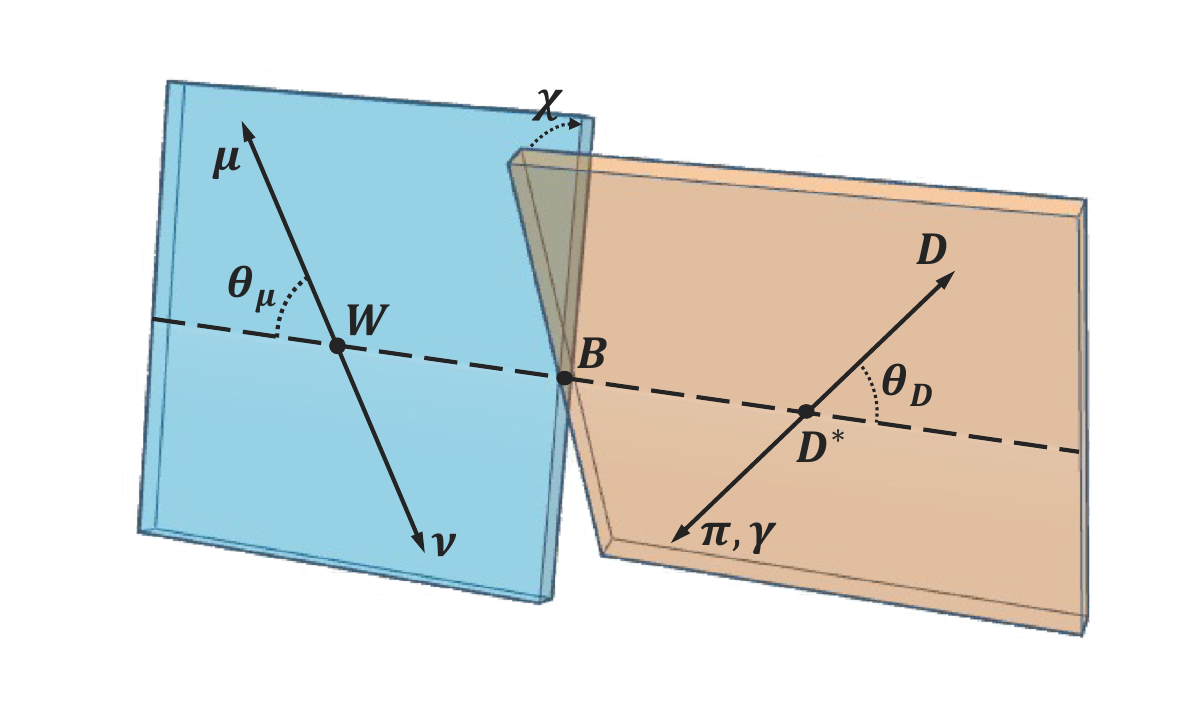}\\
\caption{Graphical representation of the helicity angles in $\B\to\DS\mu\nu$ decays. The definitions are provided in the text. \label{fig:angles}}
\end{figure}

\subsection{{\boldmath $\B\to\DS\mu\nu$} decays}
The $\B\to\DS\mu\nu$ differential decay rate can be expressed in terms of one recoil variable, $w$, and three helicity angles, $\theta_\mu$, $\theta_D$ and $\chi$, as
\begin{equation}
\label{eq:rate_vector_4d}
\frac{\diff^4 \Gamma(\B\to\DS\mu\nu)}{\diff w\, \diff\!\cos\theta_\mu\, \diff\!\cos\theta_D\, \diff\chi}=\frac{3 m_\B^3 m^2_{\DS} G^2_{\rm F}}{16(4\pi)^4} \etaEW^2 \Vcb^2 |\mathcal{A}(w,\theta_\mu,\theta_D,\chi)|^2,
\end{equation}
where $G_{\rm F}$ is the Fermi constant and the coefficient $\etaEW\approx1.0066$ accounts for the leading-order electroweak correction~\cite{Sirlin}. The recoil variable is defined as the scalar product of the four-velocities of the \B and \DS mesons, $w=v_\B\cdot v_{\DS}=(m^2_\B+m^2_{\DS}-q^2)/(2m_\B m_{\DS})$, with $m_{\B(\DS)}$ being the mass of the \B (\DS) meson. The minimum value, $w=1$, corresponds to zero recoil of the \DS meson in the \B rest frame, \ie, the largest kinematically allowed value of \qsq. The helicity angles (represented in Fig.~\ref{fig:angles}) are $\theta_\mu$, the angle between the direction of the muon in the \W rest frame and the direction of the \W boson in the \B rest frame; $\theta_D$, the angle between the direction of the \D in the \DS rest frame and the direction of the \DS in the \B rest frame; and $\chi$, the angle between the plane formed by the \DS decay products and that formed by the two leptons. In the limit of massless leptons, the decay amplitude $\mathcal{A}$ can be decomposed in terms of three amplitudes, $H_{\pm/0}(w)$, corresponding to the three possible helicity states of the \DS meson, and its squared modulus is written as
\begin{equation}
|\mathcal{A}(w,\theta_\mu,\theta_D,\chi)|^2 = \sum_i^6 \mathcal{H}_i(w) k_i(\theta_\mu,\theta_D,\chi)\,,
\end{equation}
with the $\mathcal{H}_i$ and  $k_i$ terms defined in Table~\ref{tab:angularTermsDst}. The helicity amplitudes are expressed by three form factors, $h_{A_1}(w)$, $R_1(w)$, and $R_2(w)$, as
\begin{equation}
H_{\pm/0}(w) =  2\frac{\sqrt{m_\B m_{\DS}}}{m_\B + m_{\DS}}(1-r^2)(w+1)(w^2-1)^{1/4} h_{A_1}(w) \tilde{H}_{\pm/0}(w)\,,
\end{equation}
with $r=m_{\DS}/m_\B$ and
\begin{align}
\tilde{H}_\pm(w) &= \frac{\sqrt{1 - 2wr + r^2}}{1-r }\left[1 \mp \sqrt{\frac{w-1}{w+1}} R_1(w)\right]\,,\\
\tilde{H}_0(w) &= 1 + \frac{(w-1)(1-R_2(w))}{1-r}\,.
\end{align}

\begin{table}[t]
\caption{\label{tab:angularTermsDst} Functions describing the differential decay rate of $\B\to\DS\mu\nu$ decays, separately for the cases in which the \DS meson decays to $\D\gamma$ or $\D\pi^0$.}
\centering
\resizebox{\textwidth}{!}{
\begin{tabular}{lccc}
\toprule
\multirow{2}{*}{$i$} & \multirow{2}{*}{$\mathcal{H}_i(w)$} & \multicolumn{2}{c}{$k_i(\theta_\mu,\theta_D,\chi )$} \\
\cmidrule{3-4}
       &                           &    $\DS\to\D\gamma$   &    $\DS\to\D\pi^0$           \\
\midrule
1     & $H_+^2$ &  $\frac{1}{2}(1 + \cos^2\theta_D)(1 - \cos\theta_\mu)^2$ &  
                                $\sin^2\theta_D(1 - \cos\theta_\mu)^2$                       \\
2     & $H_-^2$ &  $\frac{1}{2}(1 + \cos^2\theta_D)(1 + \cos\theta_\mu)^2$ &  
                                $\sin^2\theta_D(1 + \cos\theta_\mu)^2$\\
3     & $H_0^2$ &  $2\sin^2\theta_D\sin^2\theta_\mu$ &  
                                $4\cos^2\theta_D\sin^2\theta_\mu$ \\
4     & $H_+H_-$ &  $\sin^2\theta_D\sin^2\theta_\mu\cos2\chi$ &  
                                 $-2 \sin^2\theta_D\sin^2\theta_\mu \cos2\chi$\\
5     & $H_+H_0$ &  $\sin2\theta_D\sin\theta_\mu(1 - \cos\theta_\mu)\cos\chi$ &  
                                 $-2\sin2\theta_D\sin\theta_\mu(1-\cos\theta_\mu)\cos\chi$\\
6     & $H_-H_0$ &  $-\sin2\theta_D\sin\theta_\mu(1 + \cos\theta_\mu)\cos\chi$ &
                                 $2\sin2\theta_D\sin\theta_\mu(1+\cos\theta_\mu)\cos\chi$\\
\bottomrule
\end{tabular}}
\end{table}

The CLN parametrization uses dispersion relations and reinforced unitarity bounds based on Heavy Quark Effective Theory to derive simplified expressions for the form factors~\cite{Caprini:1997mu}. For the $\B\to\DS\mu\nu$ case, the three form factors are written as~\cite{Caprini:1997mu}
\begin{align}
h_{A_1}(w)&=h_{A_1}(1)\left[1-8\rho^2 z+(53\rho^2-15)z^2-(231\rho^2-91)z^3\right]\,,\label{eq:hA1_CLN}\\
R_1(w)&=R_1(1)-0.12(w-1)+0.05(w-1)^2\,,\label{eq:R1_CLN}\\
R_2(w)&=R_2(1)-0.11(w-1)-0.06(w-1)^2\,,\label{eq:R2_CLN}
\end{align}
where the same numerical coefficients, originally computed for \Bd decays, are considered also for \Bs decays, and where the conformal variable $z$ is defined as
\begin{equation}
\label{eq:z_def}
z\equiv\frac{\sqrt{w+1}-\sqrt{2}}{\sqrt{w+1}+\sqrt{2}}\,.
\end{equation}
The form factors depend only on four parameters: \rsq, \Rone, \Rtwo and $h_{A_1}(1)$. %

The BGL parametrization follows from more general arguments based on dispersion relations, analyticity, and crossing symmetry~\cite{Boyd:1994tt,Boyd:1995sq,Boyd:1997kz}. In the case of $\B\to\DS\mu\nu$ decays, the form factors are written in terms of three functions, $f(w)$, $g(w)$ and $\mathcal{F}_1(w)$, as follows
\begin{align}
h_{A_1}(w) &= \frac{f(w)}{\sqrt{m_\B m_{\DS}}(1+w)}\,,\label{eq:hA1_BGL} \\
R_1(w) &= (w+1) m_\B m_{\DS} \frac{g(w)}{f(w)}\,, \label{eq:R1_BGL}\\
R_2(w) &= \frac{w-r}{w-1} - \frac{\mathcal{F}_1(w)}{m_B(w-1)f(w)}\,. \label{eq:R2_BGL}
\end{align}
These functions are expanded as convergent power series of $z$ as
\begin{align}
f(z) &= \frac{1}{P_{1^+}(z)\phi_f(z)}\sum_{n=0}^\infty b_n z^n\,,\label{eq:f-series-D*} \\
g(z) &= \frac{1}{P_{1^-}(z)\phi_g(z)}\sum_{n=0}^\infty a_n z^n\,,\label{eq:g-series-D*} \\
\mathcal{F}_1(z) &= \frac{1}{P_{1^+}(z)\phi_{\mathcal{F}_1}(z)}\sum_{n=0}^\infty c_n z^n\,.\label{eq:F1-series-D*}
\end{align}
Here, the $P_{1^\pm}(z)$ functions are known as Blaschke factors for the $J^P=1^\pm$ resonances, and $\phi_{f,g,\mathcal{F}_1}(z)$ are the so-called outer functions. Adopting the formalism of Ref.~\cite{Bigi:2016mdz}, the Blaschke factors take the form
\begin{equation}\label{eq:P_i}
P_{1^\pm}(z) = C_{1^\pm} \prod^{\rm poles}_{k=1} \frac{z - z_{k}}{1 - z\,z_{k}}\,,
\end{equation}
where 
\begin{equation}
z_{k} = \frac{\sqrt{t_+ - m^2_{k}} - \sqrt{t_+ - t_-^{\phantom{2}}}}{\sqrt{t_+ - m^2_{k}} + \sqrt{t_+ - t_-^{\phantom{2}}}}\,,    
\end{equation}
$t_{\pm} = (m_\B \pm m_{\DS})^2$, and $m_{k}$ denotes the pole masses of the $k$-th excited \Bc states that are below the $\B\DS$ threshold and have the appropriate $J^P$ quantum numbers. The constants $C_{1^\pm}$ are scale factors calculated to use in \Bs decays the same Blaschke factor derived for \Bd decays. The outer functions are defined as
\begin{align}
\phi_f(z)&=\frac{4r}{m^2_B}\sqrt{\frac{n_I}{3\pi\tilde{\chi}_{1^+}(0)}}\, \frac{(1+z)\sqrt{(1-z)^3}}{[(1+r)(1-z)+2\sqrt{r}(1+z)]^4}\,, \\
\phi_g(z)&=16r^2\sqrt{\frac{n_I}{3\pi\tilde{\chi}_{1^-}(0)}}\,\frac{(1+z)^2}{\sqrt{(1-z)}[(1+r)(1-z)+2\sqrt{r}(1+z)]^4}\,, \\
\phi_{\mathcal{F}_1}(z)&=\frac{4r}{m^3_B}\sqrt{\frac{n_I}{6\pi\tilde{\chi}_{1^+}(0)}}\, \frac{(1+z)\sqrt{(1-z)^5}}{[(1+r)(1-z)+2\sqrt{r}(1+z)]^5}\,,
\end{align}
where $n_I=2.6$ is the number of spectator quarks (three), corrected for $SU(3)$-breaking effects~\cite{Bigi:2017njr}. The \Bc resonances used in the computation of the Blaschke factors, the $\tilde{\chi}_{1^\pm}(0)$ coefficients of the outer functions, and the constants $C_{1^\pm}$ are reported in Table~\ref{tab:Bcpoles}. The coefficients of the series in Eqs.~\eqref{eq:f-series-D*}--\eqref{eq:F1-series-D*} are bound by the unitarity constraints 
\begin{equation}\label{eq:unitarity_BGL_DS}
\sum_{n=0}^\infty a_n^2\leqslant 1\,,\quad\sum_{n=0}^\infty (b_n^2 + c_n^2)\leqslant 1\,.
\end{equation}
The first coefficient of $f(z)$, $b_0$, is related to \Fnorm by the expression
\begin{equation}
\label{eq:b0_F1}
b_0 = 2\sqrt{m_\B m_{\DS}}\,P_{1^+}(0)\,\phi_f(0)\,\Fnorm\,,
\end{equation}
while $c_0$ is fixed from $b_0$ through 
\begin{equation}
\label{eq:c0_b0}
c_0 = (m_\B - m_{\DS})\,\frac{\phi_{\mathcal{F}_1}(0)}{\phi_f(0)}\, b_0\,.
\end{equation}

\begin{table}[t]
\caption{\label{tab:Bcpoles} 
Pole masses for the \Bc resonances considered in the BGL parameterization of the \Bs decays, with the $\tilde{\chi}_{J^P}(0)$ constants of the outer functions and the $C_{J^P}$ constants of the Blaschke factors~\cite{Bigi:2017njr}. For \Bd decays, the Blaschke factors do not include the last $1^-$ resonance and $C_{1^\pm}$ have both unit value.}
\centering
\begin{tabular}{cccc}
\toprule
$J^P$ & Pole mass & $\tilde{\chi}_{J^P}(0)$ & $C_{J^P}$ \\
      & $[\gevcc]$& $[10^{-4}\gev^{-2}c^{4}]$ & \\
\midrule
\multirow{4}{*}{$1^-$} & 6.329 & \multirow{4}{*}{5.131} & \multirow{4}{*}{2.52733} \\ 
 & 6.920 & & \\
 & 7.020 & & \\
 & 7.280 & & \\
\midrule
\multirow{4}{*}{$1^+$} & 6.739 & \multirow{4}{*}{3.894} & \multirow{4}{*}{2.02159} \\ 
 & 6.750 & & \\
 & 7.145 & & \\
 & 7.150 & & \\
\bottomrule
\end{tabular}
\end{table}

\subsection{{\boldmath $\B\to\D\mu\nu$} decays}
In the $\B\to\D\mu\nu$ case, the decay rate only depends  upon the recoil variable $w=v_B\cdot v_D$. In the limit of negligible lepton masses, the differential decay rate can be written as~\cite{NEUBERT1991455} 
\begin{equation}\label{eq:rate_pseudo}
\frac{\diff\Gamma(\B\to\D\mu\nu)}{\diff w}=\frac{G^2_{\rm F}m^3_D}{48\pi^3}(m_B+m_D)^2\etaEW^2\Vcb^2(w^2-1)^{3/2}|\mathcal{G}(w)|^2\,.
\end{equation}

In the CLN parametrization, using the conformal variable $z(w)$ defined in Eq.~\eqref{eq:z_def}, the form factor $\mathcal{G}(z)$ is expressed in terms of its value at zero recoil, \Gnorm, and a slope parameter, \rsq, as~\cite{Caprini:1997mu}
\begin{equation}\label{eq:G_slope}
\mathcal{G}(z)=\Gnorm\left[1-8\rho^2 z+(51\rho^2-10)z^2-(252\rho^2-84)z^3\right]\,.
\end{equation}
In the BGL parametrization, it is expressed as~\cite{Boyd:1994tt,Boyd:1995sq,Boyd:1997kz}
\begin{equation}
|\mathcal{G}(z)|^2 = \frac{4r}{(1+r)^2}\,|f_+(z)|^2\,,
\end{equation}
with $r=m_\D/m_\B$ and
\begin{equation}\label{eq:BGLffP}
f_+(z)=\frac{1}{P_{1^-}(z)\phi(z)}\sum_{n=0}^\infty d_n z^n\,.
\end{equation}
The outer function $\phi(z)$ is defined as
\begin{equation}
\phi(z)=\frac{8r^2}{m_B}\sqrt{\frac{8n_I}{3\pi \tilde{\chi}_{1^-}(0)}}\frac{(1+z)^2\sqrt{1-z}}{[(1+r)(1-z)+2\sqrt{r}(1+z)]^5}\,.
\end{equation} 
The coefficients of the series in Eq.~\eqref{eq:BGLffP} are bound by unitarity, 
\begin{equation}\label{eq:unitarity_BGL_D}
\sum_{n=0}^\infty d_n^2<1\,,
\end{equation}
with the coefficient $d_0$ being related to \Gnorm through
\begin{equation}\label{eq:d0_G0}
d_0 = \frac{1+r}{2\sqrt{r}}\mathcal{G}(0)P_{1^-}(0)\phi(0)\,.
\end{equation}

 % end input ./formalism.tex
 %
% start input ./detector.tex
\section{Detector and simulation}\label{sec:detector}
The \lhcb detector~\cite{LHCb-DP-2008-001,LHCb-DP-2014-002} is a single-arm forward spectrometer covering the \mbox{pseudorapidity} range $2<\eta <5$, designed for the study of particles containing \bquark or \cquark quarks. The detector includes a high-precision tracking system consisting of a silicon-strip vertex detector surrounding the $pp$ interaction region, a large-area silicon-strip detector located upstream of a dipole magnet with a bending power of about $4{\mathrm{\,Tm}}$, and three stations of silicon-strip detectors and straw drift tubes placed downstream of the magnet. The tracking system provides a measurement of the momentum, \ptot, of charged particles with a relative uncertainty that varies from 0.5\% at low momentum to 1.0\% at 200\gevc. The minimum distance of a track to a primary vertex, the impact parameter, is measured with a resolution of $(15+29/\pt)\mum$, where \pt is the component of the momentum transverse to the beam, in\,\gevc. Different types of charged hadrons are distinguished using information from two ring-imaging Cherenkov detectors. Photons, electrons and hadrons are identified by a calorimeter system consisting of scintillating-pad and preshower detectors, an electromagnetic and a hadronic calorimeter. Muons are identified by a system composed of alternating layers of iron and multiwire proportional chambers. 

Simulation is required to model the expected sample composition and develop the selection requirements, to calculate the reconstruction and selection efficiencies, and to build templates describing the distributions of signal and background decays used in the fit that determines the parameters of interest. In the simulation, \proton\proton collisions are generated using \pythia~\cite{Sjostrand:2006za,*Sjostrand:2007gs} with a specific \lhcb configuration~\cite{LHCb-PROC-2010-056}.  Decays of unstable particles are described by \evtgen~\cite{Lange:2001uf}, in which final-state radiation is generated using \photos~\cite{Golonka:2005pn}. The interaction of the generated particles with the detector, and its response, are implemented using the \geant toolkit~\cite{Allison:2006ve, *Agostinelli:2002hh} as described in Ref.~\cite{LHCb-PROC-2011-006}. Simulation is corrected for mismodeling of the reconstruction and selection efficiency, of the response of the particle identification algorithms, and of the kinematic properties of the generated \BorBs mesons. The corrections are determined by comparing data and simulation in large samples of control decays, such as \mbox{$\DSp \to \Dz(\to \Km \pip) \pip$}, \mbox{$\Bu\to \jpsi(\to\mup\mun)\Kp$}, \mbox{$\Bs\to\jpsi(\to\mup\mun)\phi(\to\Kp\Km)$}, \mbox{$\Bd\to\Dm(\to\Kp\pim\pim)\pip$}, and \mbox{$\Bs\to\Dsm(\to\Kp\Km\pim)\pip$}. Residual small differences between data and the corrected simulation are accounted for in the systematic uncertainties. % end input ./detector.tex
 %
% start input ./selection.tex
\section{Selection and expected sample composition}\label{sec:selection}
\begin{figure}[tb]
\centering
% inside_import 
% before 
% ignored 
% args width=0.5\textwidth
% full_filename figs/mcorr_vs_pperp_0
% after \hfil
\includegraphics[width=0.5\textwidth]{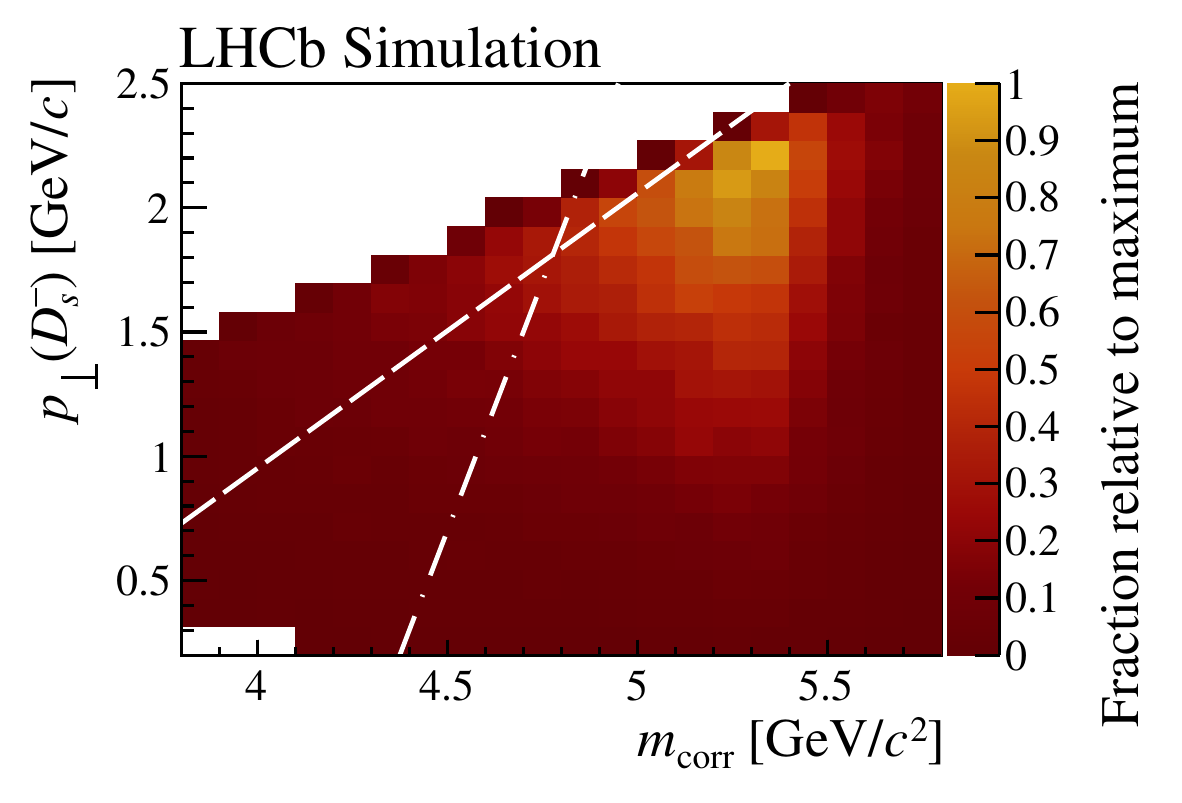}\hfil
% inside_import 
% before 
% ignored 
% args width=0.5\textwidth
% full_filename figs/mcorr_vs_pperp_1
% after \\
\includegraphics[width=0.5\textwidth]{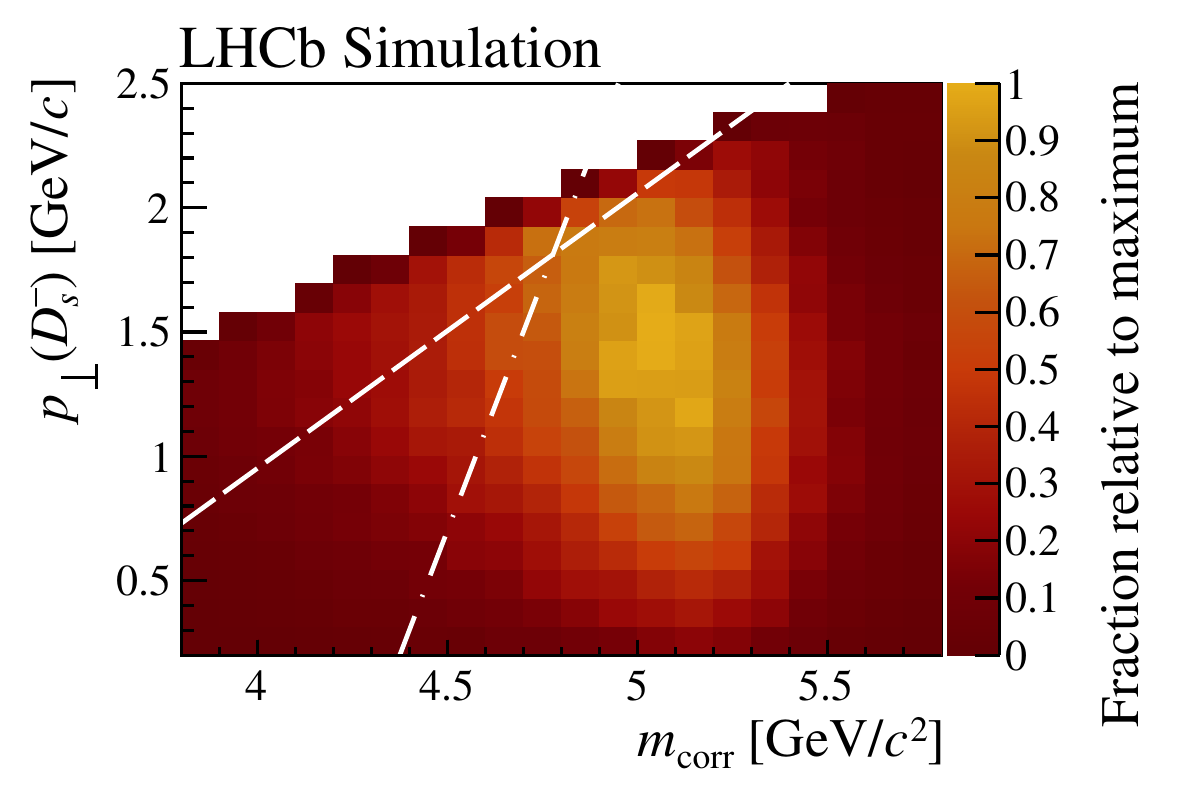}\\
% inside_import 
% before 
% ignored 
% args width=0.5\textwidth
% full_filename figs/mcorr_vs_pperp_2
% after \hfil
\includegraphics[width=0.5\textwidth]{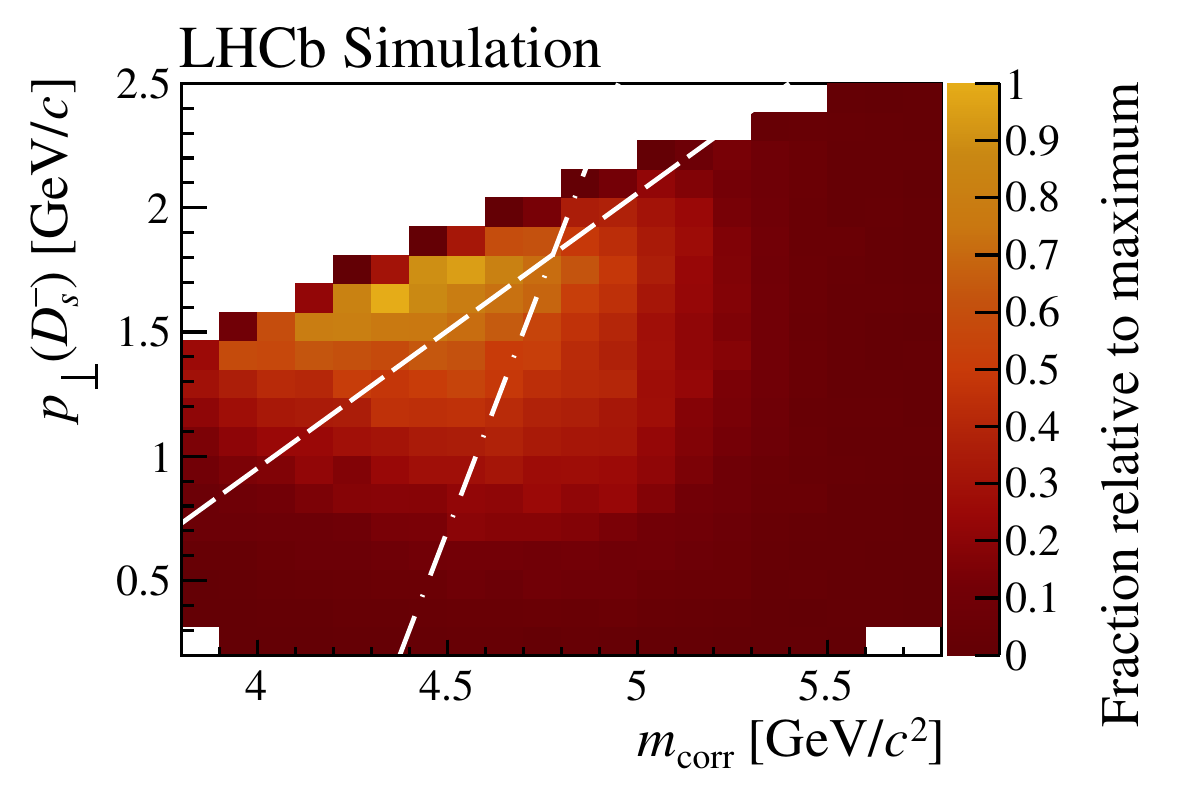}\hfil
% inside_import 
% before 
% ignored 
% args width=0.5\textwidth
% full_filename figs/mcorr_vs_pperp_3
% after \\
\includegraphics[width=0.5\textwidth]{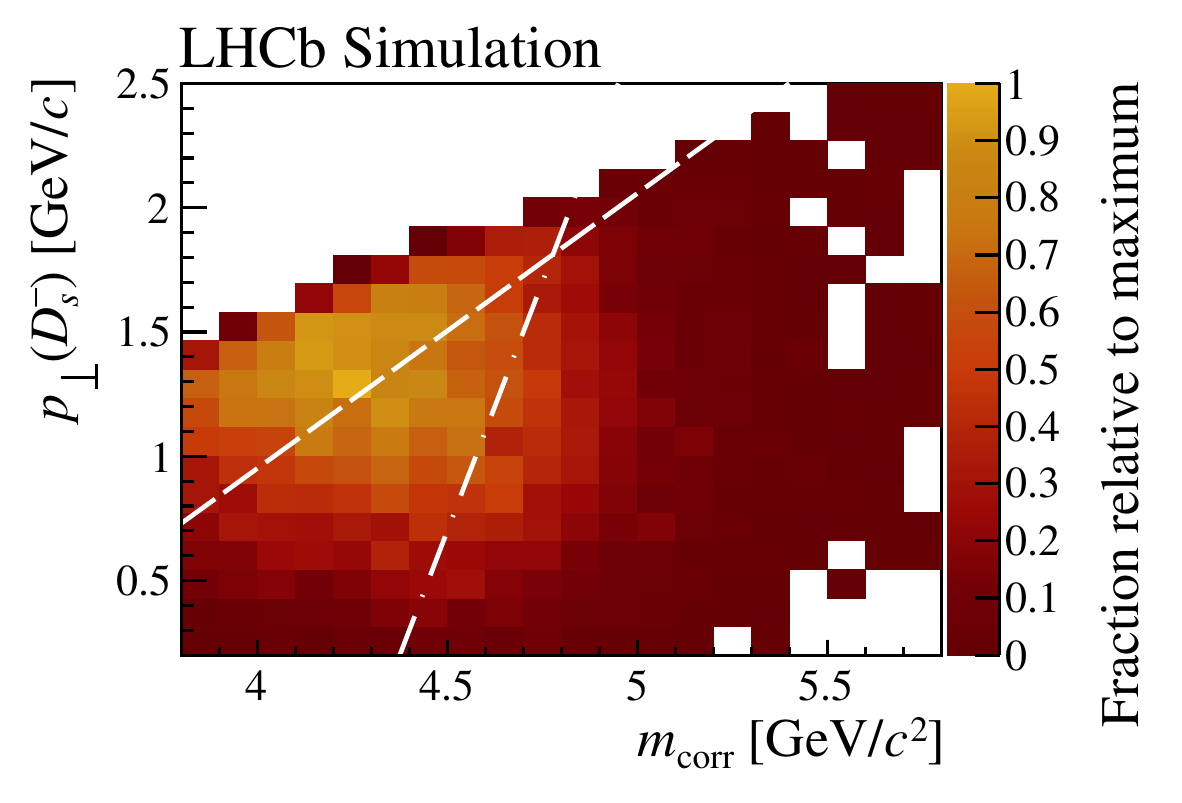}\\
\caption{Two-dimensional distributions of \pperpDs \vs \mcorr for simulated (top-left) \bsDs decays, (top-right) \bsDsS decays, (bottom-left) background decays from \Bs feed-down and \bquark-hadron decays to a doubly-charmed final state, and (bottom-right) background decays from \Bd cross-feed and semitauonic \Bs decays. The background components are grouped according to their shapes in the \mcorr \vs \pperpDs space. The requirement $\pperpDs\, [\gevc\,] < 1.5 + 1.1 \times (\mcorr\,[\gevcc\,] - 4.5)$ is drawn as a dashed line; the dot-dashed line shows the tighter requirement, applied on top of the baseline, which is used in Sec.~\ref{sec:systematics} to further suppress background and assess the systematic uncertainty due to the residual contamination.\label{fig:Bs-templates-mcorr-pperp}}
\end{figure}

The selection of the $\BorBs\to\DorDsorSm\mup\neum$ candidates closely follows that of Ref.~\cite{LHCb-PAPER-2017-004}. Online, a trigger~\cite{LHCb-DP-2012-004} selects events containing a high-\pt muon candidate associated with one, two, or three charged particles, all with origins displaced from the collision points. In the offline reconstruction, the muon candidate is combined with three charged particles consistent with the topology and kinematics of signal \mbox{$\Bs\to[\Kp\Km\pim]_{\Dsm}\mup\neum$} and reference \mbox{$\Bz\to[\Kp\Km\pim]_{\Dm}\mup\neum$} decays. The $\Kp\Km\pim$ mass is restricted to be in the ranges $[1.945,1.995]\gevcc$ and $[1.850,1.890]\gevcc$ to define the inclusive samples of $\Dsm\mup$ signal and $\Dm\mup$ reference candidates, respectively. Cross-contamination between signal and reference samples is smaller than 0.1\%, as estimated from simulation. The $\Kp\Km$ mass must be in the range $[1.008,1.032]\gevcc$, to suppress the background under the \DorDsm peaks and ensure similar kinematic distributions for signal and reference decays. Same-sign $\DorDsm\mun$ candidates are also reconstructed to model combinatorial background from accidental $\DorDsm\mup$ associations. The candidate selection is optimized towards suppressing the background under the charm signals and making same-sign candidates a reliable model for the combinatorial background: track- and vertex-quality, vertex-displacement, transverse-momentum, and particle-identification criteria are chosen to minimize shape and yield differences between same-sign and signal candidates in the $m(\DorDsm\mup) > 5.5\gevcc$ region, where genuine \bquark-hadron decays are kinematically excluded and combinatorial background dominates. Mass vetoes suppress to a negligible level background from misreconstructed decays, such as $\Bs\to \psi^{\scalebox{0.4}{(}\prime\scalebox{0.4}{)}}(\to\mup\mun)\phi(\to\Kp\Km)$ decays where a muon is misidentified as a pion, $\Lb \to \Lc(\to p\Km\pip)\mun\bar{\nu}_\mu X$ decays where the proton is misidentified as a kaon or a pion (and $X$ indicates other possible final-state particles), and $\BorBs\to\DorDsm\pip$ decays where the pion is misidentified as a muon. The requirement $\pperpDorDs\,[\gevc] < 1.5 + 1.1 \times (\mcorr\,[\gevcc] - 4.5)$ is imposed to suppress background from all other partially reconstructed \bquark-hadron decays, as shown in Fig.~\ref{fig:Bs-templates-mcorr-pperp} for \Bs decays. Tighter and looser variations of this requirement are used in Sec.~\ref{sec:systematics} to estimate the systematic uncertainty due to the residual background contamination.

A total of $2.72\times 10^5$ $\Dsm\mup$ and $0.82\times 10^5$ $\Dm\mup$ candidates satisfy the selection criteria. Simulation is used to describe all sources of \bquark-hadron decays contributing to these inclusive samples. Assuming for \bsDs and \bsDsS decays the same branching fractions as for \bdD and \bdDS, respectively, \bsDs and \bsDsS decays are expected to constitute about 30\% and 60\% of the inclusive sample of the selected $\Dsm\mup$ candidates, while  \bdD and \bdDS decays are expected to constitute about 50\% and 30\% of the $\Dm\mup$ sample. The lower expected fraction of semimuonic decays into \DSorDsS mesons for \Bd decays compared to \Bs decays is due to the branching fraction of $\DSm\to\Dm X$ decays. A significant background originates from \BorBs semimuonic decays into excited \DorDsm states other than \DSorDsSm, indicated inclusively as $\DorDs^{**-}$ hereafter, or from decays with a nonresonant combination of a \DorDsorSm with pions. All these decays are referred to as feed-down background in the following. The sum of all feed-down background sources from \Bd decays is expected to total about 9\% of the $\Dm\mup$ sample. For \Bs decays less experimental information is available  to estimate the $\D_s^{**-}$ feed-down contamination to the $\Dsm\mup$ sample. The decays considered here are those into $D_{s0}^{*}(2317)^-$ and $D_{s1}(2460)^-$ mesons, because these states have a mass below the kinematic threshold required to decay strongly into $DK$ or $D^*K$ final states. Decays into the $D_{s1}(2536)^-$ meson are also considered, even if this state is above the $D^*K$ threshold, because it has been observed to decay to a \Dsm meson~\cite{PDG2018}. Branching fractions for these \Bs decays are not known, but, based on the yields measured in Ref.~\cite{LHCb-PAPER-2017-004}, they are estimated to be a few percent of the $\Dsm\mup$ sample. Background from semileptonic \Bu decays into a $\Dm\mup X$ final state is expected to be about 9\% of the $\Dm\mup$ sample, including both semimuonic and semitauonic decays, with $\taup\to\mup\neum\neutb$. Semitauonic \BorBs decays are estimated to contribute less than 1\% to both the $\Dsm\mup$ and $\Dm\mup$ samples, comprising all decays into \DorDsorSm mesons and their excited states. In the case of \Bs decays, as no experimental information is available, assumptions based on measurements of \Bd decays are made, and the same $\D_s^{**-}$ states considered for the semimuonic decays are included. Background can also originate from \Bu, \Bd, \Bs or \Lb decaying into a pair of charm hadrons, where one hadron is the fully reconstructed \DorDsm candidate and the other decays semileptonically. While this background is expected to be negligible in the $\Dm\mup$ sample, it is estimated to be about 2\% of the $\Dsm\mup$ sample, following Ref.~\cite{LHCb-PAPER-2017-004}. Such decays include $\BorBs\to\DorDsorSm\DorDsp$, $\Bp\to\Dbar{}^{\scalebox{0.4}{(}*\scalebox{0.4}{)}0}\DorDSp$, $\Bs\to\Dz\Dsm\Kp$, $\Bs\to\Dm\Dsp\Kz$, $\Lb\to\Lc\DsorDsSm X$, and $\Lb\to\Dsp\Lz\mu^-\bar{\nu}_\mu X$. Cross-feed semileptonic \Bs decays can be neglected in the inclusive $\Dm\mup$ sample, whereas those of \Bd and \Bu decays to final states with a \Dsm candidate and an unreconstructed kaon, such as $\B\to\DsorDsSm K \mup\neum$, must be considered in the $\Dsm\mup$ sample. This contamination is estimated to be at most 2\%.

Reconstruction and selection efficiencies are determined from simulation. Given that  signal decays are measured relative to  reference \Bd decays, only efficiency ratios are needed. They are measured to be $1.568\pm0.008$ for \bsDs relative to \bdD decays, and $1.464\pm0.007$ for \bsDsS relative to \bdDS decays. They depart from unity mainly because of the requirement on $m(\Kp\Km)$ to be around the $\phi(1020)$ mass. This requirement reduces systematic uncertainties due to the modeling of trigger and particle-identification criteria. However, its efficiency relies on an accurate description in the simulation of the $\DorDsm\to\Kp\Km\pim$ amplitude model; a systematic uncertainty is assigned to cover for a possible mismodeling, as discussed in Sec.~\ref{sec:systematics}. An additional difference between the efficiency of signal and reference decays originates from the \Dm lifetime being about two times longer than the \Dsm lifetime~\cite{PDG2018}. The trigger selection is more efficient for decays with closely spaced \BorBs and \DorDsm vertices, favoring smaller \DorDsm flight distances and hence decay times~\cite{LHCb-PAPER-2017-004}. As a consequence, the efficiency for selecting $\Dsm\mup$ candidates in the trigger is about 10\% larger than that for $\Dm\mup$ candidates.

 % end input ./selection.tex
 %
% start input ./fit.tex
\section{Analysis method}\label{sec:fit}
Signal and reference yields can be precisely measured through a fit to the corrected mass distribution following the method of Ref.~\cite{LHCb-PAPER-2017-004}. To be able to access the form factors, yields are measured as a function of the recoil variable $w$ and of the helicity angles, as discussed in Sec.~\ref{sec:formalism}. However, these quantities cannot be computed precisely because of the undetected neutrino and the inability to resolve the \bquark-hadron kinematic properties by balancing it against the accompanying \bquarkbar hadron produced in the event, as done in $e^+e^-$ collisions. 

Approximate methods, based on geometric and kinematic constraints, and on the assumption that only the neutrino is undetected, allow the determination of these quantities up to a two-fold ambiguity in the neutrino momentum component parallel to the \bquark-hadron flight direction~\cite{Kodama:1993ec,Aitala:1998ey,Link:2004uk,Dambach:2006ha}. Such an ambiguity can  be resolved, \eg, by using multivariate regression algorithms~\cite{Ciezarek:2016lqu} or by imposing additional constraints on the \bquark-hadron production~\cite{Stone:2014mza}. These approximate methods have already been successfully used by several LHCb analyses of semileptonic \bquark-hadrons decays~\cite{LHCb-PAPER-2015-013,LHCB-PAPER-2017-016,LHCB-PAPER-2017-027,LHCB-PAPER-2018-024}. However, $\mathcal{O}(20\%)$ inefficiencies are introduced because, due to resolution effects, the second-order equation responsible for the two-fold ambiguity does not always have real solutions. The inability to use candidates for which no real solutions are found also restricts the candidate \mcorr values to be smaller than the nominal \BorBs mass, thus reducing the discriminating power between the different sample components.

\begin{figure}[ht]
\centering
% inside_import 
% before 
% ignored 
% args width=0.5\textwidth
% full_filename figs/Ds_w_vs_pperp
% after \hfil
\includegraphics[width=0.5\textwidth]{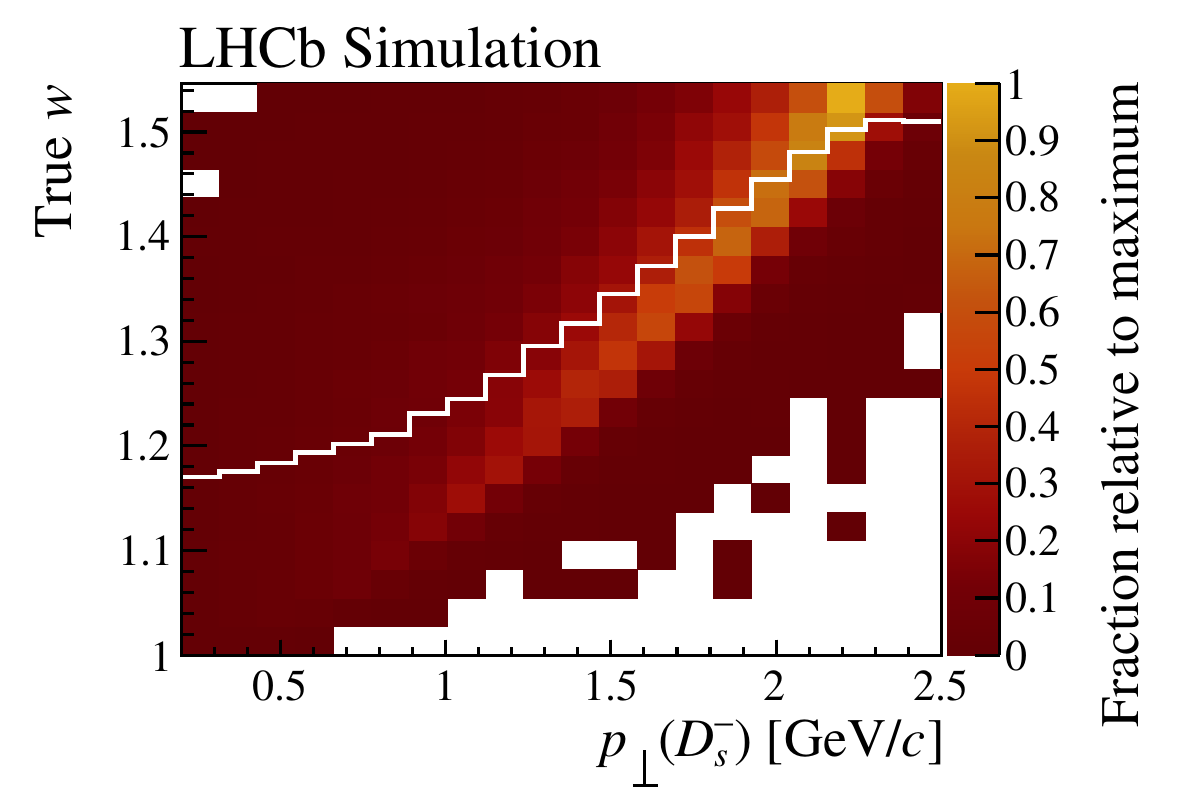}\hfil
% inside_import 
% before 
% ignored 
% args width=0.5\textwidth
% full_filename figs/DsS_w_vs_pperp
% after \\
\includegraphics[width=0.5\textwidth]{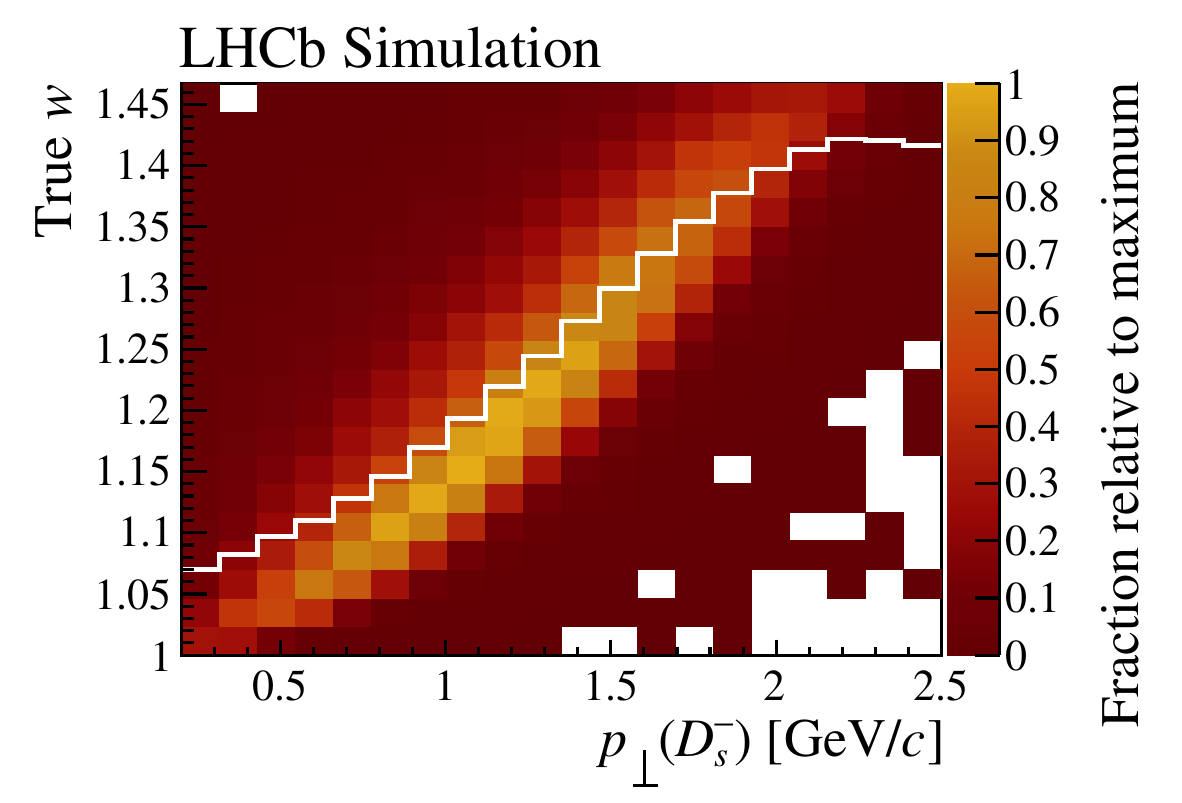}\\
% inside_import 
% before 
% ignored 
% args width=0.5\textwidth
% full_filename figs/ctd_vs_pperp
% after \hfil
\includegraphics[width=0.5\textwidth]{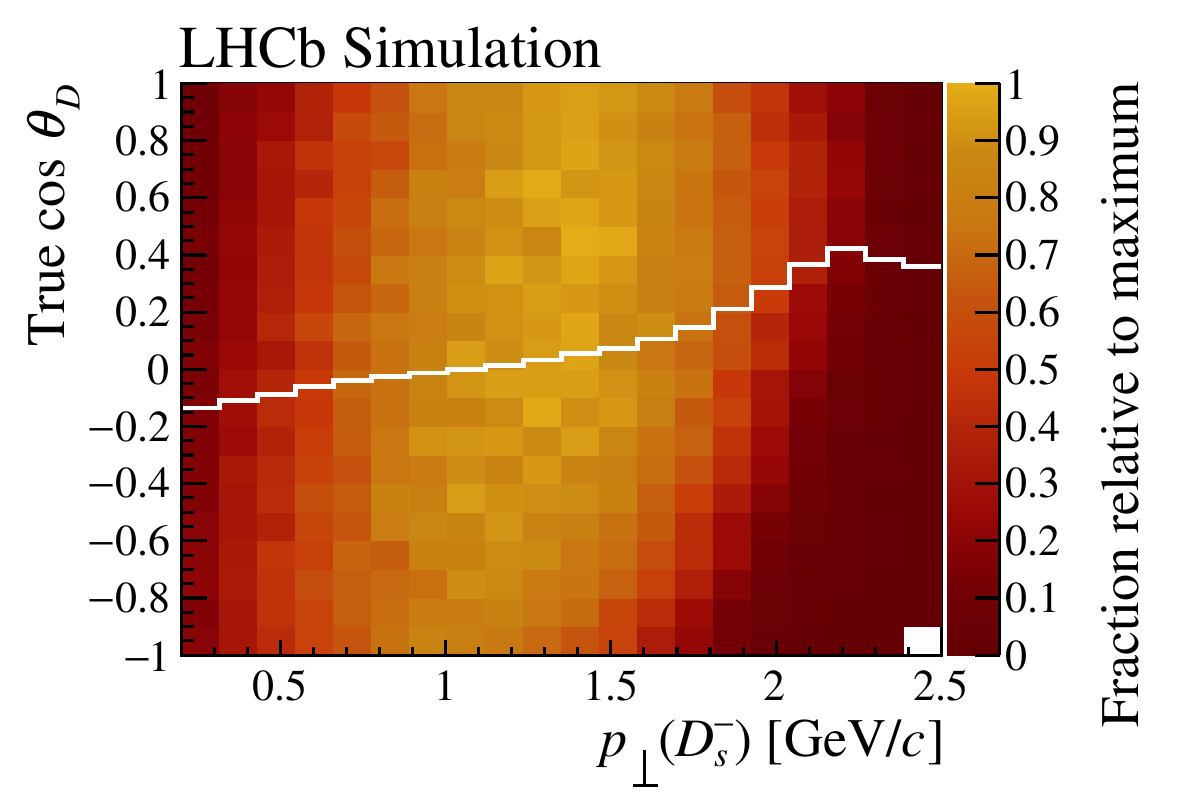}\hfil
% inside_import 
% before 
% ignored 
% args width=0.5\textwidth
% full_filename figs/ctl_vs_pperp
% after \\
\includegraphics[width=0.5\textwidth]{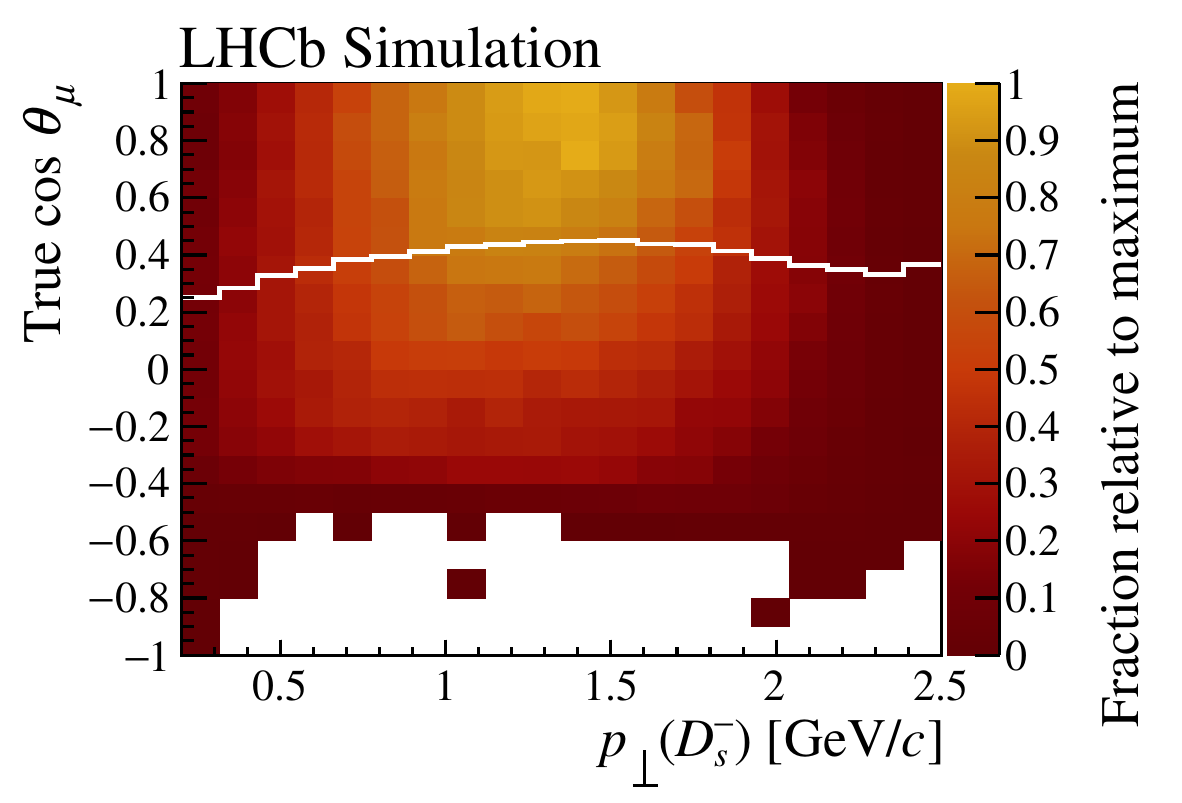}\\
\caption{\label{fig:pperp-qsq} (Top) Distribution of true value of the $w$ recoil variable versus reconstructed \pperpDs for (left) \bsDs and  (right) \bsDsS simulated decays. (Bottom) Distribution of the true values of (left) $\cos\theta_D$ and (right) $\cos\theta_\mu$ versus reconstructed \pperpDs for \bsDsS simulated decays. Only simulated candidates that fulfill the selection requirements are shown. In each histogram the solid line represents the average of the variable displayed on the vertical axis as a function of \pperpDs. The distributions of $\Bd\to \DorDsm\mup\neum$ decays show similar features.}
\end{figure}

To overcome such problems, a novel approach is adopted in this analysis. In \mbox{$\BorBs\to\DorDsm\mup\neum$} decays the component of the \DorDsm momentum perpendicular to the \BorBs flight direction, $\pperp(\DorDsm)$, is opposite and equal in magnitude to the component of the \Wp momentum vector that is perpendicular to the \BorBs flight direction. Therefore, $\pperp(\DorDsm)$ is highly correlated with $w$, as shown in the top-left distribution of Fig.~\ref{fig:pperp-qsq} for \Bs decays. In $\BorBs\to\DSorDsSm\mup\neum$ decays the correlation is kept, as shown in the top-right distribution of Fig.~\ref{fig:pperp-qsq}, because the unreconstructed photon or pion from the \DSorDsSm decay carries very little energy, which only leads to a small dilution. In these decays, the $\pperp(\DorDsm)$ variable is also correlated, albeit to a lesser extent, with the helicity angles $\theta_\mu$ and $\theta_D$, as shown in the bottom distributions of Fig.~\ref{fig:pperp-qsq} for \Bs decays. Through such correlations, the distribution of $\pperp(\DorDsm)$ has a strong dependence on the form factors, particularly on $\mathcal{G}(w)$ for the scalar case and on $h_{A_1}(w)$ for the vector case. Therefore, the form factors can be accessed by analysing the shape of the $\pperp(\DorDsm)$ distribution of the signal decays, with no need to estimate the momentum of the unreconstructed particles. The $\pperp(\DorDsm)$ variable has the experimental advantage of being reconstructed fully from the tracks of the \DorDsm decay products and from the well-measured origin and decay vertices of the \BorBs meson. It is also correlated with \mcorr, and the two variables together provide very efficient discrimination between signal and background decays, which accumulate in different regions of the two-dimensional space spanned by \mcorr and $\pperp(\DorDsm)$, as already shown in Fig.~\ref{fig:Bs-templates-mcorr-pperp} for \Bs decays.

A least-squares fit to the \mcorr--$\pperp(\DorDsm)$ distribution of the selected inclusive samples of $\DorDsm\mup$ candidates is used to simultaneously determine the form factors and (signal) reference yields that are needed for the measurement of \Vcb, or of the ratios of branching fractions \RDorDS. In the fit, the data are described by several fit components, which will be detailed later, separately for the \Bd and \Bs cases. The shape of each component in the \mcorr--$\pperp(\DorDsm)$ space is modeled with two-dimensional histogram templates derived either from simulation (for signal, reference and all physics background decays) or from same-sign data candidates (for combinatorial background). The binning of the histograms is chosen such that there are at least 15 entries per bin (for both data and templates distributions), to guarantee unbiased estimates of the least-squares fit. A few bins at the edges of the \mcorr--$\pperp(\DorDsm)$ space have a smaller number of entries, but studies performed on pseudoexperiments show that they do not introduce biases in the fit results.

Signal templates are built using a per-candidate weight calculated as the ratio between the differential decay rate featuring a given set of form-factors parameters and that with the parameters used in the generation of the simulated samples. The set of parameters of the differential decay rate at numerator is varied in the least-squares minimization. The differential decay rates are given in Eq.~\eqref{eq:rate_vector_4d} for $\BorBs\to\DSorDsSm\mup\neum$ decays, and in Eq.~\eqref{eq:rate_pseudo} for $\BorBs\to\DorDsm\mup\neum$ decays. They are evaluated at the candidate true value of $w$, and of the helicity angles for $\BorBs\to\DSorDsSm\mup\neum$. The \mcorr--$\pperp(\DorDsm)$ templates are rebuilt at each iteration of the least-squares minimization using the values of form-factors parameters probed at that iteration. With this weighting procedure, all efficiency and resolution effects are accounted for, making the templates independent of the form-factor values assumed in the generation of the simulated candidates.

In the fit, the yield of each component is a free parameter. To determine \Vcb, the signal yields, \nsigDorDS, are expressed as the integral of the differential decay rates multiplied by the \Bs lifetime, $\tau$. The signal yields are normalized to the yields, \nrefDorDS, and to the measured branching fractions of the reference \Bd modes, correcting for the efficiency ratios between signal and reference decays, \effratioDorDS. The full expression for the signal yields is
\begin{equation}\label{eq:Ns_Br}
\nsigDorDS = \nDorDS\,\tau\int\frac{\diff\Gamma(\Bs\to\DsorDsSm\mup\neum)}{\diff\zeta}\,\diff\zeta \,,
\end{equation}
where the integral is performed over \mbox{$\zeta\equiv w$} for \bsDs and \mbox{$\zeta\equiv(w,\cos\theta_\mu,\cos\theta_D,\chi)$} for \bsDsS, and where
\begin{align}
\nDorDS       &\equiv\frac{\nrefDorDS\,\effratioDorDS\,\kDorDS}{\BF(\Bd\to\DorDSm\mup\neum)}\,,\\
\mathcal{K}   &\equiv\frac{f_s}{f_d}\,\frac{\BF(\Dsm\to\Kp\Km\pim)}{\BF(\Dm\to\Kp\Km\pim)}\,,\\
\mathcal{K}^* &\equiv\frac{f_s}{f_d}\,\frac{\BF(\Dsm\to\Kp\Km\pim)}{\BF(\Dstarm\to\Dm X)\BF(\Dm\to\Kp\Km\pim)}\,,
\end{align}
with $f_s/f_d$ being the ratio of \Bs- to \Bd-meson production fractions. The dependence on \Vcb in Eq.~\eqref{eq:Ns_Br} is enclosed in the differential decay rate of Eqs.~\eqref{eq:rate_vector_4d} and~\eqref{eq:rate_pseudo}. The other parameters entering the differential decay rate are either left free to float in the fit, together with \Vcb, or constrained to external determinations by a penalty term in the least-squares function, as detailed in the following sections. A similar fit is performed to determine the ratios of branching fractions, with the difference that the expression of the signal yields simplifies to
\begin{equation}\label{eq:Ns_R}
\nsigDorDS = \nrefDorDS\,\effratioDorDS\,\kDorDS\,\RDorDS\,,
\end{equation}
and \RD and \RDS become free parameters instead of \Vcb. In the fit to the reference sample, the yields are free parameters, not expressed in terms of \Vcb. Their histogram templates are functions of the form factors and are allowed to float in the fit.
 % end input ./fit.tex
 %
% start input ./reference.tex
\section{Fit to the reference sample\label{sec:reference}}
The reference yields \nrefDorDS are determined by fitting the \mcorr--\pperpD distribution of the inclusive $\Dm\mup$ sample using the following four components: the two reference decays, \bdD and \bdDS; physics background due to the the sum of semileptonic \Bd feed-down and $\Bu\to\Dm\mup X$ decays; and combinatorial background. The \bdDS template is generated assuming a fraction of approximately 5\% for $\DSm\to\Dm\gamma$ decays and 95\% for $\DSm\to\Dm\pi^0$ decays, according to the measured \DSm branching fractions~\cite{PDG2018}. The physics background components are grouped together into a single template because their \mcorr--\pperpD distributions are too similar to be discriminated by the fit. A contribution from semitauonic decays is neglected because its yield is found to be consistent with zero in an alternate fit in which this component is included, and no significant change of the reference yields is observed. The fit parameters are \nrefDorDS, the yields of the background components and the $\Bd\to\DorDSm\mup\neum$ form-factors parameters expressed in the CLN parametrization: \rsqD, \rsqDS, \Rone and \Rtwo. Given the limited size of the $\Dm\mup$ samples, the CLN parametrization is preferred over BGL because of its reduced number of free parameters. 

\begin{figure}[b]
\centering
% inside_import 
% before 
% ignored 
% args width=0.5\textwidth
% full_filename figs/fit_mcorr_BdKKpi_1112_RFit_oldCut_weigths3_allFFfree_26092019
% after \hfil
\includegraphics[width=0.5\textwidth]{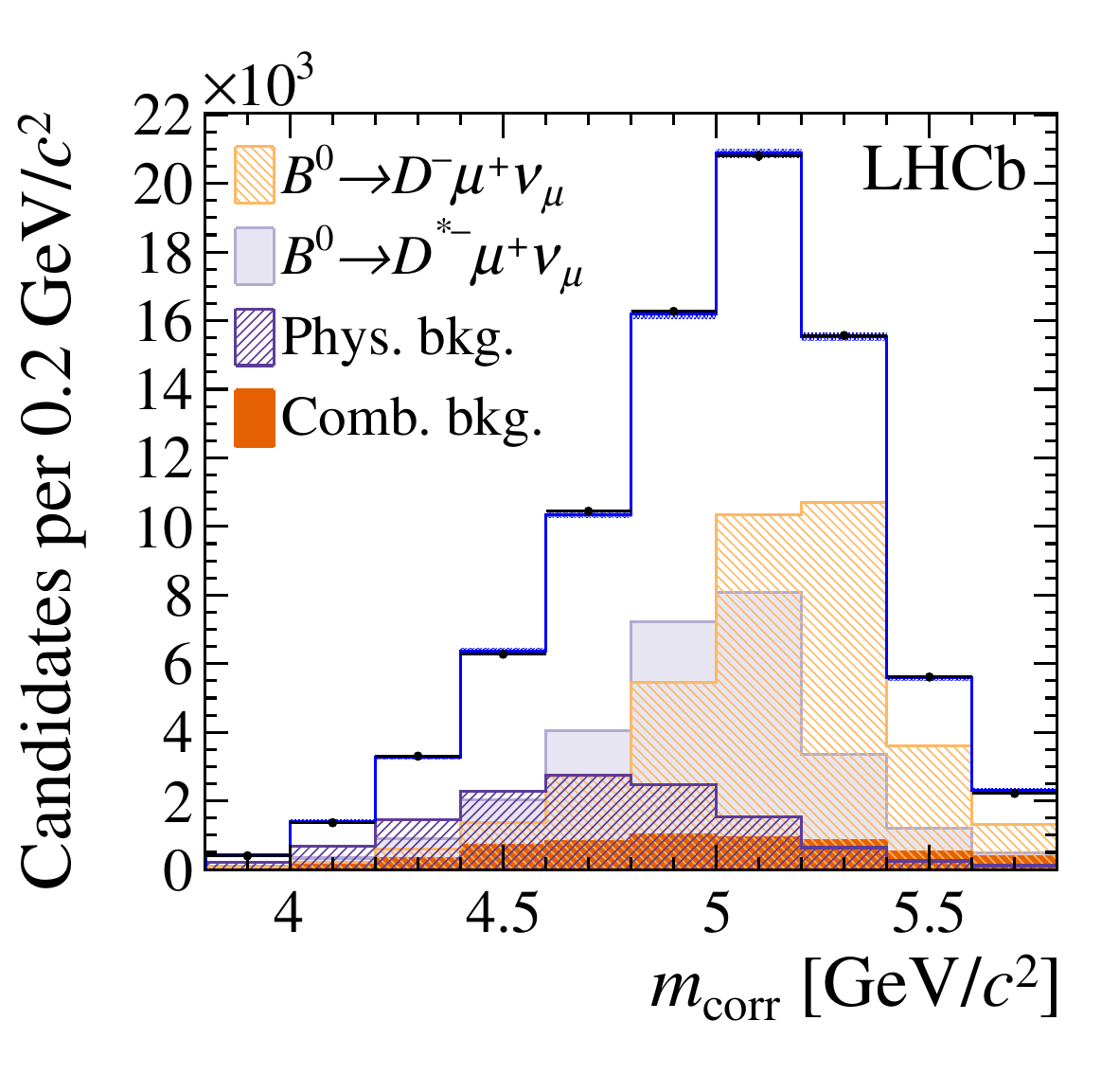}\hfil
% inside_import 
% before 
% ignored 
% args width=0.5\textwidth
% full_filename figs/fit_pperp_BdKKpi_1112_RFit_oldCut_weigths3_allFFfree_26092019
% after \\
\includegraphics[width=0.5\textwidth]{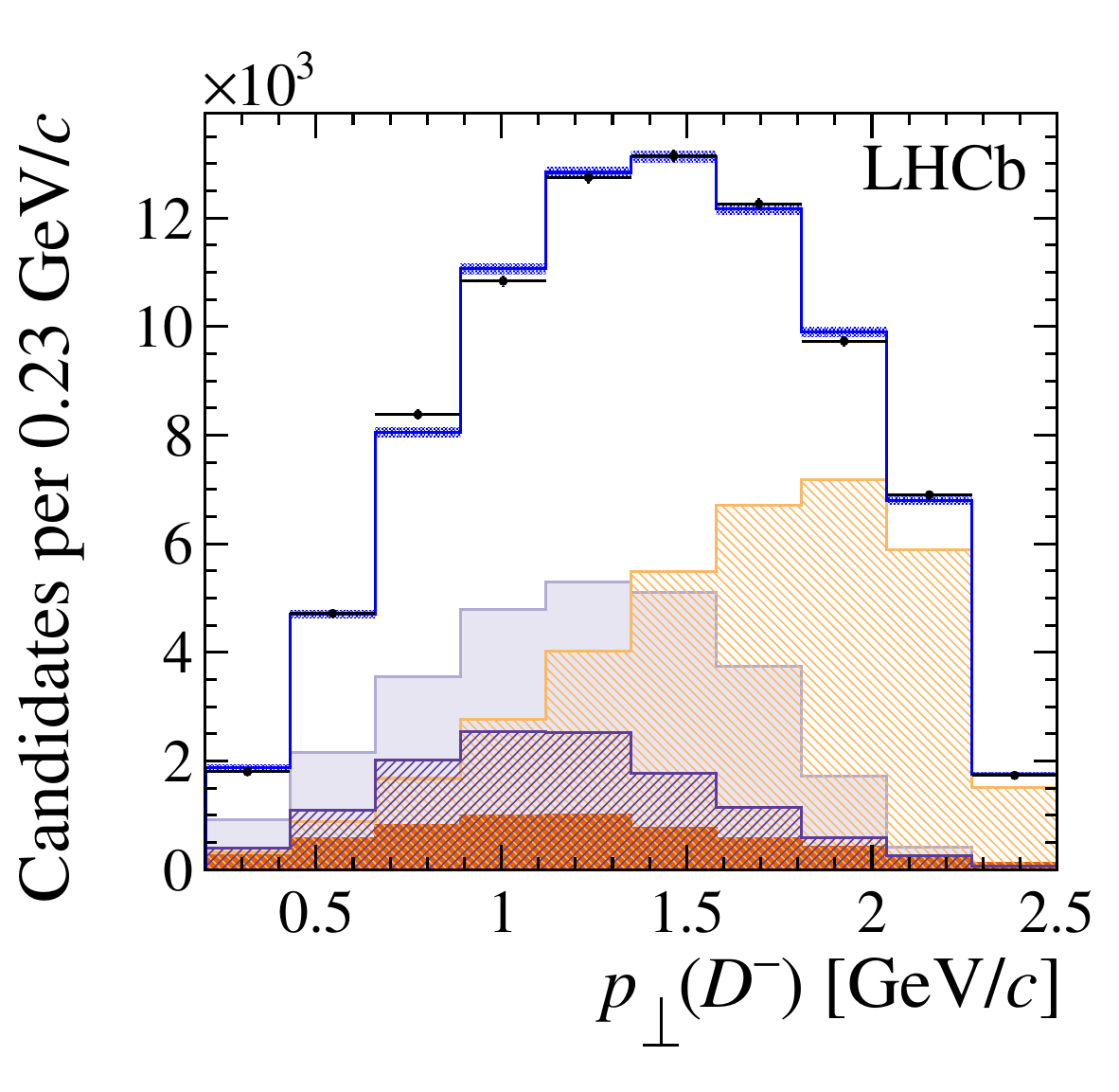}\\
\caption{Distribution of (left) \mcorr and (right) \pperpD for the inclusive sample of reference $\Dm\mup$ candidates, with fit projections overlaid.\label{fig:norm-fit}}
\end{figure}

The reference yields are determined to be $\nrefD = (36.4 \pm 1.6)\times 10^3$ and $\nrefDS = (27.8 \pm 1.2)\times 10^3$ with a correlation of $-70.3\%$. These results do not depend significantly on the choice of the form-factor parametrization. The one-dimensional projections of the fit on the \mcorr and \pperpD variables are shown in Fig.~\ref{fig:norm-fit}. The fit describes the data well with a minimum \chisq/ndf of $76/70$, corresponding to a $p$-value of $29\%$. The form-factors parameters are measured to be in agreement with their world-average values~\cite{HFLAV18}, with relative uncertainties ranging from 20\% to 50\% depending on the parameter. % end input ./reference.tex
 %
% start input ./signal.tex
\section{Fit to the signal sample\label{sec:signal}}
The fit function for the $\Dsm\mup$ sample features five components: the two signal decays, \bsDs and \bsDsS; a background component made by the sum of semimuonic \Bs feed-down decays and \bquark-hadron decays to a doubly charmed final state; a background component made by the sum of cross-feed semileptonic \Bd decays and semitauonic \Bs decays; combinatorial background. The \bsDsS template is generated assuming a fraction of approximately 94\% for $\DsSm\to\Dsm\gamma$ decays and 6\% for $\DsSm\to\Dsm\pi^0$ decays, according to the measured \DsSm branching fractions~\cite{PDG2018}. The physics background components that are merged together in the two templates have very similar shapes in the \mcorr \vs \pperpDs plane and cannot be discriminated by the fit when considered as separate components. They are therefore merged according to the expected approximate fractions. 

\begin{table}[t]
\caption{\label{tab:exp-inputs} External inputs based on experimental measurements.}
\centering
\begin{tabular}{lcc}
\toprule
Parameter & Value  & Reference \\
\midrule
$f_s/f_d\times\BF(\Dsm\to\Km\Kp\pim)\times\tau$ [\!$\ps$]  & $0.0191\pm0.0008$ & \cite{LHCb-PAPER-2018-050,LHCB-PAPER-2019-020} \\
$\BF(\Dm\to\Km\Kp\pim)$  & $0.00993 \pm 0.00024$  & \cite{PDG2018}\\
$\BF(\DSm\to\Dm X)$ & $0.323 \pm 0.006$  & \cite{PDG2018}\\
$\BF(\bdD)$         & $0.0231 \pm 0.0010$  & \cite{PDG2018}\\
$\BF(\bdDS)$        & $0.0505 \pm 0.0014$  & \cite{PDG2018}\\
\Bs mass [\!$\gevcc$]   & $5.36688 \pm 0.00017$  & \cite{PDG2018}\\
\Dsm mass [\!$\gevcc$]  & $1.96834 \pm 0.00007$  & \cite{PDG2018}\\
\DsSm mass [\!$\gevcc$] & $2.1122 \pm 0.0004$  & \cite{PDG2018}\\
\bottomrule
\end{tabular}
\end{table}

\begin{table}[t]
\caption{\label{tab:theory-inputs} External inputs based on theory calculations. The values and their correlations are derived in  Appendix~\ref{app:LQCD}, based on Ref.~\cite{McLean:2019qcx}.}
\centering
\begin{tabular}{lcc}
\toprule
Parameter & Value & Reference \\
\midrule
\etaEW    & $1.0066\pm0.0050$ & \cite{Sirlin}\\
\Fnorm    & $0.902 \pm0.013$  & \cite{McLean:2019sds} \\
\midrule
\multicolumn{3}{l}{CLN parametrization} \\
\Gnorm    & $1.07 \pm0.04$ & \cite{McLean:2019qcx} \\
\rsqDs    & $1.23 \pm0.05$ & \cite{McLean:2019qcx} \\
\midrule
\multicolumn{3}{l}{BGL parametrization} \\
\Gnorm    &  $1.07 \pm0.04$ & \cite{McLean:2019qcx} \\
$d_1$     & $-0.012\pm0.008\phantom{-}$ & \cite{McLean:2019qcx} \\
$d_2$     & $-0.24\pm0.05\phantom{-}$ & \cite{McLean:2019qcx} \\
\bottomrule
\end{tabular}
\end{table}

The yields of the five components are free parameters in the fit, with the signal yields expressed in terms of the  parameters of interest according to Eq.~\eqref{eq:Ns_Br}, when determining \Vcb, or Eq.~\eqref{eq:Ns_R}, when determining \RDorDS. The measurement relies on the external inputs reported in Tables~\ref{tab:exp-inputs} and~\ref{tab:theory-inputs}. Correlations between external inputs, \eg, between \nrefD and \nrefDS or between the LQCD inputs, are accounted for in the fit. The value of $f_s/f_d$ is derived from the measurement of Ref.~\cite{LHCb-PAPER-2018-050}, which is the most precise available. It is obtained using an independent sample of semileptonic \BorBs decays collected with the LHCb detector in $pp$ collisions at the center-of-mass energy of 13\tev. This measurement uses the branching fraction of the $\Dsm\to\Kp\Km\pim$ decay and the \Bs lifetime as external inputs~\cite{PDG2018}. To properly account for all correlations, the value of the product $f_s/f_d\times\BF(\Dsm\to\Km\Kp\pim)\times\tau$ is derived directly from Ref.~\cite{LHCb-PAPER-2018-050}. The measured dependence of $f_s/f_d$ on the collision energy~\cite{LHCB-PAPER-2019-020} is also accounted for in the computation, by scaling the 13\tev measurement to the value at 7 and $8\tev$ needed in this analysis. All other branching fractions and the particle masses are taken from Ref.~\cite{PDG2018}. The external inputs listed in Table~\ref{tab:theory-inputs} are based exclusively on theory calculations: \etaEW and \Fnorm are constrained to the values reported in Refs.~\cite{Sirlin} and \cite{McLean:2019sds}, respectively; the constraints on the \bsDs form factors are based on the LQCD calculations of Ref.~\cite{McLean:2019qcx}, which provide the form factor $f_+(z)$ over the full \qsq spectrum using the parametrization proposed by Bourrely, Caprini and Lellouch (BCL)~\cite{Bourrely:2008za}. In Appendix~\ref{app:LQCD}, the corresponding CLN and BGL parameters reported in Table~\ref{tab:theory-inputs} are derived. 

\subsection{Determination of \boldmath\Vcb with the CLN parametrization}
The analysis in the CLN parametrization uses the form factors defined in Eqs.~\eqref{eq:hA1_CLN}--\eqref{eq:R2_CLN}, for \bsDsS decays, and in Eq.~\eqref{eq:G_slope}, for \bsDs decays. The form-factor parameters \rsqDsS, \Rone, \Rtwo are free to float in the fit, while \Fnorm, \Gnorm and \rsqDs are constrained.

\begin{figure}[t]
\centering
% inside_import 
% before 
% ignored 
% args width=0.5\textwidth
% full_filename figs/fit_mcorr_BsKKpi_1112_VcbFit_oldCut_weigths3_CLNgaussLQCD_26092019
% after \hfil
\includegraphics[width=0.5\textwidth]{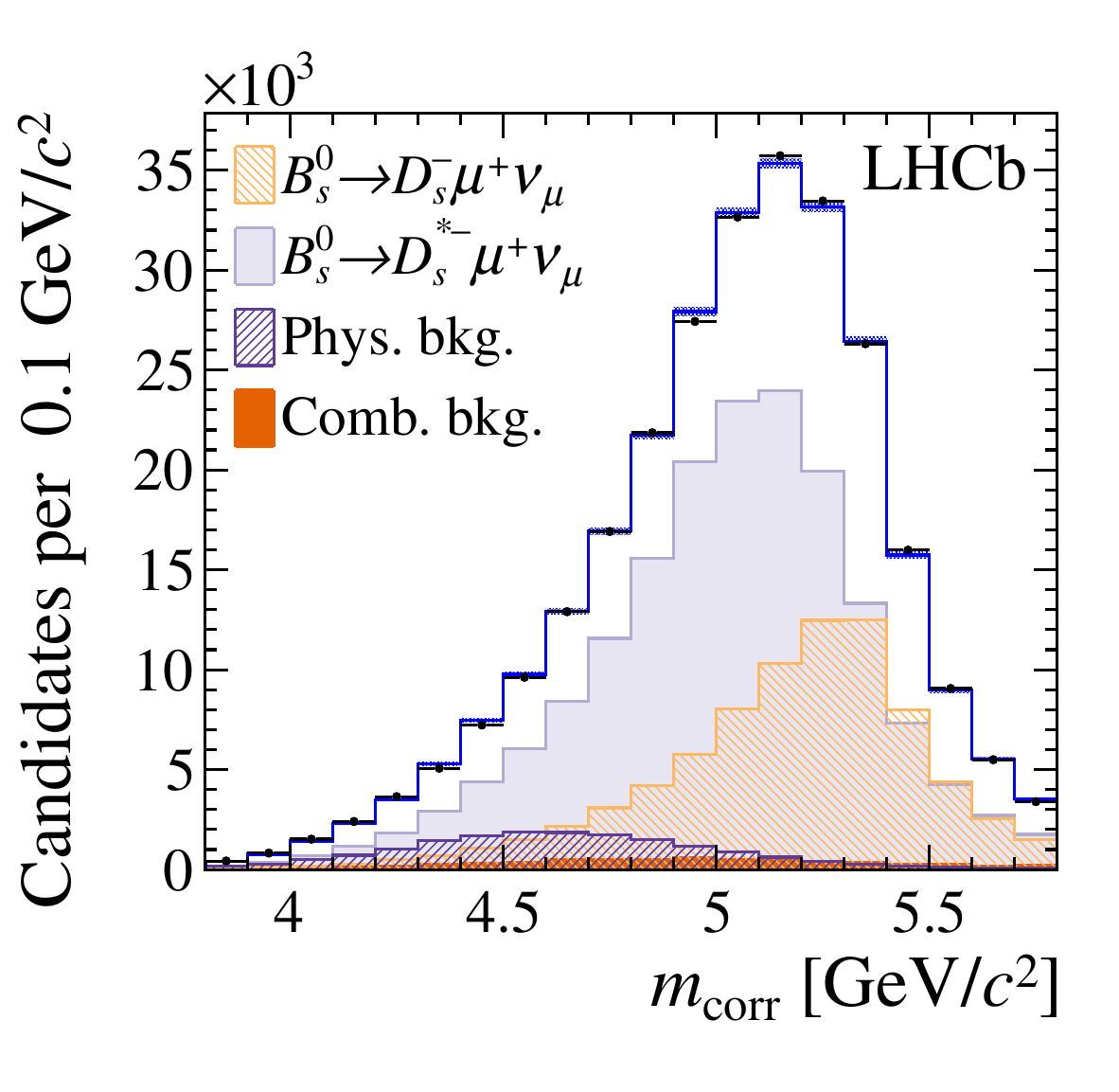}\hfil
% inside_import 
% before 
% ignored 
% args width=0.5\textwidth
% full_filename figs/fit_pperp_BsKKpi_1112_VcbFit_oldCut_weigths3_CLNgaussLQCD_26092019
% after \\
\includegraphics[width=0.5\textwidth]{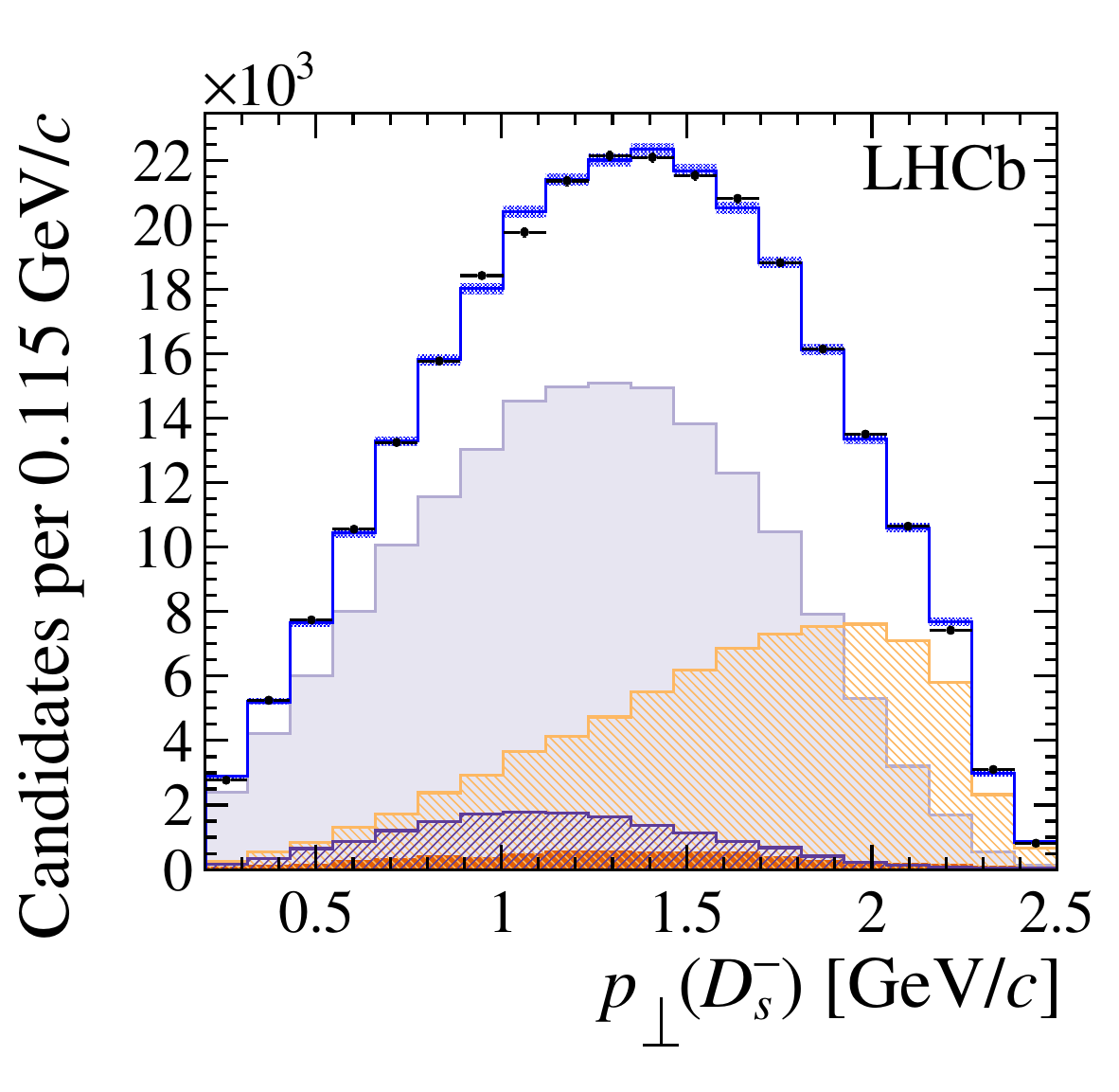}\\
\caption{Distribution of (left) \mcorr and (right) \pperpDs for the inclusive sample of signal $\Dsm\mup$ candidates, with fit projections based on the CLN parametrization overlaid. The projections of the two physics background components  are merged together  for displaying purposes.\label{fig:Bs-fit-CLN}}
\end{figure}

One-dimensional projections of the fit results on \mcorr and \pperpDs are shown in Fig.~\ref{fig:Bs-fit-CLN}. The fit has a minimum \chisq/ndf of $279/285$, corresponding to a $p$-value of $58\%$. The results for the parameters of interest are reported in Table~\ref{tab:CLN-results}. In addition to \Vcb, these include the form-factors parameters that are determined exclusively by the data, such as \rsqDsS, \Rone and \Rtwo, and those for which the precision improves compared to the external constraints, such as \Gnorm and \rsqDs. Detailed fit results for all parameters, including their correlations, are reported in Appendix~\ref{app:corr-fits}. The uncertainties returned by the fit include the statistical contribution arising from the limited size of the data and simulation samples\,\stat, and the contribution due to the external inputs\,\ext. The calculation of this latter contribution is detailed in Sec.~\ref{sec:systematics}. The value of \Vcb, $(\VcbResultCLN\pm \VcbStatErrCLN\stat\pm\VcbExtErrCLN\ext)\times\VcbUnits$, agrees with the exclusive determination from \Bu and \Bd decays~\cite{HFLAV18}. When only \Gnorm is constrained and \rsqDs is left free, the fit returns $\rsqDs = 1.30 \pm 0.06\stat$, in agreement with the LQCD estimation, and $\Vcb=(41.8 \pm 0.8\stat \pm 1.2\ext)\times\VcbUnits$. Including the constraint on \rsqDs improves the statistical precision on \Vcb by about 20\% and also that on \Gnorm by 10\%, because of the large correlation between \Gnorm and \rsqDs.

\begin{table}[t]
\caption{\label{tab:CLN-results} Fit results in the CLN parametrization. The uncertainty is split into two contributions, statistical\,\stat and that due to the external inputs\,\ext.}
\centering
\begin{tabular}{lr}
\toprule
Parameter & \multicolumn{1}{c}{Value}   \\
\midrule
\Vcb [$\VcbUnits$] & $\VcbResultCLN\phantom{00} \pm \VcbStatErrCLN\phantom{00}\stat \pm \VcbExtErrCLN\phantom{00}\ext$ \\
\Gnorm  & $\GnormCLNResult \pm \GnormCLNStatErr\stat \pm \GnormCLNExtErr\ext$ \\
\rsqDs  & $\rsqDsResult\phantom{0} \pm \rsqDsStatErr\phantom{0}\stat \pm \rsqDsExtErr\phantom{0}\ext$ \\
\rsqDsS & $\rsqDsSResult\phantom{0} \pm \rsqDsSStatErr\phantom{0}\stat \pm \rsqDsSExtErr\phantom{0}\ext$ \\
\Rone   & $\RoneResult\phantom{0} \pm \RoneStatErr\phantom{0}\stat \pm \RoneExtErr\phantom{0}\ext$ \\
\Rtwo   & $\RtwoResult\phantom{0} \pm \RtwoStatErr\phantom{0}\stat \pm \RtwoExtErr\phantom{0}\ext$ \\
\bottomrule
\end{tabular}
\end{table}

\subsection{Determination of \Vcb with the BGL parametrization}
The BGL form-factor functions are given by Eqs.~\eqref{eq:hA1_BGL}--\eqref{eq:R2_BGL}, for \bsDsS decays, and Eq.~\eqref{eq:BGLffP}, for \bsDs decays. The fit parameters are the coefficients of the series of the $z$ expansion. For \bsDsS decays, the expansion of the $f$, $g$ and $\mathcal{F}_1$ form factors is truncated after the first order in $z$. The coefficients $b_0$ and $c_0$ are constrained through \Fnorm using Eqs.~\eqref{eq:b0_F1} and \eqref{eq:c0_b0}. The coefficients $b_1$, $a_0$, $a_1$, and $c_1$ are free parameters. For \bsDs decays, the expansion of the  $f_+(z)$ form factor is truncated after the second order in $z$ and the coefficients $d_0$, $d_1$ and $d_2$, are constrained to the values obtained in Appendix~\ref{app:LQCD} using Ref.~\cite{McLean:2019qcx}, with $d_0$ expressed in terms of the parameter \Gnorm using Eq.~\eqref{eq:d0_G0}. No constraints from the unitarity bounds of Eqs.~\eqref{eq:unitarity_BGL_DS} and \eqref{eq:unitarity_BGL_D} are imposed, to avoid potential biases on the parameters or fit instabilities due to convergence at the boundary of the parameter space.

\begin{figure}[tb]
\centering
% inside_import 
% before 
% ignored 
% args width=0.5\textwidth
% full_filename figs/BGLvsCLN_bkgsubtracted_BsDs
% after \hfil
\includegraphics[width=0.5\textwidth]{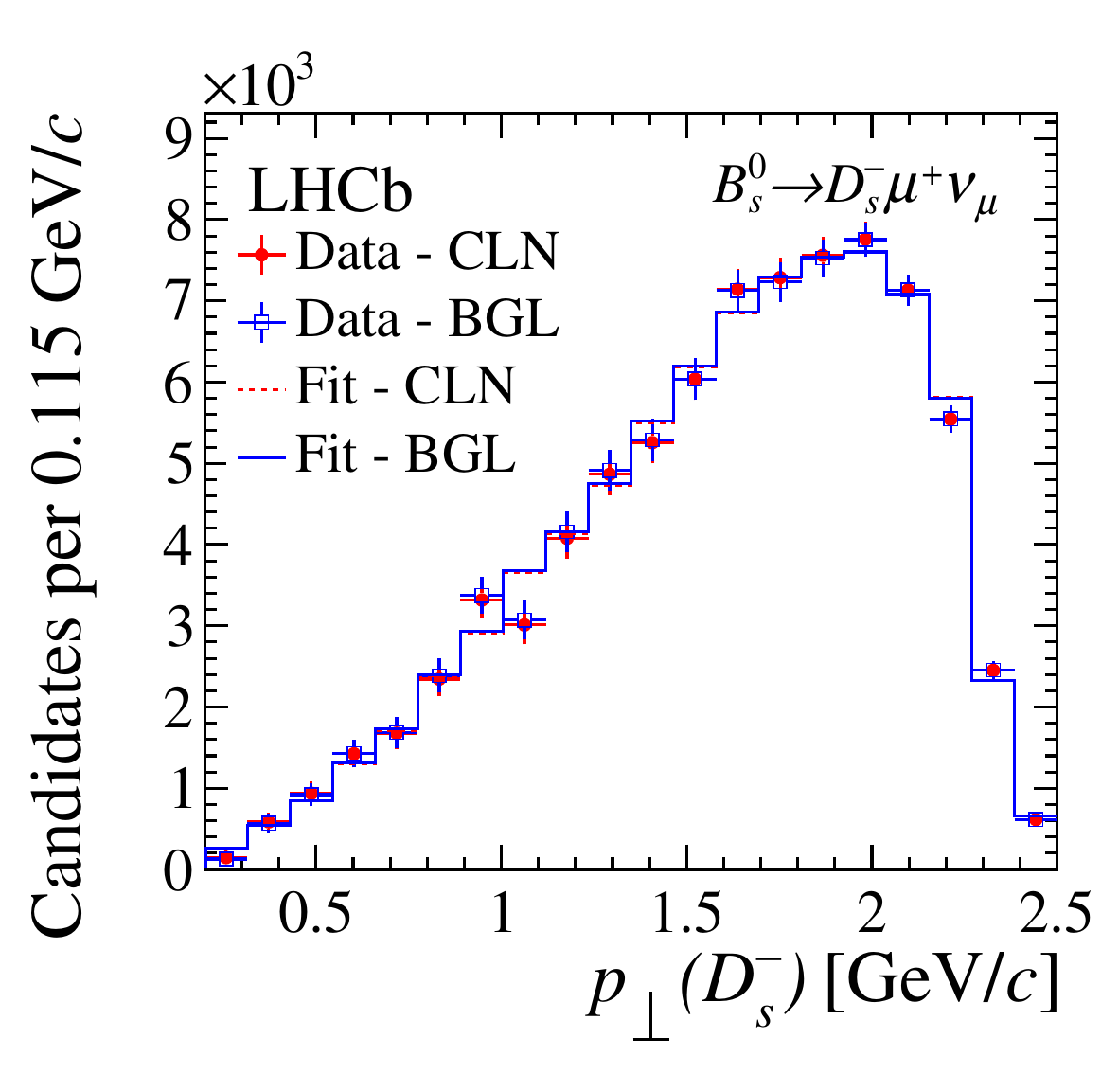}\hfil
% inside_import 
% before 
% ignored 
% args width=0.5\textwidth
% full_filename figs/BGLvsCLN_bkgsubtracted_BsDsS
% after \\
\includegraphics[width=0.5\textwidth]{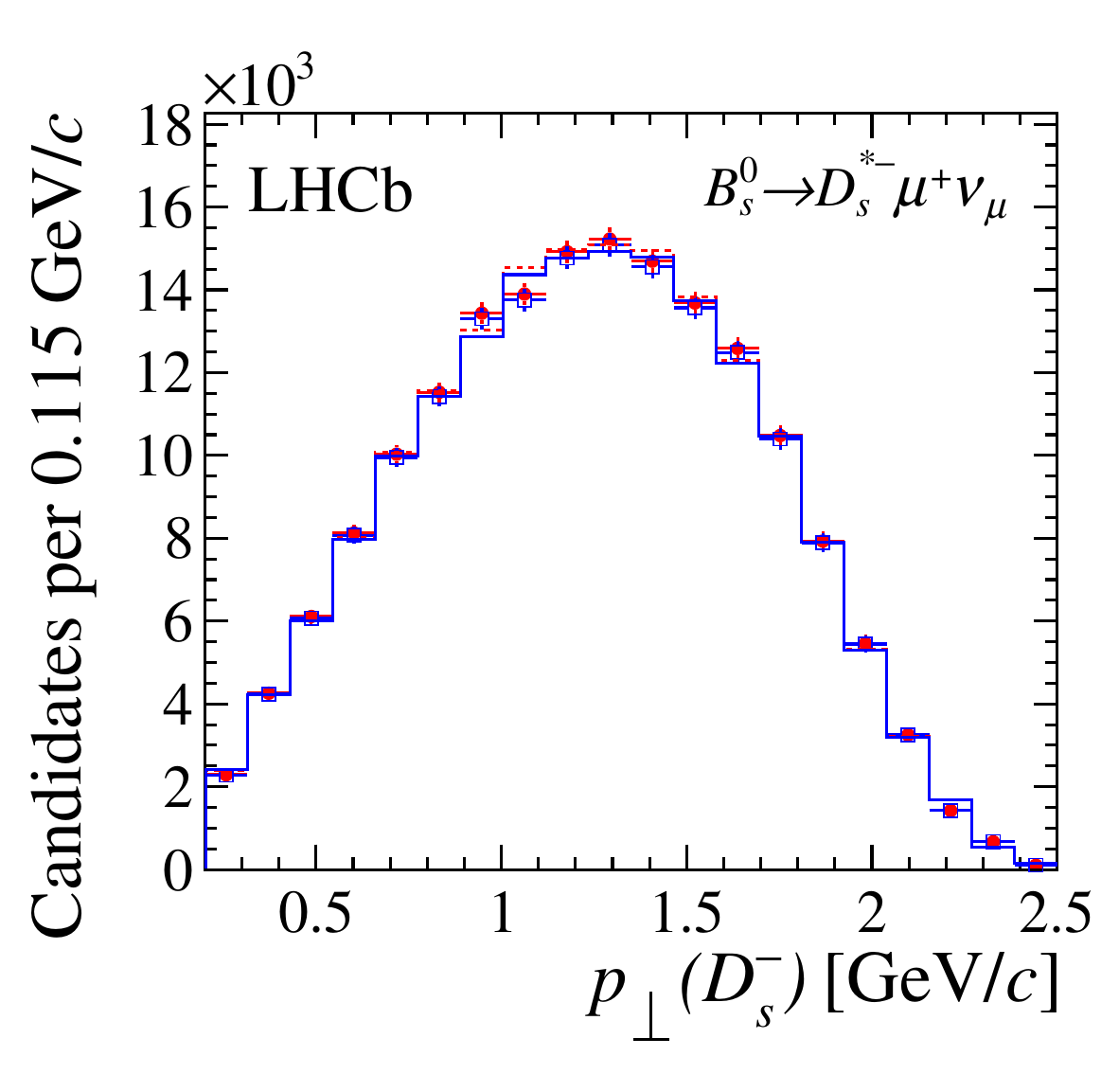}\\
\caption{Background-subtracted distribution of \pperpDs for (left) \bsDs and (right) \bsDsS decays obtained from the fit based on the (red closed points, dashed line) CLN and (blue open points, solid line) BGL parametrizations, with corresponding fit projections overlaid.\label{fig:Bs-signal-CLN-vs-BGL}}
\end{figure}

The fit has minimum \chisq/ndf of $276/284$, corresponding to a $p$-value of $63\%$. Figure~\ref{fig:Bs-signal-CLN-vs-BGL} shows a comparison of the \pperpDs background-subtracted distributions obtained with the CLN and BGL fits. No significant differences are found between the two fits for both \bsDs and \bsDsS decays. The fit results for the parameters of interest are reported in Table~\ref{tab:BGL-results}. Detailed fit results for all parameters, including their correlations, are reported in Appendix~\ref{app:corr-fits}. The values found for the form-factor coefficients satisfy the unitarity bounds of Eqs.~\eqref{eq:unitarity_BGL_DS} and \eqref{eq:unitarity_BGL_D}. The value of \Vcb is found to be $(\VcbResultBGL\pm \VcbStatErrBGL\stat\pm\VcbExtErrBGL\ext)\times\VcbUnits$, in agreement with the CLN analysis. The correlation between the BGL and CLN results is $34.0\%$. When only \Gnorm is constrained and $d_1$ and $d_2$ are left free, \Vcb is found to be $(42.2 \pm 1.5\stat \pm 1.2\ext)\times\VcbUnits$. The constraints on $d_1$ and $d_2$ improve the statistical precision on \Vcb by about 50\% and that on \Gnorm by 10\%. Without such constraints, the fit returns $d_1 = 0.02 \pm 0.05\stat$ and $d_2 = -0.9 \pm 0.8\stat$, both in agreement with the LQCD estimations, and within the unitarity bound of Eq.~\eqref{eq:unitarity_BGL_D}.

\begin{table}[tb]
\caption{\label{tab:BGL-results} Fit results in the BGL parametrization. The uncertainty is split into two contributions, statistical\,\stat and that due to the uncertainty on the external inputs\,\ext.}
\centering
\begin{tabular}{lr}
\toprule
Parameter & \multicolumn{1}{c}{Value}   \\
\midrule
\Vcb [$\VcbUnits$] & $\VcbResultBGL\phantom{000} \pm \VcbStatErrBGL\phantom{000}\stat \pm \VcbExtErrBGL\phantom{000}\ext$ \\
\Gnorm& $\GnormBGLResult\phantom{0} \pm \GnormBGLStatErr\phantom{0}\stat \pm \GnormBGLExtErr\phantom{0}\ext$ \\
$d_1$ & $\doneResult\phantom{0} \pm \doneStatErr\phantom{0} \stat \pm \doneExtErr\phantom{0}\ext$\\
$d_2$ & $\dtwoResult\phantom{00} \pm \dtwoStatErr\phantom{00} \stat \pm \dtwoExtErr\phantom{00}\ext$ \\
$b_1$ & $\boneResult\phantom{00} \pm \boneStatErr\phantom{00} \stat \pm \boneExtErr\phantom{00}\ext$\\
$a_0$ & $\azeroResult\phantom{0} \pm \azeroStatErr\phantom{0} \stat \pm \azeroExtErr\phantom{0}\ext$\\
$a_1$ & $\aoneResult\phantom{00} \pm \aoneStatErr\phantom{00} \stat \pm \aoneExtErr\phantom{00}\ext$\\ 
$c_1$ & $\coneResult \pm \coneStatErr \stat \pm \coneExtErr\ext$\\ 
\bottomrule
\end{tabular}
\end{table}

Variations of the orders of the form-factor expansions have been probed for the \bsDsS decay, while  for the \bsDs decay the expansion is kept at order $z^2$ to exploit the constraints on $d_1$ and $d_2$. A first alternative fit, where only the order zero of the $g$ series is considered by fixing $a_1$ to zero, returns a $p$-value of $62\%$ and \mbox{$\Vcb = (41.7 \pm 0.6\ext \pm 1.2\ext)\times\VcbUnits$}, in agreement with the nominal result of Table~\ref{tab:BGL-results}. The shift in the central value of \Vcb is consistent with that observed in pseudoexperiments where data are generated by using the nominal truncation and fit with the zero-order expansion of $g$. In a second alternative fit, $g$ is kept at order zero and $f$ is expanded at order $z^2$, by adding the coefficient $b_2$ as a free parameter. The fit has a $p$-value of $64\%$ and returns $\Vcb=(42.2\pm0.8\stat\pm1.2\ext)\times\VcbUnits$ and $b_2= 1.9 \pm 1.4\stat$. Configurations at lower order than those considered for $f$ and $\mathcal{F}_1$ lead to  poor fit quality and are discarded. Higher orders than those discussed here are not considered because they result in fit instabilities and degrade the sensitivity to \Vcb and to the form-factor coefficients.

\subsection{Determination of \RD and \RDS}
The ratios of \Bs to \Bd branching fractions are determined by a fit where the signal yields are expressed using Eq.~\eqref{eq:Ns_R}, with \RD and \RDS as free parameters. In the fit, the constraint on $f_s/f_d\times\BF(\Dsm\to\Kp\Km\pim)$ is obtained by dividing the value of the first row of Table~\ref{tab:exp-inputs} by the \Bs lifetime $\tau$~\cite{PDG2018}. The form factors are expressed in the CLN parametrization and a systematic uncertainty is assigned for this arbitrary choice, as discussed in Sec.~\ref{sec:systematics}. The fit returns $\RD = \RDsResult \pm \RDsStatErr\stat \pm \RDsExtErr\ext$ and $\RDS = \RDsSResult \pm \RDsSStatErr\stat \pm \RDsSExtErr\ext$, with a $p$-value of $59\%$. Detailed fit results for all fit parameters, including their correlations, are reported in Appendix~\ref{app:corr-fits}. % end input ./signal.tex
 %
% start input ./systematics.tex
\section{Systematic uncertainties\label{sec:systematics}}
Systematic uncertainties affecting the measurements can be split into two main categories: those due to external inputs, indicated with\,\ext; and those due to the experimental methods, indicated with\,\syst. The individual contributions for each category are discussed in the following and are reported in Table~\ref{tab:syst-summary}, together with the statistical uncertainties.

%
% start input ./systematics-bigtable.tex
\begin{sidewaystable}[p]
\caption{Summary of the uncertainties affecting the measured parameters. The upper section reports the systematic uncertainties due to the external inputs\,\ext, the middle section those due to the experimental methods\,\syst, and the lower section the statistical uncertainties\,\stat. For the first source of uncertainty the multiplication by $\tau$ holds only for the \Vcb fits.\label{tab:syst-summary}}
\centering
\resizebox{\textheight}{!}{
\begin{tabular}{lcccccccccccccccccc}
\toprule
\multirow{4}{*}{Source} & \multicolumn{18}{c}{Uncertainty} \\
\cmidrule{2-19}
 & \multicolumn{6}{c}{CLN parametrization} & & \multicolumn{8}{c}{BGL parametrization} & & \\
\cmidrule{2-7}\cmidrule{9-16}
 & \Vcb        & \rsqDs      & \Gnorm      & \rsqDsS     & \Rone       & \Rtwo       &  & \Vcb        & $d_1$       & $d_2$       & \Gnorm      & $b_1$       & $c_1$       & $a_0$       & $a_1$       & & \RD & \RDS \\
 & [$10^{-3}$] & [$10^{-1}$] & [$10^{-2}$] & [$10^{-1}$] & [$10^{-1}$] & [$10^{-1}$] &  & [$10^{-3}$] & [$10^{-2}$] & [$10^{-1}$] & [$10^{-2}$] & [$10^{-1}$] & [$10^{-3}$] & [$10^{-2}$] & [$10^{-1}$] & & [$10^{-1}$] & [$10^{-1}$]\\
 \midrule
$f_s/f_d\times\BF(\Dsm\to\Kp\Km\pim)(\times\tau)$
 & $0.8$ 	 & $0.0$ 	 & $0.0$ 	 & $0.0$ 	 & $0.0$ 	 & $0.0$   & & $0.8$ & $0.0$ & $0.0$ & $0.0$ & $0.0$ & $0.0$ & $0.0$ & $0.1$ & & $0.4$ 	 & $0.4$ \\
$\BF(\Dm\to\Km\Kp\pim)$
 & $0.5$ & $0.0$ & $0.0$ & $0.0$ & $0.0$ & $0.0$ & & $0.5$ & $0.0$ & $0.0$ & $0.0$ & $0.0$ & $0.0$ & $0.0$ & $0.1$ & & $0.3$ & $0.3$ \\
$\BF(\DSm\to\Dm X)$
 & $0.2$ & $0.0$ & $0.1$ & $0.0$ & $0.1$ & $0.0$ & & $0.1$ & $0.0$ & $0.0$ & $0.1$ & $0.0$ & $0.2$ & $0.0$ & $0.3$ & & --    & $0.2$ \\
$\BF(\bdD)$
 & $0.4$ & $0.0$ & $0.3$ & $0.1$ & $0.2$ & $0.1$ & & $0.5$ & $0.1$ & $0.0$ & $0.1$ & $0.1$ & $0.4$ & $0.1$ & $0.7$ & & --    & -- \\
$\BF(\bdDS)$
 & $0.3$ & $0.0$ & $0.2$ & $0.1$ & $0.1$ & $0.1$ & & $0.2$ & $0.0$ & $0.0$ & $0.1$ & $0.1$ & $0.3$ & $0.1$ & $0.4$ & & --    & -- \\
$m(\Bs)$, $m(\DorDSm)$
 & $0.0$ & $0.0$ & $0.0$ & $0.0$ & $0.0$ & $0.0$ & & $0.0$ & $0.0$ & $0.0$ & $0.0$ & $0.0$ & $0.0$ & $0.0$ & $0.1$ & & --    & -- \\
\etaEW
 & $0.2$ & $0.0$ & $0.0$ & $0.0$ & $0.0$ & $0.0$ & & $0.2$ & $0.0$ & $0.0$ & $0.0$ & $0.0$ & $0.0$ & $0.0$ & $0.1$ & & --    & -- \\
\Fnorm
 & $0.3$ & $0.0$ & $0.2$ & $0.1$ & $0.1$ & $0.1$ & & $0.3$ & $0.0$ & $0.0$ & $0.1$ & $0.1$ & $0.3$ & $0.1$ & $0.5$ & & --    & -- \\
\cmidrule{2-19}
External inputs\,\ext
 & $1.2$ & $0.0$ & $0.4$ & $0.1$ & $0.2$ & $0.1$ & & $1.2$ & $0.1$ & $0.0$ & $0.1$ & $0.1$ & $0.6$ & $0.1$ & $0.8$ & & $0.5$ & $0.5$ \\
\midrule
$\DorDsm\to\Kp\Km\pim$ model
 & $0.8$ & $0.0$ & $0.0$ & $0.0$ & $0.0$ & $0.0$ & & $0.8$ & $0.0$ & $0.0$ & $0.0$ & $0.0$ & $0.0$ & $0.0$ & $0.0$ & & $0.5$ & $0.4$ \\
Background
 & $0.4$ & $0.3$ & $2.2$ & $0.5$ & $0.9$ & $0.7$ & & $0.1$ & $0.5$ & $0.2$ & $2.3$ & $0.7$ & $2.0$ & $0.5$ & $2.0$ & & $0.4$ & $0.6$ \\
Fit bias
 & $0.0$ & $0.0$ & $0.0$ & $0.0$ & $0.0$ & $0.0$ & & $0.2$ & $0.0$ & $0.0$ & $0.0$ & $0.2$ & $0.4$ & $0.2$ & $0.4$ & & $0.0$ & $0.0$ \\
Corrections to simulation
 & $0.0$ & $0.0$ & $0.5$ & $0.0$ & $0.1$ & $0.0$ & & $0.0$ & $0.1$ & $0.0$ & $0.1$ & $0.0$ & $0.0$ & $0.0$ & $0.1$ & & $0.0$ & $0.0$ \\
Form-factor parametrization
 & --    & --    & --    & --    & --    & --    & & --    & --    & --    & --    & --    & --    & --    & --    & & $0.0$ & $0.1$ \\
\cmidrule{2-19}
Experimental\,\syst
 & $0.9$ & $0.3$ & $2.2$ & $0.5$ & $0.9$ & $0.7$ & & $0.9$ & $0.5$ & $0.2$ & $2.3$ & $0.7$ & $2.1$ & $0.5$ & $2.0$ & & $0.6$ & $0.7$ \\
\midrule               
Statistical\,\stat
 & $0.6$ & $0.5$ & $3.4$ & $1.7$ & $2.5$ & $1.6$ & & $0.8$ & $0.7$ & $0.5$ & $3.4$ & $0.7$ & $2.2$ & $0.9$ & $2.6$ & & $0.5$ & $0.5$ \\ 
\bottomrule
\end{tabular}}
\end{sidewaystable}
 % end input ./systematics-bigtable.tex
 
The uncertainties returned by the fit include the statistical contribution arising from the finite size of the data and simulation samples, and the contribution due to the external inputs that constrain some of the fit parameters through penalty terms in the least-squares function. To evaluate the purely statistical component, a second fit is performed with all external parameters fixed to the values determined by the first fit. The contribution due to the external inputs is then obtained by subtracting in quadrature the uncertainties from the two sets of results. The procedure is repeated for each individual input to estimate its contribution to the uncertainty. The results are reported in the upper section of Table~\ref{tab:syst-summary}. 
Here the uncertainty on $f_s/f_d\times\BF(\Dsm\to\Kp\Km\pim)(\times\tau)$ comprises also that due to a difference in the distribution of the transverse momentum of the $\DorDsm\mup$ system with respect to Ref.~\cite{LHCb-PAPER-2018-050}, which results in a relative 1\% change of the value of  $f_s/f_d$. The branching fractions of the \Bd decays taken in input are obtained from averages that assume isospin symmetry in decays of the \Y4S meson~\cite{PDG2018}. This symmetry is observed to hold with a precision of 1 to 2\%, and no uncertainty is assigned. However, it is noted that considering the correction suggested in Ref.~\cite{Jung:2015yma} increases the value of \Vcb by $0.2 \times \VcbUnits$ both in the CLN and BGL parametrizations.

The efficiency of the requirement that limits $m(\Kp\Km)$ to be around the $\phi(1020)$ mass is evaluated using simulation. Given that the simulated model of the intermediate amplitudes contributing to the $\DorDsm\to\Kp\Km\pim$ decays may be inaccurate, a systematic uncertainty is estimated by comparing the efficiency of the $m(\Kp\Km)$ requirement derived from simulation with that based on data from an independent control sample of \mbox{$\DorDsm\to\Kp\Km\pim$} decays. The efficiency ratios \effratioDorDS change by a relative $-4\%$ when substituting the simulation-based efficiency of the $m(\Kp\Km)$ requirement with that determined from data. This variation modifies the values of \Vcb, \RD and \RDS found in the fit, while producing negligible shifts to the form-factor parameters. The differences with respect to the nominal values are assigned as systematic uncertainties.

The knowledge of the physics backgrounds contributing to the inclusive $\Dsm\mup$ sample is limited by the lack of experimental measurements of exclusive semileptonic \Bs decays. These background components are, however, well separated in the \mcorr \vs \pperpDs plane and their contribution is reduced to a few percent by the requirement \mbox{$\pperpDs\,[\gevc] < 1.5 + 1.1 \times (\mcorr\,[\gevcc] - 4.5)$} (dashed line in Fig.~\ref{fig:Bs-templates-mcorr-pperp}). To quantify by how much the assumed background composition can affect the determination of the parameters of interest, the fit is repeated by varying the requirements on the \mcorr \vs \pperpDorDs plane for both signal and reference samples. In the first variation, the more restrictive requirement \mbox{$\pperpDorDs\,[\gevc] < 0.7 + 4.0\times(\mcorr\,[\gevcc] - 4.5)$} is added on top of the baseline selection to further halve the expected background fractions. This requirement is shown as a dot-dashed line in Fig.~\ref{fig:Bs-templates-mcorr-pperp}, for the \Bs case. In the second variation, the baseline requirement is removed to allow maximum background contamination, which doubles with respect to that of the nominal selection. For both variations, the resulting samples are fit accounting for changes in the templates and in the efficiencies. The residuals for each parameter are computed as the difference between the values obtained in the alternative and baseline fits. The root-mean-square deviation of the residuals is taken as systematic uncertainty.

The analysis method is validated using large ensembles of pseudoexperiments, generated by resampling with repetitions (bootstrapping~\cite{efron:1979}) the samples of simulated signal and background decays and the same-sign data that model the combinatorial background. The relative proportions of signal and background components of the nominal fit to data are reproduced. Signal decays are generated by using both the CLN and BGL parametrizations with the form factors determined in the fit to data. Each sample is fit with the same form-factor parametrization used in the generation, and residuals between the fit and the generation values of each parameter are computed. The residuals that are observed to be at least two standard deviations different from zero are assigned as systematic uncertainties.

The simulated samples are corrected for mismodeling of the reconstruction and selection efficiency, of the response of the particle identification algorithms, and of the kinematic properties of the generated \BorBs meson. A systematic uncertainty is assigned by varying the corrections within their uncertainties.

The measurement of \RDorDS is performed only in the CLN parametrization, because, as shown in Fig.~\ref{fig:Bs-signal-CLN-vs-BGL}, the signal templates are marginally affected by the choice of the form-factor parametrization. Nevertheless, a systematic uncertainty is assigned as the shift in the \RDorDS central values when fitting the data with the BGL parametrization.

The experimental systematic uncertainties are combined together, accounting for their correlations, in the middle section of Table~\ref{tab:syst-summary}. The correlations are reported in Appendix~\ref{app:corr-fits}.

As a consistency test, the fit is repeated by expressing the signal yields of the \bsDs and \bsDsS decays in terms of two different \Vcb parameters. The fit returns values of the two parameters in agreement with each other within one standard deviation.

Finally, a data-based null test of the analysis method is performed using a control sample of $\Bd\to\DorDSm\mup\neum$ decays where the \Dm decays to the Cabibbo-favored $\Kp\pim\pim$ final state. These decays are normalized to the same $\Bd\to\DorDSm\mup\neum$ decays, with $\Dm\to[\Kp\Km]_\phi\pim$, used in the default analysis to measure ratios of branching fractions between control and reference decays consistent with unity. The control sample is selected with criteria very similar to those of the reference sample, but the different \Dm final state introduces differences between the efficiencies of the control and reference decays that are $40\%$ larger than those between signal and reference decays. The control sample features the same fit components as described in Sec.~\ref{sec:reference} for the reference sample, with signal and background decays modeled with simulation and combinatorial background with same-sign data. External inputs are changed to reflect the replacement of the signal with the control decays. Fits are performed using both the CLN and the BGL parametrizations. In both cases, the ratios of branching fractions between control and reference decays are all measured to be compatible with unity with 5 to 6\% relative precision. % end input ./systematics.tex
 %
% start input ./results.tex
\section{Final results and conclusions\label{sec:results}}
A study of the \bsDs and \bsDsS decays is performed using proton-proton collision data collected with the LHCb detector at center-of-mass energies of 7 and 8\tev, corresponding to an integrated luminosity of 3\invfb. A novel analysis method is used to identify the two exclusive decay modes from the inclusive sample of selected $\Dsm\mup$ candidates, and measure the CKM matrix element \Vcb using \bdD and \bdDS decays as normalization. The analysis is performed with both the CLN~\cite{Caprini:1997mu} and BGL~\cite{Boyd:1994tt,Boyd:1995sq,Boyd:1997kz} parametrizations to determine
\begin{align*} 
\Vcb_{\rm CLN} &= (\VcbResultCLN \pm \VcbStatErrCLN \stat\pm \VcbSystErrCLN \syst \pm \VcbExtErrCLN\ext)\times \VcbUnits\,,\\ 
\Vcb_{\rm BGL} &= (\VcbResultBGL \pm \VcbStatErrBGL \stat\pm \VcbSystErrBGL \syst \pm \VcbExtErrBGL\ext )\times \VcbUnits\,,
\end{align*}
where the first uncertainties are statistical (including contributions from both data and simulation), the second systematic, and the third due to the limited knowledge of the external inputs. The two results are compatible, when accounting for their correlation. These are the first determinations of \Vcb from exclusive decays at a hadron collider and the first using \Bs decays. The results are in agreement with the exclusive measurements based on \Bd and \Bu decays, and as well with the inclusive determination~\cite{HFLAV18}.

The ratios of the branching fractions of the exclusive $\Bs\to\DsorDsSm\mup\neum$ decays relative to those of the exclusive $\Bd\to\DorDSm\mup\neum$ decays are measured to be 
\begin{align*}
\RD\equiv\frac{\BF(\bsDs)}{\BF(\bdD)} &= \RDsResult \pm \RDsStatErr\stat\pm \RDsSystErr\syst \pm \RDsExtErr\ext\,,\\
\RDS\equiv\frac{\BF(\bsDsS)}{\BF(\bdDS)} &= \RDsSResult \pm \RDsSStatErr\stat\pm \RDsSSystErr\syst \pm \RDsSExtErr\ext\,.
\end{align*}
Taking the measured values of $\BF(\bdD)$ and $\BF(\bdDS)$ as additional inputs~\cite{PDG2018}, the following exclusive branching fractions are determined for the first time
\begin{align*}
\BF(\bsDs) &= \left(\BFDsResult   \pm \BFDsStatErr\stat\pm \BFDsSystErr\syst \pm \BFDsExtErr\ext\right)\times\BFUnits\,,\\
\BF(\bsDsS) &= \left(\BFDsSResult \pm \BFDsSStatErr\stat\pm \BFDsSSystErr\syst \pm \BFDsSExtErr\ext\right)\times\BFUnits\,,
\end{align*}
where the third uncertainties also include the contribution due to the limited knowledge of the normalization branching fractions. Finally, the ratio of \bsDs to \bsDsS branching fractions is determined to be
\begin{equation*}
\frac{\BF(\bsDs)}{\BF(\bsDsS)}=0.464\pm0.013\stat\pm0.043\syst\,.
\end{equation*}

The novel method employed in this analysis can also be used to measure \Vcb with semileptonic \Bd decays at LHCb. In this case, the uncertainty from the external inputs can be substantially decreased, as the dominant contribution in the current measurement is due to the knowledge of the \Bs- to \Bd-meson production ratio $f_s/f_d$. The limiting factor for \Bd decays stems from the knowledge of the reference decays branching fractions, but these are expected to improve from new measurements at the Belle~II experiment~\cite{Kou:2018nap}.
 % end input ./results.tex
 %
% start input ./acknowledgements.tex
\section*{Acknowledgements}
\noindent We express our gratitude to our colleagues in the CERN
accelerator departments for the excellent performance of the LHC. We
thank the technical and administrative staff at the LHCb
institutes.
We acknowledge support from CERN and from the national agencies:
CAPES, CNPq, FAPERJ and FINEP (Brazil); 
MOST and NSFC (China); 
CNRS/IN2P3 (France); 
BMBF, DFG and MPG (Germany); 
INFN (Italy); 
NWO (Netherlands); 
MNiSW and NCN (Poland); 
MEN/IFA (Romania); 
MSHE (Russia); 
MinECo (Spain); 
SNSF and SER (Switzerland); 
NASU (Ukraine); 
STFC (United Kingdom); 
DOE NP and NSF (USA).
We acknowledge the computing resources that are provided by CERN, IN2P3
(France), KIT and DESY (Germany), INFN (Italy), SURF (Netherlands),
PIC (Spain), GridPP (United Kingdom), RRCKI and Yandex
LLC (Russia), CSCS (Switzerland), IFIN-HH (Romania), CBPF (Brazil),
PL-GRID (Poland) and OSC (USA).
We are indebted to the communities behind the multiple open-source
software packages on which we depend.
Individual groups or members have received support from
AvH Foundation (Germany);
EPLANET, Marie Sk\l{}odowska-Curie Actions and ERC (European Union);
ANR, Labex P2IO and OCEVU, and R\'{e}gion Auvergne-Rh\^{o}ne-Alpes (France);
Key Research Program of Frontier Sciences of CAS, CAS PIFI, and the Thousand Talents Program (China);
RFBR, RSF and Yandex LLC (Russia);
GVA, XuntaGal and GENCAT (Spain);
the Royal Society
and the Leverhulme Trust (United Kingdom).

 % end input ./acknowledgements.tex
 
\clearpage
\appendix
%
% start input ./appLQCD.tex
\section{Lattice QCD calculation for \boldmath\bsDs form factors\label{app:LQCD}}

References~\cite{Monahan:2017uby,McLean:2019qcx} report LQCD calculations of the form-factor function over the full \qsq spectrum for \bsDs decays. The calculations differ in the methodology and in the treatment of the sea quarks, with Ref.~\cite{Monahan:2017uby} using ensembles that include \mbox{2+1} flavors and Ref.~\cite{McLean:2019qcx} using \mbox{2+1+1} flavors. The two calculations agree.

\begin{table}[h]
\caption{\label{tab:LQCD-BCL} Coefficients of the $f_+(w)$ form factor in the BCL parametrization from Ref.~\cite{McLean:2019qcx}.}
\centering
\begin{tabular}{lcccc}
\toprule
\multirow{2}{*}{Parameter} & \multirow{2}{*}{Value} & \multicolumn{3}{c}{Covariance} \\
\cmidrule{3-5}
          &       & $a^{\rm BCL}_0$ & $a^{\rm BCL}_1$ & $a^{\rm BCL}_2$ \\
\midrule
$a^{\rm BCL}_0$ & $\phantom{-}0.66574$ & $0.00015$ & $0.00022$ & $0.00003$ \\
$a^{\rm BCL}_1$ & $-3.23599$ &           & $0.20443$ & $0.10080$ \\ 
$a^{\rm BCL}_2$ & $-0.07478$ &           &           & $4.04413$ \\ 
\bottomrule
\end{tabular}
\end{table}

The results reported in Ref.~\cite{McLean:2019qcx} are expressed in the BCL parametrization~\cite{Bourrely:2008za}, with the series expanded up to order $z^2$ (see Appendix A of Ref.~\cite{McLean:2019qcx}). The parameters describing the $f_+(w)$ form factor are reported in Table~\ref{tab:LQCD-BCL}. To be used in this analysis, they need to be translated into the CLN and BGL parametrizations. For this purpose, one thousand ensembles, each consisting of ten million \qsq values distributed according to $f_+(w)$, are generated by sampling the BCL parameters within their covariance. Each sample is then fit with the CLN and BGL equations of Sec.~\ref{sec:formalism} to derive the corresponding set of parameters. Each fit parameter features a Gaussian distribution. The central value and uncertainty of each parameter are defined as the mean and the width of these distributions, respectively. In the CLN parametrization, the derived parameters are \mbox{$\Gnorm = 1.07 \pm 0.04$} and \mbox{$\rsqDs = 1.23 \pm 0.05$}, with a correlation of $84.2\%$. Both values are in agreement with the results reported in Ref.~\cite{Monahan:2017uby}, $\Gnorm = 1.068 \pm 0.040$ and \mbox{$\rsqDs = 1.244 \pm 0.076$}. (A combination is not attempted because of the unknown correlation between the two LQCD calculations.) In the BGL parametrization, the derived parameters are \mbox{$\Gnorm = 1.07 \pm 0.04$}, \mbox{$d_1 = -0.012 \pm 0.008$} and \mbox{$d_2 = -0.24 \pm 0.05$}, with correlation coefficients \mbox{$\varrho(\Gnorm, d_1) = -82.4\%$}, \mbox{$\varrho(\Gnorm, d_2)=-37.2\%$} and \mbox{$\varrho(d_1,d_2) = 10.0\%$}.

 % end input ./appLQCD.tex
 %
% start input ./appDetailed.tex
\section{Detailed fit results\label{app:corr-fits}}

Detailed results for the \Vcb fits, in both the CLN and BGL parametrizations, are reported in
Table~\ref{tab:detailed-Vcb}. The full correlation matrices are given in Tables~\ref{tab:corr-CLN} and \ref{tab:corr-BGL}, separately for the CLN and BGL configurations. Detailed results for the \RD and \RDS fit are given in Table~\ref{tab:detailed-R}, with correlations in Table~\ref{tab:corr-R}.

\begin{table}[h]
\caption{Detailed results for the \Vcb fits. The uncertainties on the free parameters include the statistical contribution and that due to the external inputs.\label{tab:detailed-Vcb}}
\centering
\resizebox{\textwidth}{!}{
\begin{tabular}{lccc}
\toprule
\multirow{2}{*}{Parameter} & \multicolumn{2}{c}{Value} & \multirow{2}{*}{Constraint} \\
\cmidrule{2-3}
                        & CLN fit             & BGL fit                       &  \\
\midrule
\Vcb [$\VcbUnits$]        & $\VcbResultCLN \pm  1.3$     & $\VcbResultBGL\pm1.4\phantom{0}$       & free \\
\Gnorm & $1.102  \pm0.034$   & $1.094 \pm 0.034$                           & $1.07\pm 0.04$\\
\rsqDs                  & $\rsqDsResult\pm0.05$     &  --                           & $1.23\pm 0.05$\\
$d_1$                   & --                  & $\doneResult\pm 0.008\phantom{-}$ & $-0.012\pm0.008\phantom{-}$\\
$d_2$                   & --                  & $\dtwoResult\pm 0.05\phantom{-}$  & $-0.24\pm0.05\phantom{-}$\\
\Fnorm                  & $0.899  \pm0.013$   & $0.900 \pm 0.013$             & $0.902\pm0.013$ \\
\rsqDsS                 & $\rsqDsSResult\pm 0.17$     & --                            & free \\
\Rone                   & $\RoneResult\pm 0.25$     & --                            & free \\
\Rtwo                   & $\RtwoResult\pm 0.16$     & --                            & free \\
$a_0$                   & --                  & $\azeroResult\pm 0.009$           & free \\
$a_1$                   & --                  & $\aoneResult\pm 0.27$               & free \\
$b_1$                   & --                  & $\boneResult\pm 0.07\phantom{-}$   & free \\
$c_1$                   & --                  & $\coneResult\pm 0.0023$           & free \\
$f_s/f_d\times\BF(\Dsm\to\Km\Kp\pim)\times\tau$ [\!$\ps$] & $0.0191\pm0.0008$  & $0.0191\pm0.0008$           & $0.0191\pm0.0008$ \\
$\BF(\Dm\to\Kp\Km\pim)$ & $0.00993\pm0.00024$ & $0.00993 \pm 0.00024$         & $0.00993 \pm 0.00024$ \\
$\BF(\DSm\to\Dm X)$     & $0.323  \pm0.006$   & $0.323 \pm 0.006$             & $0.323 \pm 0.006$\\
$\BF(\bdD)$             & $0.0228 \pm0.0010$  & $0.0230 \pm 0.0010$           & $0.0231 \pm 0.0010$ \\
$\BF(\bdDS)$            & $0.0507 \pm0.0014$  & $0.0506 \pm 0.0014$           & $0.0505 \pm 0.0014$ \\
\Bs mass [$\!\gevcc$]     & $5.36688\pm0.00017$ & $5.36688 \pm 0.00017$         & $5.36688 \pm 0.00017$ \\
\Dsm mass [$\!\gevcc$]    & $1.96834\pm0.00007$ & $1.96834 \pm 0.00007$         & $1.96834 \pm 0.00007$ \\
\DsSm mass [$\!\gevcc$]   & $2.1122 \pm0.0004$  & $2.1122 \pm 0.0004$           & $2.1122 \pm 0.0004$ \\
\bottomrule
\end{tabular}}
\end{table}

\begin{sidewaystable}
\caption{\label{tab:corr-CLN} Correlations (in \%) for the \Vcb fit in the CLN parametrization. The top section includes contributions from statistical sources and external inputs, the bottom section contributions from the experimental systematic uncertainties.}
\centering
\resizebox{0.9\textheight}{!}{
\begin{tabular}{lcccccccccccc}
\toprule
 & \rotatebox{90}{$\BF(\Dm\!\to\Kp\Km\pim)$} & \rotatebox{90}{$\BF(\DSm\!\to\Dm X)$} & \rotatebox{90}{$\BF(\bdD)$} & \rotatebox{90}{$\BF(\bdDS)$} & \rotatebox{90}{\Rone} & \rotatebox{90}{\Rtwo} & \rotatebox{90}{\rsqDs} & \rotatebox{90}{\Gnorm} & \rotatebox{90}{\rsqDsS} & \rotatebox{90}{\Fnorm} & \rotatebox{90}{\etaEW} & \rotatebox{90}{\Vcb} \\ 
\midrule%
$f_s/f_d\times\BF(\Dsm\to\Km\Kp\pim)\times\tau$ & $0.2$ & $\phantom{-}0.1$ & $\phantom{-}0.2$ & $\phantom{-}0.1$ & $-0.1$ & $\phantom{-}0.1$ & $\phantom{-}0.0$ & $-0.1$ & $\phantom{-}0.0$ & $-0.1$ & $-0.1$ & $-62.1\phantom{0}$ \\
$\BF(\Dm\to\Kp\Km\pim)$ & & $\phantom{-}0.0$ & $-0.1$ & $-0.1$ & $-0.1$ & $\phantom{-}0.0$ & $\phantom{-}0.0$ & $\phantom{-}0.0$ & $\phantom{-}0.0$ & $\phantom{-}0.1$ & $\phantom{-}0.1$ &$\phantom{-}38.0\phantom{0}$ \\
$\BF(\DSm\to\Dm X)$ & & & $\phantom{-}7.9$ & $-4.7$ & $\phantom{-}2.4$ & $-1.8$ & $\phantom{-}0.8$ & $-2.8$ & $\phantom{-}1.9$ & $\phantom{-}5.0$ & $\phantom{-}0.0$ &$\phantom{-}17.4\phantom{0}$ \\
$\BF(\bdD)$ & & & & $\phantom{-}11.2\phantom{0}$ & $-5.7$ & $\phantom{-}4.2$ & $-1.9$ & $\phantom{-}6.8$ & $-4.5$ & $-11.8\phantom{0}$ & $\phantom{-}0.0$ &$\phantom{-}33.9\phantom{0}$ \\
$\BF(\bdDS)$ & & & & & $\phantom{-}3.4$ & $-2.5$ & $\phantom{-}1.1$ & $-4.0$ & $\phantom{-}2.7$ & $\phantom{-}7.1$ & $\phantom{-}0.0$ &$\phantom{-}24.6\phantom{0}$ \\
\Rone & & & & & & $-87.5\phantom{0}$& $-2.5$ & $\phantom{-}0.4$ &$\phantom{-}89.1\phantom{0}$ & $-3.6$ & $\phantom{-}0.0$ & $-10.3\phantom{0}$ \\
\Rtwo & & & & & & & $-18.0\phantom{0}$ & $-16.8\phantom{0}$ & $-96.6\phantom{0}$ & $\phantom{-}2.7$ & $\phantom{-}0.0$ & $\phantom{-}10.1\phantom{0}$ \\
\rsqDs & & & & & & & & $\phantom{-}83.8\phantom{0}$ & $\phantom{-}15.9\phantom{0}$ & $-1.1$ & $\phantom{-}0.0$ & $-3.1$ \\
\Gnorm & & & & & & & & & $\phantom{-}15.1\phantom{0}$ & $\phantom{-}4.2$ & $\phantom{-}0.0$ & $-17.3\phantom{0}$ \\
\rsqDsS & & & & & & & & & & $-2.8$ & $\phantom{-}0.0$ & $-5.9$ \\ 
\Fnorm & & & & & & & & & & & $\phantom{-}0.0$ & $-26.0\phantom{0}$ \\ 
\etaEW & & & & & & & & & & & & $-15.6\phantom{0}$ \\ 
\midrule
\Rone & & & & & & $-82.0$ &$\phantom{-}82.0\phantom{0}$ &$\phantom{-}91.9\phantom{0}$ & $\phantom{-}92.0\phantom{0}$ & $\phantom{-}0.0$ & $\phantom{-}0.0$ & $-26.8\phantom{0}$ \\
\Rtwo & & & & & & & $-100.0\phantom{00}$ & $-97.9\phantom{0}$ & $-53.1\phantom{0}$ & $\phantom{-}0.0$ & $\phantom{-}0.0$ & $\phantom{-}3.3$ \\
\rsqDs & & & & & & & & $\phantom{-}97.9\phantom{0}$ &$\phantom{-}53.0\phantom{0}$ & $\phantom{-}0.0$ & $\phantom{-}0.0$ & $-3.2$ \\
\Gnorm & & & & & & & & & $\phantom{-}69.2\phantom{0}$ & $\phantom{-}0.0$ & $\phantom{-}0.0$ & $-11.7\phantom{0}$ \\
\rsqDsS & & & & & & & & & & $\phantom{-}0.0$ & $\phantom{-}0.0$ & $-37.5\phantom{0}$ \\
\bottomrule
\end{tabular}
}
\end{sidewaystable}

\begin{sidewaystable}
\caption{\label{tab:corr-BGL} Correlations (in \%) for the \Vcb fit in the BGL parametrization. The top section includes contributions from statistical sources and external inputs, the bottom section contributions from the experimental systematic uncertainties.}
\centering
\resizebox{\textheight}{!}{
\begin{tabular}{lcccccccccccccc}
\toprule
 & \rotatebox{90}{$\BF(\Dm\!\to\Kp\Km\pim)$} & \rotatebox{90}{$\BF(\DSm\!\to\Dm X)$} & \rotatebox{90}{$\BF(\bdD)$} & \rotatebox{90}{$\BF(\bdDS)$} & \rotatebox{90}{$a_0$} & \rotatebox{90}{$c_1$} & \rotatebox{90}{$d_1$} & \rotatebox{90}{$d_2$} & \rotatebox{90}{\Gnorm} & \rotatebox{90}{$b_1$} & \rotatebox{90}{$a_1$} & \rotatebox{90}{\Fnorm} & \rotatebox{90}{\etaEW} & \rotatebox{90}{\Vcb} \\ 
\midrule
$f_s/f_d\times\BF(\Dsm\to\Km\Kp\pim)\times\tau$ & $\phantom{-}0.2$ & $\phantom{-}0.1$ & $\phantom{-}0.2$ & $\phantom{-}0.1$ & $-0.1$ & $-0.2$ & $\phantom{-}0.0$ & $\phantom{-}0.0$ & $-0.1$ & $-0.2$ & $\phantom{-}0.2$ & $-0.1$ & $-0.1$ & $-57.6\phantom{0}$ \\ 
$\BF(\Dm\to\Kp\Km\pim)$ & & $\phantom{-}0.0$ & $-0.1$ & $-0.1$ & $\phantom{-}0.1$ & $\phantom{-}0.1$ & $\phantom{-}0.0$ & $\phantom{-}0.0$ & $\phantom{-}0.0$ & $\phantom{-}0.1$ & $-0.1$ & $\phantom{-}0.1$ & $\phantom{-}0.1$ & $\phantom{-}35.3\phantom{0}$ \\ 
$\BF(\DSm\to\Dm X)$ & & & $\phantom{-}6.1$ & $-3.8$ & $\phantom{-}2.7$ & $\phantom{-}6.8$ & $-2.2$ & $-0.7$ & $-0.3$ & $\phantom{-}5.2$ & $-8.3$ & $\phantom{-}4.0$ & $\phantom{-}0.0$ & $\phantom{-}12.8\phantom{0}$ \\
$\BF(\bdD)$	& & & & $\phantom{-}8.7$ & $-6.1$ & $-15.6\phantom{0}$ & $\phantom{-}5.1$ & $\phantom{-}1.7$ & $\phantom{-}0.9$ & $-12.1\phantom{0}$ & $\phantom{-}19.1\phantom{0}$ & $-9.1$ & $\phantom{-}0.1$ & $\phantom{-}38.3\phantom{0}$ \\
$\BF(\bdDS)$ & & & & & $\phantom{-}3.8$ & $\phantom{-}9.6$ & $-3.1$ & $-1.0$ & $-0.5$ & $\phantom{-}7.4$ & $-11.7\phantom{0}$ & $\phantom{-}5.6$ & $\phantom{-}0.0$ & $\phantom{-}18.0\phantom{0}$ \\ 
$a_0$ & & & & & & $\phantom{-}43.3\phantom{0}$ & $-1.4$ & $-3.4$ & $\phantom{-}4.4$ & $-56.6\phantom{0}$ & $-3.5$ & $-4.0$ & $\phantom{-}0.0$ & $-10.4\phantom{0}$ \\
$c_1$ & & & & & & & $-16.9\phantom{0}$ & $-11.0\phantom{0}$	& $\phantom{-}22.3\phantom{0}$ & $\phantom{-}28.9\phantom{0}$ & $-70.7\phantom{0}$ & $-10.1\phantom{0}$ & $-0.1$ & $-37.5\phantom{0}$ \\ 
$d_1$ & & & & & & & & $\phantom{-}1.8$ & $-80.8\phantom{0}$ & $\phantom{-}4.7$ & $\phantom{-}5.0$ & $\phantom{-}3.2$ & $\phantom{-}0.0$ & $-0.5$ \\
$d_2$ & & & & & & & & & $-32.5\phantom{0}$ & $\phantom{-}1.6$ & $\phantom{-}6.2$ & $\phantom{-}1.0$ & $\phantom{-}0.0$ & $\phantom{-}3.0$ \\
\Gnorm & & & & & & & & & & $-0.2$ & $-12.8\phantom{0}$ & $\phantom{-}0.5$ & $\phantom{-}0.0$ & $-14.5\phantom{0}$ \\
$b_1$ & & & & & & & & & & & $-77.9\phantom{0}$ & $-7.8$ & $\phantom{-}0.0$ & $-26.1\phantom{0}$ \\ 
$a_1$ & & & & & & & & & & & & $\phantom{-}12.3\phantom{0}$ & $\phantom{-}0.1$ & $\phantom{-}37.2\phantom{0}$ \\ 
\Fnorm & & & & & & & & & & & & & $\phantom{-}0.0$ & $-18.9\phantom{0}$ \\ 
\etaEW & & & & & & & & & & & & & & $-14.5\phantom{0}$ \\ 
\midrule
$a_0$ & & & & & & $-76.5\phantom{0}$ & $\phantom{-}14.5\phantom{0}$ & $\phantom{-}10.7\phantom{0}$ & $-5.8$ & $-92.4\phantom{0}$ & $\phantom{-}93.4\phantom{0}$ & $\phantom{-}0.0$ & $\phantom{-}0.0$ & $-6.4$ \\
$c_1$ & & & & & & & $-68.5\phantom{0}$ & $-65.6\phantom{0}$ & $\phantom{-}61.7\phantom{0}$ & $\phantom{-}78.8\phantom{0}$ & $-75.0\phantom{0}$ & $\phantom{-}0.0$ & $\phantom{-}0.0$ & $\phantom{-}13.5\phantom{0}$ \\
$d_1$ & & & & & & & & $\phantom{-}99.9\phantom{0}$ & $-99.6\phantom{0}$ & $-17.0\phantom{0}$ & $\phantom{-}9.1$ & $\phantom{-}0.0$ & $\phantom{-}0.0$ & $-15.1\phantom{0}$ \\
$d_2$ & & & & & & & & & $-99.9\phantom{0}$ & $-13.2\phantom{0}$ & $\phantom{-}5.2$ & $\phantom{-}0.0$ & $\phantom{-}0.0$ & $-14.9\phantom{0}$ \\
\Gnorm & & & & & & & & & & $\phantom{-}8.2$ & $-0.1$ & $\phantom{-}0.0$ & $\phantom{-}0.0$ & $\phantom{-}14.6\phantom{0}$ \\
$b_1$ & & & & & & & & & & & $-94.5\phantom{0}$ & $\phantom{-}0.0$ & $\phantom{-}0.0$ & $\phantom{-}6.8$ \\ 
$a_1$ & & & & & & & & & & & & $\phantom{-}0.0$ & $\phantom{-}0.0$ & $-5.7$   \\
\bottomrule
\end{tabular}
}
\end{sidewaystable}

\begin{table}[h]
\caption{Detailed results for the \RD and \RDS fit. The uncertainties on the free parameters include the statistical contribution and that due to the external inputs.\label{tab:detailed-R}}
\centering
\begin{tabular}{lcc}
\toprule
Parameter & Value & Constraint \\
\midrule
\RD                     & $1.093 \pm  0.074$  & free \\
\RDS                    & $1.059  \pm0.071$   & free \\
$f_s/f_d\times\BF(\Dsm\to\Km\Kp\pim)$ & $0.0127\pm0.0005$ & $0.0127\pm0.0005$ \\
$\BF(\Dm\to\Kp\Km\pim)$ & $0.00993\pm0.00024$ & $0.00993 \pm 0.00024$ \\
$\BF(\DSm\!\to\Dm X)$     & $0.323  \pm0.006$   & $0.323 \pm 0.006$ \\
\toprule
\end{tabular}
\end{table}

\begin{table}[hb]
\caption{\label{tab:corr-R} Correlations (in \%) for the \RD and \RDS fit. The top section includes contributions from statistical sources and external inputs, the bottom section contributions from the experimental systematic uncertainties.}
\centering
\begin{tabular}{lcccc}
\toprule
                         & $\BF(\Dm\!\to\Kp\Km\pim)$ & $\BF(\DSm\!\to\Dm X)$ &                \RD &               \RDS \\
\midrule
$f_s/f_d\times\BF(\Dsm\to\Km\Kp\pim)$ &  $\phantom{-}0.1$ &  $\phantom{-}0.1$ &           $-58.1$ &           $-58.8$ \\
$\BF(\Dm\to\Kp\Km\pim)$	 & 	                     &  $\phantom{-}0.0$ & $\phantom{-}35.5$ & $\phantom{-}36.0$ \\
$\BF(\DSm\to\Dm X)$	  	 &                       &                   & $\phantom{-0}0.0$ & $\phantom{-}29.4$ \\
\RD	  	                 &                       &                   &                   & $\phantom{-}14.0$ \\
\midrule
\RD	  	                 &                       &                   &                   & $\phantom{-}18.9$ \\
\bottomrule
\end{tabular}
\end{table}
 % end input ./appDetailed.tex
 
\clearpage
\addcontentsline{toc}{section}{References}
\bibliographystyle{LHCb}

\begin{mcitethebibliography}{10}
\mciteSetBstSublistMode{n}
\mciteSetBstMaxWidthForm{subitem}{\alph{mcitesubitemcount})}
\mciteSetBstSublistLabelBeginEnd{\mcitemaxwidthsubitemform\space}
{\relax}{\relax}

\bibitem{HFLAV18}
Heavy Flavor Averaging Group, Y.~Amhis {\em et~al.},
  \ifthenelse{\boolean{articletitles}}{\emph{{Averages of $b$-hadron,
  $c$-hadron, and $\tau$-lepton properties as of summer 2018}},
  }{}\href{http://arxiv.org/abs/1909.12524}{{\normalfont\ttfamily
  arXiv:1909.12524}}, {updated results and plots available at
  \href{https://hflav.web.cern.ch}{{\texttt{https://hflav.web.cern.ch}}}}\relax
\mciteBstWouldAddEndPuncttrue
\mciteSetBstMidEndSepPunct{\mcitedefaultmidpunct}
{\mcitedefaultendpunct}{\mcitedefaultseppunct}\relax
\EndOfBibitem
\bibitem{Caprini:1997mu}
I.~Caprini, L.~Lellouch, and M.~Neubert,
  \ifthenelse{\boolean{articletitles}}{\emph{{Dispersive bounds on the shape of
  $\Bbar\to\D^{(*)}\ell\bar{\nu}$ form factors}},
  }{}\href{https://doi.org/10.1016/S0550-3213(98)00350-2}{Nucl.\ Phys.\
  \textbf{B530} (1998) 153},
  \href{http://arxiv.org/abs/hep-ph/9712417}{{\normalfont\ttfamily
  arXiv:hep-ph/9712417}}\relax
\mciteBstWouldAddEndPuncttrue
\mciteSetBstMidEndSepPunct{\mcitedefaultmidpunct}
{\mcitedefaultendpunct}{\mcitedefaultseppunct}\relax
\EndOfBibitem
\bibitem{Boyd:1994tt}
C.~G. Boyd, B.~Grinstein, and R.~F. Lebed,
  \ifthenelse{\boolean{articletitles}}{\emph{{Constraints on form factors for
  exclusive semileptonic heavy to light meson decays}},
  }{}\href{https://doi.org/10.1103/PhysRevLett.74.4603}{Phys.\ Rev.\ Lett.\
  \textbf{74} (1995) 4603},
  \href{http://arxiv.org/abs/hep-ph/9412324}{{\normalfont\ttfamily
  arXiv:hep-ph/9412324}}\relax
\mciteBstWouldAddEndPuncttrue
\mciteSetBstMidEndSepPunct{\mcitedefaultmidpunct}
{\mcitedefaultendpunct}{\mcitedefaultseppunct}\relax
\EndOfBibitem
\bibitem{Boyd:1995sq}
C.~G. Boyd, B.~Grinstein, and R.~F. Lebed,
  \ifthenelse{\boolean{articletitles}}{\emph{{Model-independent determinations
  of $\Bbar\to\D\ell\bar{\nu}, \D^*\ell\bar{\nu}$ form factors}},
  }{}\href{https://doi.org/10.1016/0550-3213(95)00653-2}{Nucl.\ Phys.\
  \textbf{B461} (1996) 493},
  \href{http://arxiv.org/abs/hep-ph/9508211}{{\normalfont\ttfamily
  arXiv:hep-ph/9508211}}\relax
\mciteBstWouldAddEndPuncttrue
\mciteSetBstMidEndSepPunct{\mcitedefaultmidpunct}
{\mcitedefaultendpunct}{\mcitedefaultseppunct}\relax
\EndOfBibitem
\bibitem{Boyd:1997kz}
C.~G. Boyd, B.~Grinstein, and R.~F. Lebed,
  \ifthenelse{\boolean{articletitles}}{\emph{{Precision corrections to
  dispersive bounds on form factors}},
  }{}\href{https://doi.org/10.1103/PhysRevD.56.6895}{Phys.\ Rev.\  \textbf{D56}
  (1997) 6895},
  \href{http://arxiv.org/abs/hep-ph/9705252}{{\normalfont\ttfamily
  arXiv:hep-ph/9705252}}\relax
\mciteBstWouldAddEndPuncttrue
\mciteSetBstMidEndSepPunct{\mcitedefaultmidpunct}
{\mcitedefaultendpunct}{\mcitedefaultseppunct}\relax
\EndOfBibitem
\bibitem{Bourrely:2008za}
C.~Bourrely, I.~Caprini, and L.~Lellouch,
  \ifthenelse{\boolean{articletitles}}{\emph{{Model-independent description of
  $\B\to\pi\ell\nu$ decays and a determination of $|V_{ub}|$}},
  }{}\href{https://doi.org/10.1103/PhysRevD.82.099902}{Phys.\ Rev.\
  \textbf{D79} (2009) 013008}, Erratum
  \href{https://doi.org/10.1103/PhysRevD.79.013008}{ibid.\   \textbf{D82}
  (2010) 099902}, \href{http://arxiv.org/abs/0807.2722}{{\normalfont\ttfamily
  arXiv:0807.2722}}\relax
\mciteBstWouldAddEndPuncttrue
\mciteSetBstMidEndSepPunct{\mcitedefaultmidpunct}
{\mcitedefaultendpunct}{\mcitedefaultseppunct}\relax
\EndOfBibitem
\bibitem{Bernlochner:2017xyx}
F.~U. Bernlochner, Z.~Ligeti, M.~Papucci, and D.~J. Robinson,
  \ifthenelse{\boolean{articletitles}}{\emph{{Tensions and correlations in \Vcb
  determinations}}, }{}\href{https://doi.org/10.1103/PhysRevD.96.091503}{Phys.\
  Rev.\  \textbf{D96} (2017) 091503},
  \href{http://arxiv.org/abs/1708.07134}{{\normalfont\ttfamily
  arXiv:1708.07134}}\relax
\mciteBstWouldAddEndPuncttrue
\mciteSetBstMidEndSepPunct{\mcitedefaultmidpunct}
{\mcitedefaultendpunct}{\mcitedefaultseppunct}\relax
\EndOfBibitem
\bibitem{Bigi:2017njr}
D.~Bigi, P.~Gambino, and S.~Schacht,
  \ifthenelse{\boolean{articletitles}}{\emph{{A fresh look at the determination
  of $|V_{cb}|$ from $B\to \DS \ell \nu$}},
  }{}\href{https://doi.org/10.1016/j.physletb.2017.04.022}{Phys.\ Lett.\
  \textbf{B769} (2017) 441},
  \href{http://arxiv.org/abs/1703.06124}{{\normalfont\ttfamily
  arXiv:1703.06124}}\relax
\mciteBstWouldAddEndPuncttrue
\mciteSetBstMidEndSepPunct{\mcitedefaultmidpunct}
{\mcitedefaultendpunct}{\mcitedefaultseppunct}\relax
\EndOfBibitem
\bibitem{Grinstein:2017nlq}
B.~Grinstein and A.~Kobach,
  \ifthenelse{\boolean{articletitles}}{\emph{{Model-independent extraction of
  $\Vcb$ from $\Bbar\to \DS \ell \neub$}},
  }{}\href{https://doi.org/10.1016/j.physletb.2017.05.078}{Phys.\ Lett.\
  \textbf{B771} (2017) 359},
  \href{http://arxiv.org/abs/1703.08170}{{\normalfont\ttfamily
  arXiv:1703.08170}}\relax
\mciteBstWouldAddEndPuncttrue
\mciteSetBstMidEndSepPunct{\mcitedefaultmidpunct}
{\mcitedefaultendpunct}{\mcitedefaultseppunct}\relax
\EndOfBibitem
\bibitem{Jaiswal:2017rve}
S.~Jaiswal, S.~Nandi, and S.~K. Patra,
  \ifthenelse{\boolean{articletitles}}{\emph{{Extraction of \Vcb from
  \decay{\B}{\DorDS\ell\nu_\ell} and the Standard Model predictions of
  $R(\DorDS)$}}, }{}\href{https://doi.org/10.1007/JHEP12(2017)060}{JHEP
  \textbf{12} (2017) 060},
  \href{http://arxiv.org/abs/1707.09977}{{\normalfont\ttfamily
  arXiv:1707.09977}}\relax
\mciteBstWouldAddEndPuncttrue
\mciteSetBstMidEndSepPunct{\mcitedefaultmidpunct}
{\mcitedefaultendpunct}{\mcitedefaultseppunct}\relax
\EndOfBibitem
\bibitem{Colangelo:2018cnj}
P.~Colangelo and F.~De~Fazio,
  \ifthenelse{\boolean{articletitles}}{\emph{{Scrutinizing $ \overline{B}\to
  {D}^{\ast}\left(D\pi \right){\ell}^{-}{\overline{\nu}}_{\ell } $ and $
  \overline{B}\to {D}^{\ast}\left(D\gamma
  \right){\ell}^{-}{\overline{\nu}}_{\ell } $ in search of new physics
  footprints}}, }{}\href{https://doi.org/10.1007/JHEP06(2018)082}{JHEP
  \textbf{06} (2018) 082},
  \href{http://arxiv.org/abs/1801.10468}{{\normalfont\ttfamily
  arXiv:1801.10468}}\relax
\mciteBstWouldAddEndPuncttrue
\mciteSetBstMidEndSepPunct{\mcitedefaultmidpunct}
{\mcitedefaultendpunct}{\mcitedefaultseppunct}\relax
\EndOfBibitem
\bibitem{Abdesselam:2018nnh}
Belle collaboration, E.~Waheed {\em et~al.},
  \ifthenelse{\boolean{articletitles}}{\emph{{Measurement of the CKM matrix
  element $\Vcb$ from $\Bd\to \DSm\ell^ {+} \nu_\ell$ at Belle}},
  }{}\href{https://doi.org/10.1103/PhysRevD.100.052007}{Phys.\ Rev.\
  \textbf{D100} (2019) 052007},
  \href{http://arxiv.org/abs/1809.03290}{{\normalfont\ttfamily
  arXiv:1809.03290}}\relax
\mciteBstWouldAddEndPuncttrue
\mciteSetBstMidEndSepPunct{\mcitedefaultmidpunct}
{\mcitedefaultendpunct}{\mcitedefaultseppunct}\relax
\EndOfBibitem
\bibitem{Dey:2019bgc}
BaBar collaboration, J.~P. Lees {\em et~al.},
  \ifthenelse{\boolean{articletitles}}{\emph{{Extraction of form factors from a
  four-dimensional angular analysis of $\overline{B} \rightarrow D^\ast \ell^-
  \overline{\nu}_\ell$}},
  }{}\href{https://doi.org/10.1103/PhysRevLett.123.091801}{Phys.\ Rev.\ Lett.\
  \textbf{123} (2019) 091801},
  \href{http://arxiv.org/abs/1903.10002}{{\normalfont\ttfamily
  arXiv:1903.10002}}\relax
\mciteBstWouldAddEndPuncttrue
\mciteSetBstMidEndSepPunct{\mcitedefaultmidpunct}
{\mcitedefaultendpunct}{\mcitedefaultseppunct}\relax
\EndOfBibitem
\bibitem{Gambino:2019sif}
P.~Gambino, M.~Jung, and S.~Schacht,
  \ifthenelse{\boolean{articletitles}}{\emph{{The $V_{cb}$ puzzle: an update}},
  }{}\href{https://doi.org/10.1016/j.physletb.2019.06.039}{Phys.\ Lett.\
  \textbf{B795} (2019) 386},
  \href{http://arxiv.org/abs/1905.08209}{{\normalfont\ttfamily
  arXiv:1905.08209}}\relax
\mciteBstWouldAddEndPuncttrue
\mciteSetBstMidEndSepPunct{\mcitedefaultmidpunct}
{\mcitedefaultendpunct}{\mcitedefaultseppunct}\relax
\EndOfBibitem
\bibitem{Bordone:2019vic}
M.~Bordone, M.~Jung, and D.~van Dyk,
  \ifthenelse{\boolean{articletitles}}{\emph{{Theory determination of
  $\bar{B}\to D^{(*)}\ell^-\bar\nu$ form factors at $\mathcal{O}(1/m_c^2)$}},
  }{}\href{http://arxiv.org/abs/1908.09398}{{\normalfont\ttfamily
  arXiv:1908.09398}}\relax
\mciteBstWouldAddEndPuncttrue
\mciteSetBstMidEndSepPunct{\mcitedefaultmidpunct}
{\mcitedefaultendpunct}{\mcitedefaultseppunct}\relax
\EndOfBibitem
\bibitem{Bordone:2019guc}
M.~Bordone, N.~Gubernari, M.~Jung, and D.~van Dyk,
  \ifthenelse{\boolean{articletitles}}{\emph{{Heavy-Quark Expansion for
  $\Bbar_s\to\DsorDsS$ form factors and unitarity bounds beyond the $SU(3)_F$
  limit}}, }{}\href{http://arxiv.org/abs/1912.09335}{{\normalfont\ttfamily
  arXiv:1912.09335}}\relax
\mciteBstWouldAddEndPuncttrue
\mciteSetBstMidEndSepPunct{\mcitedefaultmidpunct}
{\mcitedefaultendpunct}{\mcitedefaultseppunct}\relax
\EndOfBibitem
\bibitem{King:2019rvk}
D.~King, M.~Kirk, A.~Lenz, and T.~Rauh,
  \ifthenelse{\boolean{articletitles}}{\emph{{$\Vcb$ and $\gamma$ from
  $B$-mixing}}, }{}\href{http://arxiv.org/abs/1911.07856}{{\normalfont\ttfamily
  arXiv:1911.07856}}\relax
\mciteBstWouldAddEndPuncttrue
\mciteSetBstMidEndSepPunct{\mcitedefaultmidpunct}
{\mcitedefaultendpunct}{\mcitedefaultseppunct}\relax
\EndOfBibitem
\bibitem{McLean:2019sds}
HPQCD collaboration, E.~McLean, C.~T.~H. Davies, A.~T. Lytle, and J.~Koponen,
  \ifthenelse{\boolean{articletitles}}{\emph{{Lattice QCD form factor for
  $\Bs\to\D_s^*\ell\nu$ at zero recoil with non-perturbative current
  renormalisation}},
  }{}\href{http://arxiv.org/abs/1904.02046}{{\normalfont\ttfamily
  arXiv:1904.02046}}\relax
\mciteBstWouldAddEndPuncttrue
\mciteSetBstMidEndSepPunct{\mcitedefaultmidpunct}
{\mcitedefaultendpunct}{\mcitedefaultseppunct}\relax
\EndOfBibitem
\bibitem{Harrison:2017fmw}
HPQCD collaboration, J.~Harrison, C.~Davies, and M.~Wingate,
  \ifthenelse{\boolean{articletitles}}{\emph{{Lattice QCD calculation of the
  ${{B}_{(s)}\to D_{(s)}^{*}\ell{\nu}}$ form factors at zero recoil and
  implications for ${|V_{cb}|}$}},
  }{}\href{https://doi.org/10.1103/PhysRevD.97.054502}{Phys.\ Rev.\
  \textbf{D97} (2018) 054502},
  \href{http://arxiv.org/abs/1711.11013}{{\normalfont\ttfamily
  arXiv:1711.11013}}\relax
\mciteBstWouldAddEndPuncttrue
\mciteSetBstMidEndSepPunct{\mcitedefaultmidpunct}
{\mcitedefaultendpunct}{\mcitedefaultseppunct}\relax
\EndOfBibitem
\bibitem{Atoui:2013zza}
M.~Atoui, V.~Mor\'{e}nas, D.~Be\v{c}irevic, and F.~Sanfilippo,
  \ifthenelse{\boolean{articletitles}}{\emph{{$B_{s} \to D_{s}\ell\nu_\ell$
  near zero recoil in and beyond the Standard Model}},
  }{}\href{https://doi.org/10.1140/epjc/s10052-014-2861-z}{Eur.\ Phys.\ J.\
  \textbf{C74} (2014) 2861},
  \href{http://arxiv.org/abs/1310.5238}{{\normalfont\ttfamily
  arXiv:1310.5238}}\relax
\mciteBstWouldAddEndPuncttrue
\mciteSetBstMidEndSepPunct{\mcitedefaultmidpunct}
{\mcitedefaultendpunct}{\mcitedefaultseppunct}\relax
\EndOfBibitem
\bibitem{Flynn:2016vej}
J.~Flynn {\em et~al.}, \ifthenelse{\boolean{articletitles}}{\emph{{Form factors
  for semi-leptonic $B$ decays}},
  }{}\href{https://doi.org/10.22323/1.256.0296}{PoS \textbf{LATTICE2016} (2016)
  296}, \href{http://arxiv.org/abs/1612.05112}{{\normalfont\ttfamily
  arXiv:1612.05112}}\relax
\mciteBstWouldAddEndPuncttrue
\mciteSetBstMidEndSepPunct{\mcitedefaultmidpunct}
{\mcitedefaultendpunct}{\mcitedefaultseppunct}\relax
\EndOfBibitem
\bibitem{Monahan:2017uby}
C.~J. Monahan {\em et~al.}, \ifthenelse{\boolean{articletitles}}{\emph{{$B_s
  \to D_s \ell \nu$ form factors and the fragmentation fraction ratio
  $f_s/f_d$}}, }{}\href{https://doi.org/10.1103/PhysRevD.95.114506}{Phys.\
  Rev.\  \textbf{D95} (2017) 114506},
  \href{http://arxiv.org/abs/1703.09728}{{\normalfont\ttfamily
  arXiv:1703.09728}}\relax
\mciteBstWouldAddEndPuncttrue
\mciteSetBstMidEndSepPunct{\mcitedefaultmidpunct}
{\mcitedefaultendpunct}{\mcitedefaultseppunct}\relax
\EndOfBibitem
\bibitem{McLean:2019qcx}
E.~McLean, C.~T.~H. Davies, J.~Koponen, and A.~T. Lytle,
  \ifthenelse{\boolean{articletitles}}{\emph{{$B_s\to D_s \ell\nu$ form factors
  for the full $q^2$ range from Lattice QCD with non-perturbatively normalized
  currents}}, }{}\href{http://arxiv.org/abs/1906.00701}{{\normalfont\ttfamily
  arXiv:1906.00701}}\relax
\mciteBstWouldAddEndPuncttrue
\mciteSetBstMidEndSepPunct{\mcitedefaultmidpunct}
{\mcitedefaultendpunct}{\mcitedefaultseppunct}\relax
\EndOfBibitem
\bibitem{LHCb-PAPER-2018-050}
LHCb collaboration, R.~Aaij {\em et~al.},
  \ifthenelse{\boolean{articletitles}}{\emph{{Measurement of $b$-hadron
  fractions in $13\tev$ $\proton\proton$ collisions}},
  }{}\href{https://doi.org/10.1103/PhysRevD.100.031102}{Phys.\ Rev.\
  \textbf{D100} (2019) 031102},
  \href{http://arxiv.org/abs/1902.06794}{{\normalfont\ttfamily
  arXiv:1902.06794}}\relax
\mciteBstWouldAddEndPuncttrue
\mciteSetBstMidEndSepPunct{\mcitedefaultmidpunct}
{\mcitedefaultendpunct}{\mcitedefaultseppunct}\relax
\EndOfBibitem
\bibitem{Bigi:2011gf}
I.~I. Bigi, T.~Mannel, and N.~Uraltsev,
  \ifthenelse{\boolean{articletitles}}{\emph{{Semileptonic width ratios among
  beauty hadrons}}, }{}\href{https://doi.org/10.1007/JHEP09(2011)012}{JHEP
  \textbf{09} (2011) 012},
  \href{http://arxiv.org/abs/1105.4574}{{\normalfont\ttfamily
  arXiv:1105.4574}}\relax
\mciteBstWouldAddEndPuncttrue
\mciteSetBstMidEndSepPunct{\mcitedefaultmidpunct}
{\mcitedefaultendpunct}{\mcitedefaultseppunct}\relax
\EndOfBibitem
\bibitem{Sirlin}
A.~Sirlin, \ifthenelse{\boolean{articletitles}}{\emph{{Large $m_{\rm W}$,
  $m_{\rm Z}$ behaviour of the O($\alpha$) corrections to semileptonic
  processes mediated by W}},
  }{}\href{https://doi.org/10.1016/0550-3213(82)90303-0}{Nucl.\ Phys.\
  \textbf{B196} (1982) 83}\relax
\mciteBstWouldAddEndPuncttrue
\mciteSetBstMidEndSepPunct{\mcitedefaultmidpunct}
{\mcitedefaultendpunct}{\mcitedefaultseppunct}\relax
\EndOfBibitem
\bibitem{Bigi:2016mdz}
D.~Bigi and P.~Gambino, \ifthenelse{\boolean{articletitles}}{\emph{{Revisiting
  $B\to D \ell \nu$}},
  }{}\href{https://doi.org/10.1103/PhysRevD.94.094008}{Phys.\ Rev.\
  \textbf{D94} (2016) 094008},
  \href{http://arxiv.org/abs/1606.08030}{{\normalfont\ttfamily
  arXiv:1606.08030}}\relax
\mciteBstWouldAddEndPuncttrue
\mciteSetBstMidEndSepPunct{\mcitedefaultmidpunct}
{\mcitedefaultendpunct}{\mcitedefaultseppunct}\relax
\EndOfBibitem
\bibitem{NEUBERT1991455}
M.~Neubert, \ifthenelse{\boolean{articletitles}}{\emph{{Model-independent
  extraction of \Vcb from semi-leptonic decays}},
  }{}\href{https://doi.org/10.1016/0370-2693(91)90377-3}{Phys.\ Lett.\
  \textbf{B264} (1991) 455}\relax
\mciteBstWouldAddEndPuncttrue
\mciteSetBstMidEndSepPunct{\mcitedefaultmidpunct}
{\mcitedefaultendpunct}{\mcitedefaultseppunct}\relax
\EndOfBibitem
\bibitem{LHCb-DP-2008-001}
LHCb collaboration, A.~A. Alves~Jr.\ {\em et~al.},
  \ifthenelse{\boolean{articletitles}}{\emph{{The \lhcb detector at the LHC}},
  }{}\href{https://doi.org/10.1088/1748-0221/3/08/S08005}{JINST \textbf{3}
  (2008) S08005}\relax
\mciteBstWouldAddEndPuncttrue
\mciteSetBstMidEndSepPunct{\mcitedefaultmidpunct}
{\mcitedefaultendpunct}{\mcitedefaultseppunct}\relax
\EndOfBibitem
\bibitem{LHCb-DP-2014-002}
LHCb collaboration, R.~Aaij {\em et~al.},
  \ifthenelse{\boolean{articletitles}}{\emph{{LHCb detector performance}},
  }{}\href{https://doi.org/10.1142/S0217751X15300227}{Int.\ J.\ Mod.\ Phys.\
  \textbf{A30} (2015) 1530022},
  \href{http://arxiv.org/abs/1412.6352}{{\normalfont\ttfamily
  arXiv:1412.6352}}\relax
\mciteBstWouldAddEndPuncttrue
\mciteSetBstMidEndSepPunct{\mcitedefaultmidpunct}
{\mcitedefaultendpunct}{\mcitedefaultseppunct}\relax
\EndOfBibitem
\bibitem{Sjostrand:2006za}
T.~Sj\"{o}strand, S.~Mrenna, and P.~Skands,
  \ifthenelse{\boolean{articletitles}}{\emph{{PYTHIA 6.4 physics and manual}},
  }{}\href{https://doi.org/10.1088/1126-6708/2006/05/026}{JHEP \textbf{05}
  (2006) 026}, \href{http://arxiv.org/abs/hep-ph/0603175}{{\normalfont\ttfamily
  arXiv:hep-ph/0603175}}\relax
\mciteBstWouldAddEndPuncttrue
\mciteSetBstMidEndSepPunct{\mcitedefaultmidpunct}
{\mcitedefaultendpunct}{\mcitedefaultseppunct}\relax
\EndOfBibitem
\bibitem{Sjostrand:2007gs}
T.~Sj\"{o}strand, S.~Mrenna, and P.~Skands,
  \ifthenelse{\boolean{articletitles}}{\emph{{A brief introduction to PYTHIA
  8.1}}, }{}\href{https://doi.org/10.1016/j.cpc.2008.01.036}{Comput.\ Phys.\
  Commun.\  \textbf{178} (2008) 852},
  \href{http://arxiv.org/abs/0710.3820}{{\normalfont\ttfamily
  arXiv:0710.3820}}\relax
\mciteBstWouldAddEndPuncttrue
\mciteSetBstMidEndSepPunct{\mcitedefaultmidpunct}
{\mcitedefaultendpunct}{\mcitedefaultseppunct}\relax
\EndOfBibitem
\bibitem{LHCb-PROC-2010-056}
I.~Belyaev {\em et~al.}, \ifthenelse{\boolean{articletitles}}{\emph{{Handling
  of the generation of primary events in Gauss, the LHCb simulation
  framework}}, }{}\href{https://doi.org/10.1088/1742-6596/331/3/032047}{J.\
  Phys.\ Conf.\ Ser.\  \textbf{331} (2011) 032047}\relax
\mciteBstWouldAddEndPuncttrue
\mciteSetBstMidEndSepPunct{\mcitedefaultmidpunct}
{\mcitedefaultendpunct}{\mcitedefaultseppunct}\relax
\EndOfBibitem
\bibitem{Lange:2001uf}
D.~J. Lange, \ifthenelse{\boolean{articletitles}}{\emph{{The EvtGen particle
  decay simulation package}},
  }{}\href{https://doi.org/10.1016/S0168-9002(01)00089-4}{Nucl.\ Instrum.\
  Meth.\  \textbf{A462} (2001) 152}\relax
\mciteBstWouldAddEndPuncttrue
\mciteSetBstMidEndSepPunct{\mcitedefaultmidpunct}
{\mcitedefaultendpunct}{\mcitedefaultseppunct}\relax
\EndOfBibitem
\bibitem{Golonka:2005pn}
P.~Golonka and Z.~Was, \ifthenelse{\boolean{articletitles}}{\emph{{PHOTOS Monte
  Carlo: A precision tool for QED corrections in $Z$ and $W$ decays}},
  }{}\href{https://doi.org/10.1140/epjc/s2005-02396-4}{Eur.\ Phys.\ J.\
  \textbf{C45} (2006) 97},
  \href{http://arxiv.org/abs/hep-ph/0506026}{{\normalfont\ttfamily
  arXiv:hep-ph/0506026}}\relax
\mciteBstWouldAddEndPuncttrue
\mciteSetBstMidEndSepPunct{\mcitedefaultmidpunct}
{\mcitedefaultendpunct}{\mcitedefaultseppunct}\relax
\EndOfBibitem
\bibitem{Allison:2006ve}
Geant4 collaboration, J.~Allison {\em et~al.},
  \ifthenelse{\boolean{articletitles}}{\emph{{Geant4 developments and
  applications}}, }{}\href{https://doi.org/10.1109/TNS.2006.869826}{IEEE
  Trans.\ Nucl.\ Sci.\  \textbf{53} (2006) 270}\relax
\mciteBstWouldAddEndPuncttrue
\mciteSetBstMidEndSepPunct{\mcitedefaultmidpunct}
{\mcitedefaultendpunct}{\mcitedefaultseppunct}\relax
\EndOfBibitem
\bibitem{Agostinelli:2002hh}
Geant4 collaboration, S.~Agostinelli {\em et~al.},
  \ifthenelse{\boolean{articletitles}}{\emph{{Geant4: A simulation toolkit}},
  }{}\href{https://doi.org/10.1016/S0168-9002(03)01368-8}{Nucl.\ Instrum.\
  Meth.\  \textbf{A506} (2003) 250}\relax
\mciteBstWouldAddEndPuncttrue
\mciteSetBstMidEndSepPunct{\mcitedefaultmidpunct}
{\mcitedefaultendpunct}{\mcitedefaultseppunct}\relax
\EndOfBibitem
\bibitem{LHCb-PROC-2011-006}
M.~Clemencic {\em et~al.}, \ifthenelse{\boolean{articletitles}}{\emph{{The
  \lhcb simulation application, Gauss: Design, evolution and experience}},
  }{}\href{https://doi.org/10.1088/1742-6596/331/3/032023}{J.\ Phys.\ Conf.\
  Ser.\  \textbf{331} (2011) 032023}\relax
\mciteBstWouldAddEndPuncttrue
\mciteSetBstMidEndSepPunct{\mcitedefaultmidpunct}
{\mcitedefaultendpunct}{\mcitedefaultseppunct}\relax
\EndOfBibitem
\bibitem{LHCb-PAPER-2017-004}
LHCb collaboration, R.~Aaij {\em et~al.},
  \ifthenelse{\boolean{articletitles}}{\emph{{Measurement of \Bs and \Dsm meson
  lifetimes}}, }{}\href{https://doi.org/10.1103/PhysRevLett.119.101801}{Phys.\
  Rev.\ Lett.\  \textbf{119} (2017) 101801},
  \href{http://arxiv.org/abs/1705.03475}{{\normalfont\ttfamily
  arXiv:1705.03475}}\relax
\mciteBstWouldAddEndPuncttrue
\mciteSetBstMidEndSepPunct{\mcitedefaultmidpunct}
{\mcitedefaultendpunct}{\mcitedefaultseppunct}\relax
\EndOfBibitem
\bibitem{LHCb-DP-2012-004}
R.~Aaij {\em et~al.}, \ifthenelse{\boolean{articletitles}}{\emph{{The \lhcb
  trigger and its performance in 2011}},
  }{}\href{https://doi.org/10.1088/1748-0221/8/04/P04022}{JINST \textbf{8}
  (2013) P04022}, \href{http://arxiv.org/abs/1211.3055}{{\normalfont\ttfamily
  arXiv:1211.3055}}\relax
\mciteBstWouldAddEndPuncttrue
\mciteSetBstMidEndSepPunct{\mcitedefaultmidpunct}
{\mcitedefaultendpunct}{\mcitedefaultseppunct}\relax
\EndOfBibitem
\bibitem{PDG2018}
Particle Data Group, M.~Tanabashi {\em et~al.},
  \ifthenelse{\boolean{articletitles}}{\emph{{\href{http://pdg.lbl.gov/}{Review
  of particle physics}}},
  }{}\href{https://doi.org/10.1103/PhysRevD.98.030001}{Phys.\ Rev.\
  \textbf{D98} (2018) 030001}\relax
\mciteBstWouldAddEndPuncttrue
\mciteSetBstMidEndSepPunct{\mcitedefaultmidpunct}
{\mcitedefaultendpunct}{\mcitedefaultseppunct}\relax
\EndOfBibitem
\bibitem{Kodama:1993ec}
Fermilab E653 collaboration, K.~Kodama {\em et~al.},
  \ifthenelse{\boolean{articletitles}}{\emph{{Limits for four- and five-prong
  semimuonic charm meson decays}},
  }{}\href{https://doi.org/10.1016/0370-2693(93)91222-9}{Phys.\ Lett.\
  \textbf{B313} (1993) 260}\relax
\mciteBstWouldAddEndPuncttrue
\mciteSetBstMidEndSepPunct{\mcitedefaultmidpunct}
{\mcitedefaultendpunct}{\mcitedefaultseppunct}\relax
\EndOfBibitem
\bibitem{Aitala:1998ey}
E791 collaboration, E.~M. Aitala {\em et~al.},
  \ifthenelse{\boolean{articletitles}}{\emph{{Measurement of the form-factor
  ratios for $\Dp\to\Kbar^{*0}\ell^+\nu_\ell$}},
  }{}\href{https://doi.org/10.1016/S0370-2693(98)01243-X}{Phys.\ Lett.\
  \textbf{B440} (1998) 435},
  \href{http://arxiv.org/abs/hep-ex/9809026}{{\normalfont\ttfamily
  arXiv:hep-ex/9809026}}\relax
\mciteBstWouldAddEndPuncttrue
\mciteSetBstMidEndSepPunct{\mcitedefaultmidpunct}
{\mcitedefaultendpunct}{\mcitedefaultseppunct}\relax
\EndOfBibitem
\bibitem{Link:2004uk}
FOCUS collaboration, J.~M. Link {\em et~al.},
  \ifthenelse{\boolean{articletitles}}{\emph{{Analysis of the semileptonic
  decay \mbox{$\Dz\to\Kzb\pim\mup\neum$}}},
  }{}\href{https://doi.org/10.1016/j.physletb.2004.12.037}{Phys.\ Lett.\
  \textbf{B607} (2005) 67},
  \href{http://arxiv.org/abs/hep-ex/0410067}{{\normalfont\ttfamily
  arXiv:hep-ex/0410067}}\relax
\mciteBstWouldAddEndPuncttrue
\mciteSetBstMidEndSepPunct{\mcitedefaultmidpunct}
{\mcitedefaultendpunct}{\mcitedefaultseppunct}\relax
\EndOfBibitem
\bibitem{Dambach:2006ha}
S.~Dambach, U.~Langenegger, and A.~Starodumov,
  \ifthenelse{\boolean{articletitles}}{\emph{{Neutrino reconstruction with
  topological information}},
  }{}\href{https://doi.org/10.1016/j.nima.2006.08.144}{Nucl.\ Instrum.\ Meth.\
  \textbf{A569} (2006) 824},
  \href{http://arxiv.org/abs/hep-ph/0607294}{{\normalfont\ttfamily
  arXiv:hep-ph/0607294}}\relax
\mciteBstWouldAddEndPuncttrue
\mciteSetBstMidEndSepPunct{\mcitedefaultmidpunct}
{\mcitedefaultendpunct}{\mcitedefaultseppunct}\relax
\EndOfBibitem
\bibitem{Ciezarek:2016lqu}
G.~Ciezarek, A.~Lupato, M.~Rotondo, and M.~Vesterinen,
  \ifthenelse{\boolean{articletitles}}{\emph{{Reconstruction of
  semileptonically decaying beauty hadrons produced in high energy $pp$
  collisions}}, }{}\href{https://doi.org/10.1007/JHEP02(2017)021}{JHEP
  \textbf{02} (2017) 021},
  \href{http://arxiv.org/abs/1611.08522}{{\normalfont\ttfamily
  arXiv:1611.08522}}\relax
\mciteBstWouldAddEndPuncttrue
\mciteSetBstMidEndSepPunct{\mcitedefaultmidpunct}
{\mcitedefaultendpunct}{\mcitedefaultseppunct}\relax
\EndOfBibitem
\bibitem{Stone:2014mza}
S.~Stone and L.~Zhang, \ifthenelse{\boolean{articletitles}}{\emph{{Method of
  studying \Lb decays with one missing particle}},
  }{}\href{https://doi.org/10.1155/2014/931257}{Adv.\ High Energy Phys.\
  \textbf{2014} (2014) 931257},
  \href{http://arxiv.org/abs/1402.4205}{{\normalfont\ttfamily
  arXiv:1402.4205}}\relax
\mciteBstWouldAddEndPuncttrue
\mciteSetBstMidEndSepPunct{\mcitedefaultmidpunct}
{\mcitedefaultendpunct}{\mcitedefaultseppunct}\relax
\EndOfBibitem
\bibitem{LHCb-PAPER-2015-013}
LHCb collaboration, R.~Aaij {\em et~al.},
  \ifthenelse{\boolean{articletitles}}{\emph{{Determination of the quark
  coupling strength $|\Vub|$ using baryonic decays}},
  }{}\href{https://doi.org/10.1038/nphys3415}{Nature Physics \textbf{11} (2015)
  743}, \href{http://arxiv.org/abs/1504.01568}{{\normalfont\ttfamily
  arXiv:1504.01568}}\relax
\mciteBstWouldAddEndPuncttrue
\mciteSetBstMidEndSepPunct{\mcitedefaultmidpunct}
{\mcitedefaultendpunct}{\mcitedefaultseppunct}\relax
\EndOfBibitem
\bibitem{LHCB-PAPER-2017-016}
LHCb collaboration, R.~Aaij {\em et~al.},
  \ifthenelse{\boolean{articletitles}}{\emph{{Measurement of the shape of the
  \mbox{\decay{\Lb}{\Lc\mun\neumb}} differential decay rate}},
  }{}\href{https://doi.org/10.1103/PhysRevD.96.112005}{Phys.\ Rev.\
  \textbf{D96} (2017) 112005},
  \href{http://arxiv.org/abs/1709.01920}{{\normalfont\ttfamily
  arXiv:1709.01920}}\relax
\mciteBstWouldAddEndPuncttrue
\mciteSetBstMidEndSepPunct{\mcitedefaultmidpunct}
{\mcitedefaultendpunct}{\mcitedefaultseppunct}\relax
\EndOfBibitem
\bibitem{LHCB-PAPER-2017-027}
LHCb collaboration, R.~Aaij {\em et~al.},
  \ifthenelse{\boolean{articletitles}}{\emph{{Test of lepton flavor
  universality by the measurement of the \mbox{\decay{\Bz}{D^{\ast-} \taup
  \nu_{\tau}}} branching fraction using three-prong $\tau$ decays}},
  }{}\href{https://doi.org/10.1103/PhysRevD.97.072013}{Phys.\ Rev.\
  \textbf{D97} (2018) 072013},
  \href{http://arxiv.org/abs/1711.02505}{{\normalfont\ttfamily
  arXiv:1711.02505}}\relax
\mciteBstWouldAddEndPuncttrue
\mciteSetBstMidEndSepPunct{\mcitedefaultmidpunct}
{\mcitedefaultendpunct}{\mcitedefaultseppunct}\relax
\EndOfBibitem
\bibitem{LHCB-PAPER-2018-024}
LHCb collaboration, R.~Aaij {\em et~al.},
  \ifthenelse{\boolean{articletitles}}{\emph{{Measurement of the relative
  \mbox{\decay{\Bm}{\Dz / \Dstarz / D^{**0} \mun\neumb}} branching fractions
  using \Bm mesons from $\Bbar{}_{s2}^{*0}$ decays}},
  }{}\href{https://doi.org/10.1103/PhysRevD.99.092009}{Phys.\ Rev.\
  \textbf{D99} (2019) 092009},
  \href{http://arxiv.org/abs/1807.10722}{{\normalfont\ttfamily
  arXiv:1807.10722}}\relax
\mciteBstWouldAddEndPuncttrue
\mciteSetBstMidEndSepPunct{\mcitedefaultmidpunct}
{\mcitedefaultendpunct}{\mcitedefaultseppunct}\relax
\EndOfBibitem
\bibitem{LHCB-PAPER-2019-020}
LHCb collaboration, R.~Aaij {\em et~al.},
  \ifthenelse{\boolean{articletitles}}{\emph{{Measurement of $f_s / f_u$
  variation with proton-proton collision energy and kinematics}},
  }{}\href{https://doi.org/https://doi.org/10.1103/PhysRevLett.124.122002}{Phys.\
  Rev.\ Lett.\  \textbf{124} (2020) 122002},
  \href{http://arxiv.org/abs/1910.09934}{{\normalfont\ttfamily
  arXiv:1910.09934}}\relax
\mciteBstWouldAddEndPuncttrue
\mciteSetBstMidEndSepPunct{\mcitedefaultmidpunct}
{\mcitedefaultendpunct}{\mcitedefaultseppunct}\relax
\EndOfBibitem
\bibitem{Jung:2015yma}
M.~Jung, \ifthenelse{\boolean{articletitles}}{\emph{{Branching ratio
  measurements and isospin violation in B-meson decays}},
  }{}\href{https://doi.org/10.1016/j.physletb.2015.12.024}{Phys.\ Lett.\
  \textbf{B753} (2016) 187},
  \href{http://arxiv.org/abs/1510.03423}{{\normalfont\ttfamily
  arXiv:1510.03423}}\relax
\mciteBstWouldAddEndPuncttrue
\mciteSetBstMidEndSepPunct{\mcitedefaultmidpunct}
{\mcitedefaultendpunct}{\mcitedefaultseppunct}\relax
\EndOfBibitem
\bibitem{efron:1979}
B.~Efron, \ifthenelse{\boolean{articletitles}}{\emph{Bootstrap methods: Another
  look at the jackknife},
  }{}\href{https://doi.org/10.1214/aos/1176344552}{Ann.\ Statist.\  \textbf{7}
  (1979) 1}\relax
\mciteBstWouldAddEndPuncttrue
\mciteSetBstMidEndSepPunct{\mcitedefaultmidpunct}
{\mcitedefaultendpunct}{\mcitedefaultseppunct}\relax
\EndOfBibitem
\bibitem{Kou:2018nap}
Belle~II collaboration, W.~Altmannshofer {\em et~al.},
  \ifthenelse{\boolean{articletitles}}{\emph{{The Belle~II Physics Book}},
  }{}\href{https://doi.org/10.1093/ptep/ptz106}{PTEP \textbf{2019} (2018)
  123C01}, \href{http://arxiv.org/abs/1808.10567}{{\normalfont\ttfamily
  arXiv:1808.10567}}, [Erratum: PTEP 2020, 029201 (2020)]\relax
\mciteBstWouldAddEndPuncttrue
\mciteSetBstMidEndSepPunct{\mcitedefaultmidpunct}
{\mcitedefaultendpunct}{\mcitedefaultseppunct}\relax
\EndOfBibitem
\end{mcitethebibliography}
\ifx\mcitethebibliography\mciteundefinedmacro
\PackageError{LHCb.bst}{mciteplus.sty has not been loaded}
{This bibstyle requires the use of the mciteplus package.}\fi
\providecommand{\href}[2]{#2}

\clearpage
%
% start input ./LHCb_Authorship_29-Oct-2019.tex
%
%
\centerline
{\large\bf LHCb collaboration}
\begin
{flushleft}
\small
R.~Aaij$^{31}$,
C.~Abell{\'a}n~Beteta$^{49}$,
T.~Ackernley$^{59}$,
B.~Adeva$^{45}$,
M.~Adinolfi$^{53}$,
H.~Afsharnia$^{9}$,
C.A.~Aidala$^{80}$,
S.~Aiola$^{25}$,
Z.~Ajaltouni$^{9}$,
S.~Akar$^{66}$
% ...

{\footnotesize \it

% ...
$ ^{9}$Universit{\'e} Clermont Auvergne, CNRS/IN2P3, LPC, Clermont-Ferrand, France\\
% ...
$ ^{25}$INFN Sezione di Milano, Milano, Italy\\
% ...
$ ^{31}$Nikhef National Institute for Subatomic Physics, Amsterdam, Netherlands\\
% ...
$ ^{45}$Instituto Galego de F{\'\i}sica de Altas Enerx{\'\i}as (IGFAE), Universidade de Santiago de Compostela, Santiago de Compostela, Spain\\
% ...
$ ^{49}$Physik-Institut, Universit{\"a}t Z{\"u}rich, Z{\"u}rich, Switzerland\\
% ...
$ ^{53}$H.H. Wills Physics Laboratory, University of Bristol, Bristol, United Kingdom\\
% ...
$ ^{59}$Oliver Lodge Laboratory, University of Liverpool, Liverpool, United Kingdom\\
% ...
$ ^{66}$Los Alamos National Laboratory (LANL), Los Alamos, United States\\
$ ^{67}$Syracuse University, Syracuse, NY, United States\\
% ...
$ ^{80}$University of Michigan, Ann Arbor, United States, associated to $^{67}$\\
\bigskip
% ...
\medskip
}
\end{flushleft} % end input ./LHCb_Authorship_29-Oct-2019.tex
 
\end{document}